\documentclass[12pt]{article}
\pdfoutput=1

\usepackage{array} 
\usepackage{amssymb}
\usepackage{graphics,graphpap}
\usepackage{graphicx}
\usepackage{color}
\usepackage{graphicx}
\usepackage{dcolumn}
\usepackage{epsfig}
\usepackage{epstopdf}
\usepackage{bbm}
\usepackage{amsmath}
\usepackage{amsfonts}
\usepackage{textcomp}
\usepackage{subfigure}
\usepackage{setspace}
\usepackage{slashed}

\usepackage{url}

\usepackage{booktabs}

\usepackage{xcolor,colortbl}
\usepackage[bookmarks,linktocpage]{hyperref}
\usepackage{multirow}

\usepackage{cite}

\newcommand{\beq}{\begin{eqnarray}}
\newcommand{\eeq}{\end{eqnarray}}

\newcommand{\bmp}{\noindent\begin{minipage}{16cm}}
\newcommand{\emp}{\end{minipage}\vskip 7mm} 

\def\drawbox#1#2{\hrule height#2pt
        \hbox{\vrule width#2pt height#1pt \kern#1pt
              \vrule width#2pt}
              \hrule height#2pt}

\def\Asym#1#2{\vcenter{\vbox{\drawbox{#1}{#2}
              \kern-#2pt 
              \drawbox{#1}{#2}}}}

\def\simge{\mathrel{%
   \rlap{\raise 0.511ex \hbox{$>$}}{\lower 0.511ex \hbox{$\sim$}}}}

\def\simle{\mathrel{
   \rlap{\raise 0.511ex \hbox{$<$}}{\lower 0.511ex \hbox{$\sim$}}}}

\def\s#1{\setbox0=\hbox{$#1$}%
\rlap{\ifdim\wd0>.7em\kern.22\wd0\else\kern.1\wd0\fi /}#1}

\newcommand{\rep}[1]{\mathbf{\underline{#1}}}

\newcommand{\SPP}{S^{++}}
\newcommand{\SMM}{S^{--}}

\definecolor{LightPurple}{rgb}{1,0.6,1}
\definecolor{LightRed}{rgb}{1,0.6,0.2}
\definecolor{LightBlue}{rgb}{0.6,0.6,1}

\def\simlt{\stackrel{<}{{}_\sim}}

\begin{document}

\begin{titlepage}
\title{\vspace*{-2.0cm}
\bf\Large
Effective theory of a doubly charged singlet scalar:
complementarity of neutrino physics and the LHC 
\\[5mm]\ }

\author{
Stephen F.~King$^a$\thanks{email: \tt S.F.King@soton.ac.uk},~~~Alexander Merle$^{a,b}$\thanks{email: \tt amerle@mpp.mpg.de},~~~and~~Luca Panizzi$^a$\thanks{email: \tt L.Panizzi@soton.ac.uk}
\\ \\
$^a${\normalsize \it Physics and Astronomy, University of Southampton,}\\
{\normalsize \it Southampton, SO17 1BJ, United Kingdom}\\
\\
$^b${\normalsize \it Max-Planck-Institut f\"ur Physik (Werner-Heisenberg-Institut),}\\
{\normalsize \it F\"ohringer Ring 6, 80805 M\"unchen, Germany}\\
}
\date{\today}
\maketitle
\thispagestyle{empty}

\begin{abstract}
\noindent
We consider a rather minimal extension of the Standard Model involving just one extra particle, namely a single $SU(2)_L$ singlet scalar $S^{++}$ and its antiparticle $S^{--}$. We propose a model independent effective operator, which yields an effective coupling of $S^{\pm \pm}$ to pairs of same sign weak gauge bosons, $W^{\pm} W^{\pm}$. We also allow tree-level couplings of $S^{\pm \pm}$ to pairs of same sign right-handed charged leptons $l^{\pm}_Rl'^{\pm}_R$ of the same or different flavour. We calculate explicitly the resulting two-loop diagrams in the effective theory responsible for neutrino mass and mixing. We propose sets of benchmark points for various $S^{\pm \pm}$ masses and couplings which can yield successful neutrino masses and mixing, consistent with limits on charged lepton flavour violation (LFV) and neutrinoless double beta decay. We discuss the prospects for $S^{\pm \pm}$ discovery at the LHC, for these benchmark points, including single and pair production and decay into same sign leptons plus jets and missing energy. The model represents a minimal example of the complementarity between neutrino physics (including LFV) and the LHC, involving just one new particle, the $S^{\pm \pm}$.
\end{abstract}

\end{titlepage}

\tableofcontents

\section{\label{sec:intro}Introduction}

The origin of neutrino mass and mixing remains one of the most important unanswered questions facing the Standard Model (SM)~\cite{Altarelli:2010gt,Ishimori:2010au,King:2013eh,King:2014nza}. It seems likely that the charged quark and lepton masses originate from Yukawa couplings to one or more Higgs doublets, a belief bolstered by the recent discovery of a Higgs boson with SM properties at around $125.5$~GeV~\cite{Aad:2012tfa,Chatrchyan:2012ufa}. However the exceedingly small values of neutrino masses, $\simlt 1$~eV, and the unique possibility of having a Majorana mass for neutrinos, raises doubts that the same mechanism is responsible for the neutrino mass. Although such Yukawa couplings might play a role in the framework of the seesaw mechanism, where heavy right-handed neutrinos with large Majorana masses are responsible for small effective left-handed neutrino masses, this mechanism is notoriously difficult to test experimentally. Other mechanisms which have been proposed for neutrino mass include $R$-parity violating supersymmetry~\cite{Hirsch:2000ef}, Higgs triplet models~\cite{Magg:1980ut,Lazarides:1980nt,Schechter:1980gr}, or loop models involving additional Higgs doublets and singlets (e.g.~\cite{Ma:2006km,Zee:1985id,Babu:1988ki}), all of which can be tested experimentally (for a review of these different mechanisms see for example~\cite{Bandyopadhyay:2007kx}, and Ref.~\cite{Law:2013dya} for a very systematic study). In particular, such settings can yield very interesting connections between lepton number violating physics and collider phenomenology~\cite{Angel:2012ug,Babu:2001ex,Bonnet:2012kh}, especially if doubly charged scalars are involved (as in the Higgs triplet case~\cite{Perez:2008ha,Chun:2013vma}).

Loop neutrino mass models are often characterised by additional Higgs doublets and singlets. These extra scalar states can in principle be detected indirectly, via low energy high precision experiments due to their contribution to charged lepton flavour violation (LFV) or neutrinoless double beta decay  ($0\nu\beta\beta$), providing a test of the underlying theory of neutrino mass. For example, in the original Zee-Babu model~\cite{Zee:1985id,Babu:1988ki,Babu:2002uu}, involving one singly ($H^+$) and one doubly charged ($S^{++}$) extra scalar singlet, neutrino masses arise via a two-loop diagram. The loop model of Ma~\cite{Ma:2006km} involves an inert Higgs doublet, odd under a discrete symmetry, which does not develop a vacuum expectation value (VEV) but has Yukawa couplings to some of the leptons (involving right-handed neutrinos) and in turn couples to another Higgs doublet which gets a VEV, allowing neutrino mass via a one-loop diagram. The inert Higgs doublet is a Dark Matter candidate, hence the name ``Scotogenic''~\cite{Ma:2006km}. More recently a ``Cocktail'' of the Zee-Babu and Ma models has been proposed~\cite{Gustafsson:2012vj,Gustafsson:2014vpa} involving an extra inert Higgs doublet and a doubly charged Higgs singlet $S^{++}$ but no right-handed neutrinos, where neutrino masses arise due to a three-loop diagram involving also $W$-bosons. In summary, although such loop models do provide a natural explanation for the smallness of neutrino mass and are phenomenologically rich, having predictions for LFV as well new Higgs discovery at the LHC~\cite{Nebot:2007bc,Herrero-Garcia:2014hfa}, they do involve rather many new particles and parameters and are rather computationally complicated, as compared for example to seesaw models. 

\begin{figure}
\centering
\includegraphics[scale=0.5]{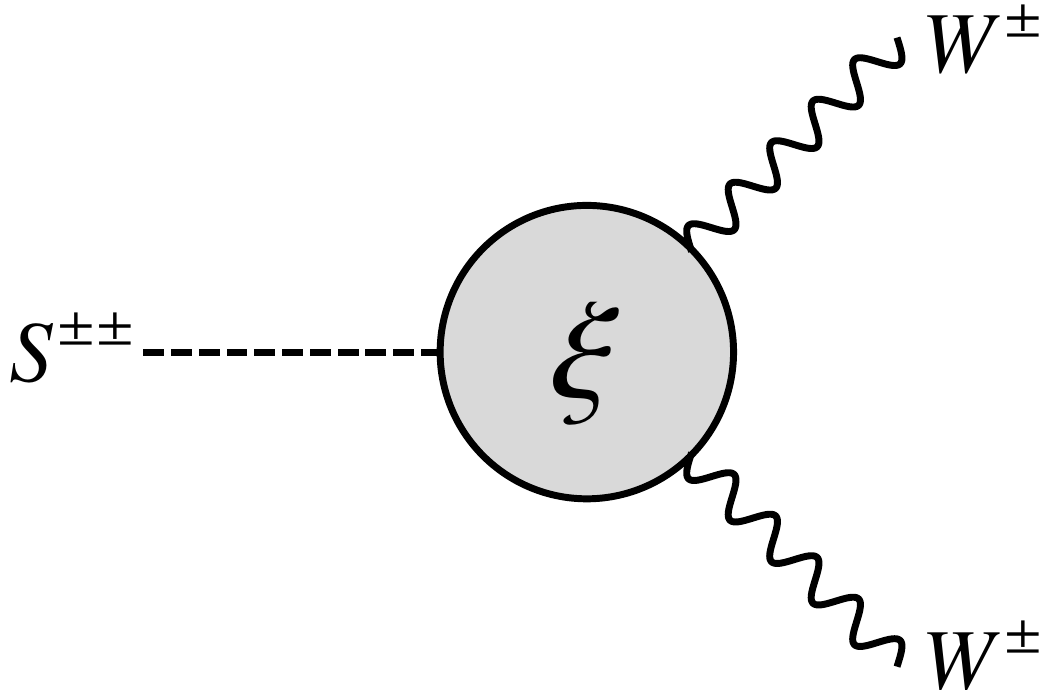}
\caption{\label{fig:vertex}
Effective vertex that connects the doubly charged singlet scalar to two $W$-bosons.
}
\end{figure}

\begin{figure}
\centering
\begin{tabular}{lr}
\includegraphics[scale=0.45]{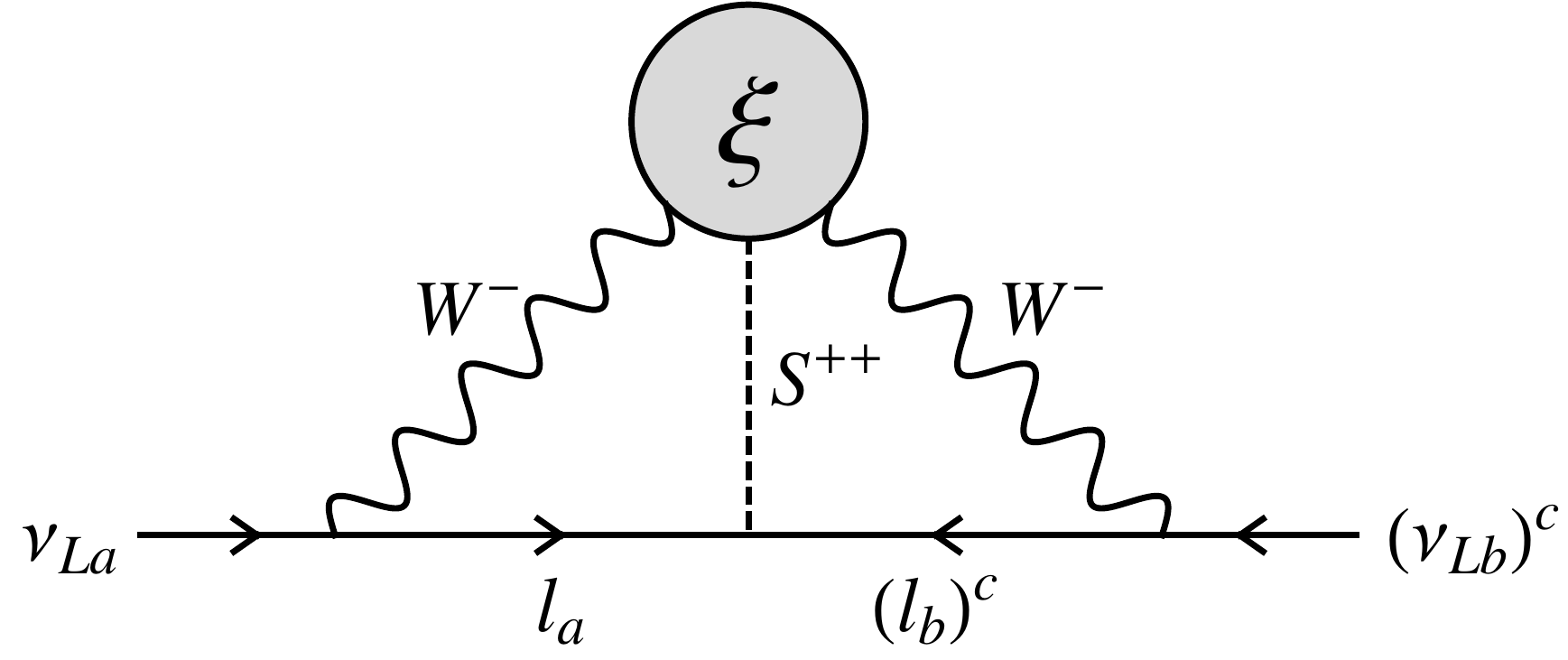} &
\end{tabular}
\caption{\label{fig:neutrino_mass0}
Two-loop diagram for the neutrino mass in our effective model.
}
\end{figure}

In this paper we shall consider an effective theory of neutrino mass involving a rather minimal extension of the Standard Model with just one extra particle, namely a single $SU(2)_L$ singlet scalar $S^{++}$ (and its antiparticle $S^{--}$). We shall propose an effective operator, which yields an effective coupling of $S^{\pm \pm}$ to pairs of same sign weak gauge bosons, $W^{\pm} W^{\pm}$, as in Fig.~\ref{fig:vertex}. As in the Zee-Babu model, for example, we also allow tree-level couplings of $S^{\pm \pm}$ to pairs of same sign right-handed charged leptons $l_a(l_b)^c$ (where $l_a$ is a right-handed charged lepton of flavour $a$ and $(l_b)^c$ is the charge-conjugate of a right-handed charged lepton of flavour $b$). We calculate explicitly the resulting two-loop diagrams in the effective theory responsible for neutrino mass and mixing as shown in Fig.~\ref{fig:neutrino_mass0}. The effective mechanism we propose has several known UV completions~\cite{Chen:2006vn,delAguila:2011gr,Gustafsson:2012vj} corresponding to various possible heavy particles responsible for generating the effective vertex in Fig.~\ref{fig:vertex}. Although our effective theory does not account for Dark Matter, since the only new particle is the $S^{\pm \pm}$ which is electrically charged and unstable, it is entirely possible that a heavy particle appearing in one of these ultraviolet extensions could provide such a stable Dark Matter candidate, such as~\cite{Gustafsson:2012vj,Gustafsson:2014vpa}.


In our approach, the key assumption is that the $S^{\pm \pm}$ is lighter than all the other new particles which appear in the effective vertex, enabling us to develop an effective theory involving just this one particle in addition to the SM, similar to what had been done e.g.\ in Ref.~\cite{Angel:2013hla} for an effective operator connecting two leptons with four quarks. This assumption dramatically reduces the number of input parameters, and enables us to obtain analytic expressions for the two-loop diagrams in Fig.~\ref{fig:neutrino_mass0}. The resulting neutrino mass at the two-loop level thereby arises from a very simple setting featuring only \emph{one} additional particle compared to the SM, plus one effective coupling, together with the Yukawa couplings to charged leptons. Due to the minimality of the extension of the SM, this is one of the simplest neutrino mass models of all, in the sense that it only involves one new particle. We shall focus on the essential connections between different areas of phenomenology (i.e., low energy leptonic physics and high energy colliders), deriving solid conclusions which hold for a large class of models. We shall propose sets of benchmark points for various $S^{\pm \pm}$ masses and couplings which can yield successful neutrino masses and mixing, consistent with limits on charged LFV and neutrinoless double beta decay. We discuss the prospects for $S^{\pm \pm}$ discovery at the LHC for these benchmark points, including single and pair production and decay into same sign leptons plus jets and missing energy. The model represents a minimal example of the complementarity between neutrino physics (including LFV) and the LHC, involving just one new particle, the $S^{\pm \pm}$. The complementarity arises from the fact that one needs data from experiments at the high intensity and high energy frontiers, taken together, in order to fully probe the model.

The layout of the remainder of the paper is as follows. In Sec.~\ref{sec:EFT} we introduce the effective on which we base our study. The resulting neutrino mass matrix is discussed in Sec.~\ref{sec:nu-mass}, before we discuss lepton number violation and lepton flavour violation in Secs.~\ref{sec:0nbb} and~\ref{sec:LFV}, respectively. Afterwards, in Sec.~\ref{sec:benchmarks}, thirty example benchmark points are discussed which are consistent with all experimental bounds available at low energies. We turn to the collider phenomenology of the effective vertex in Sec.~\ref{sec:collider}, where we will show the complementarity between high and low energy tests of the class of models investigated in this work. We finally conclude in Sec.~\ref{sec:conc}. The appendices contain further technical details: in Appendix~\hyperref[sec:app_A]{A} we explain how to calculate the 2-loop neutrino mass resulting from the effective vertex in detail. In Appendix~\hyperref[sec:app_B]{B} we present a new correlation for elements of the light neutrino mass matrix, which arises for a certain category of benchmark points and which is, in principle, also testable. In the final Appendix~\hyperref[sec:app_C]{C}, we give the numerical values of the branching ratios of the doubly charged scalar for all classes of benchmark points found.

\section{\label{sec:EFT}Obtaining the vertex $SWW$ from effective field theory}

We aim to construct the vertex $SWW$ in terms of an effective field theory involving only SM-fields as well as the $SU(2)$ singlet scalar $S$,
\begin{equation}
 S = S^{--} \sim (\rep{1}, -2),
 \label{eq:S}
\end{equation}
using the Gell-Mann/Nishijma relation in the form $Q = T_3 + Y$. It turns out that the lowest mass dimension at which the desired vertex can be realised is by operators of mass dimension $7$,\footnote{Note that the $\mathcal{O}_7$ operator discussed here is implicitly contained in the $\mathcal{O}_9$ effective operator discussed in Refs.~\cite{delAguila:2012nu,Gustafsson:2014vpa}, where the doubly charged scalar $S$ is replaced by the appropriate combination of charged lepton SM fields. However, as will become clear in the text later on, it is interesting for phenomenology \emph{not} to integrate out the scalar.}
\begin{eqnarray}
 \mathcal{O}_7^{(a)} &=& S\ (H \otimes H)_\rep{3}\ [(D_\mu H) \otimes (D^\mu H)]_\rep{3} \supset \rep{1}\,\,\,,\nonumber\\
 \mathcal{O}_7^{(b)} &=& S\ [(D_\mu H) \otimes H]_\rep{1}\ [(D^\mu H) \otimes H]_\rep{1} \supset \rep{1}\,\,\,,\nonumber\\
 \mathcal{O}_7^{(c)} &=& S\ [(D_\mu H) \otimes H]_\rep{3}\ [(D^\mu H) \otimes H]_\rep{3} \supset \rep{1}\,\,\,,
 \label{eq:realizations}
\end{eqnarray}
where $H = (H^+, H^0)^T \sim (\rep2, +1/2)$ is the SM Higgs field, and the subscripts indicate the $SU(2)$ contractions of the respective terms. Note that the hypercharge is automatically conserved, due to $-2 + 4\times 1/2 = 0$. Using well-known $SU(2)$ group theory, in connection with the corresponding Clebsch-Gordan coefficients, one can show that the resulting explicit expressions from all three operators $\mathcal{O}_7^{(a,b,c)}$ are identical. It is easy to see that the decisive Lagrangian term is given by
\begin{equation}
 \frac{\xi}{\Lambda^3} S^{--} [H^+ H^+ (D_\mu H^0) (D^\mu H^0) - 2 H^+ H^0 (D_\mu H^+) (D^\mu h^0) + H^0 H^0 (D_\mu H^+) (D^\mu H^+)] + h.c.,
 \label{eq:L_SWW_general}
\end{equation}
where $\Lambda$ denotes, as usual, the scale where the effective description breaks down. In this paper, we will always assume that this scale $\Lambda$ represents a physical cutoff, which in particular means that any particles in the full theory beyond the scalar $S$ are assumed to be heavier than $\Lambda$. Note, however, that this might not necessarily be the case if the UV-completion is done at loop-level. In that case, further ``light'' particles other than  $S$ might exist which could lead to further diagrams. For our treatment to be applicable, it may thus be necessary to pose some further assumptions on the particle spectrum of the full theory.

We can write the covariant derivative explicitly~\cite{Romao:2012pq},
\begin{equation}
 D_\mu H = [ \partial_\mu + i \frac{g}{\sqrt2} (\tau^+ W^+_\mu + \tau^- W^-_\mu) + ...] H =
 \begin{pmatrix}
 \partial_\mu H^+ + i g W^+_\mu H^0\\
 \partial_\mu H^0 + i g W^-_\mu H^+
 \end{pmatrix} + ...,
 \label{eq:cov_der}
\end{equation}
with $\tau^\pm = (\tau^1 + i \tau^2)/\sqrt2$. Parametrising the SM Higgs field as
\begin{equation}
 H =
 \begin{pmatrix}
 H^+\\
 H^0
 \end{pmatrix} =
 \begin{pmatrix}
 G^+\\
 \frac{v}{\sqrt2} + \frac{h^0 + i G_Z}{\sqrt2}
 \end{pmatrix},
 \label{eq:Higgs}
\end{equation}
where $(G^+, G^-, G_Z)$ are the longitudinal components of the massive vectors $(W^+, W^-, Z^0)$ and the VEV is given by $v = 246$~GeV, it is easy to see that the only relevant component of Eq.~\eqref{eq:cov_der} is given by
\begin{equation}
 D_\mu H^+ = \partial_\mu G^+ + i \frac{g}{\sqrt2} v W^+_\mu.
 \label{eq:int_rel}
\end{equation}
Inserting this into Eq.~\eqref{eq:L_SWW_general} and realising that only the third term $H^0 H^0 (D_\mu H^+) (D^\mu H^+)$ plays a role, one can derive three relevant vertices:
\begin{equation}
 \mathcal{L}_{\rm relevant} = \mathcal{L}_{SGG} + \mathcal{L}_{SGW} + \mathcal{L}_{SWW},
 \label{eq:L_rel}
\end{equation}
with the respective pieces given by
\begin{eqnarray}
 \mathcal{L}_{SGG} &=& - \frac{\xi v^2}{2\Lambda^3} S^{--} (\partial_\mu G^+) (\partial^\mu G^+) + h.c.,\nonumber\\
 \mathcal{L}_{SGW} &=& + \frac{i g \xi v^3}{\sqrt2\Lambda^3} S^{--}  (\partial_\mu G^+) {W^+}^\mu + h.c.,\nonumber \\
 \mathcal{L}_{SWW} &=& - \frac{g^2 \xi v^4}{4\Lambda^3} S^{--} W^+_\mu {W^+}^\mu + h.c.
 \label{eq:L_rel_expl}
\end{eqnarray}
Using the path integral method, it is easy to derive the corresponding Feynman rules:
\begin{center}
\begin{tabular}{m{5cm}l}
\includegraphics[scale=0.45]{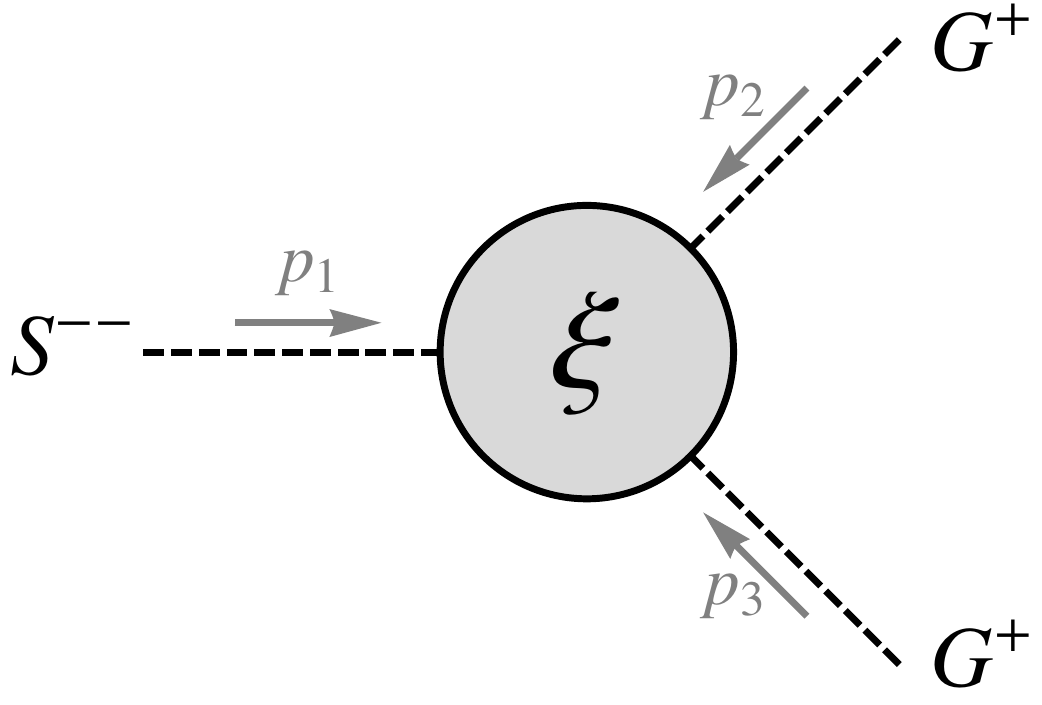} & $= -\frac{i \xi v^2}{\Lambda^3} (p_2 p_3)$\\
\includegraphics[scale=0.45]{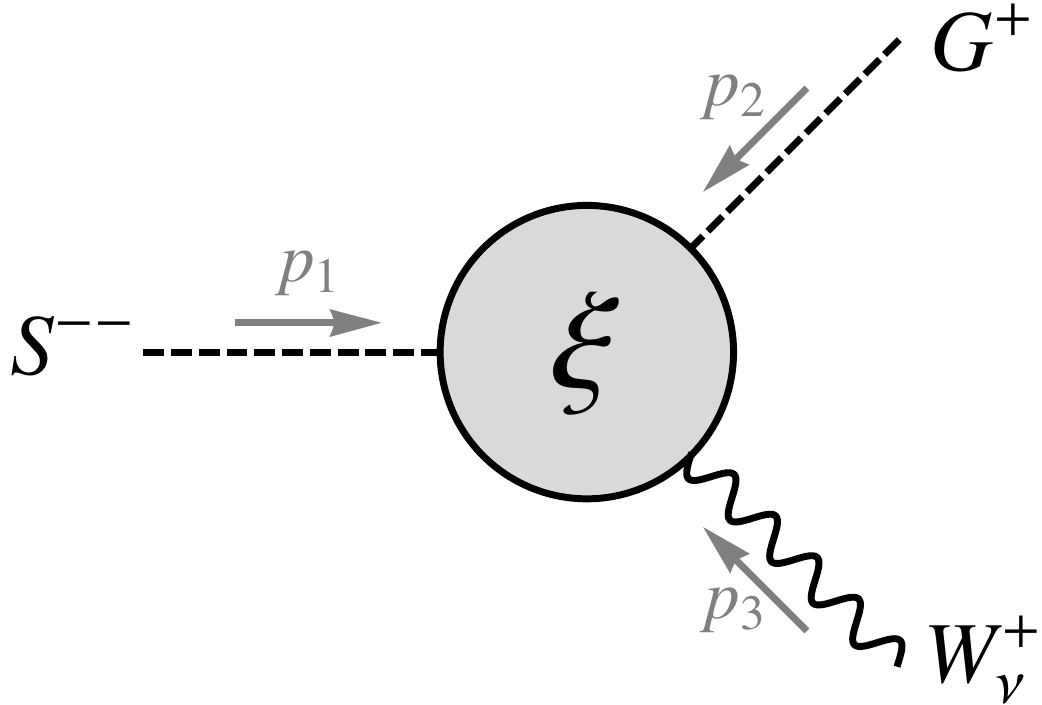} & $= -\frac{i g \xi v^3}{\sqrt2\Lambda^3} p_{2\nu}$\\
\includegraphics[scale=0.45]{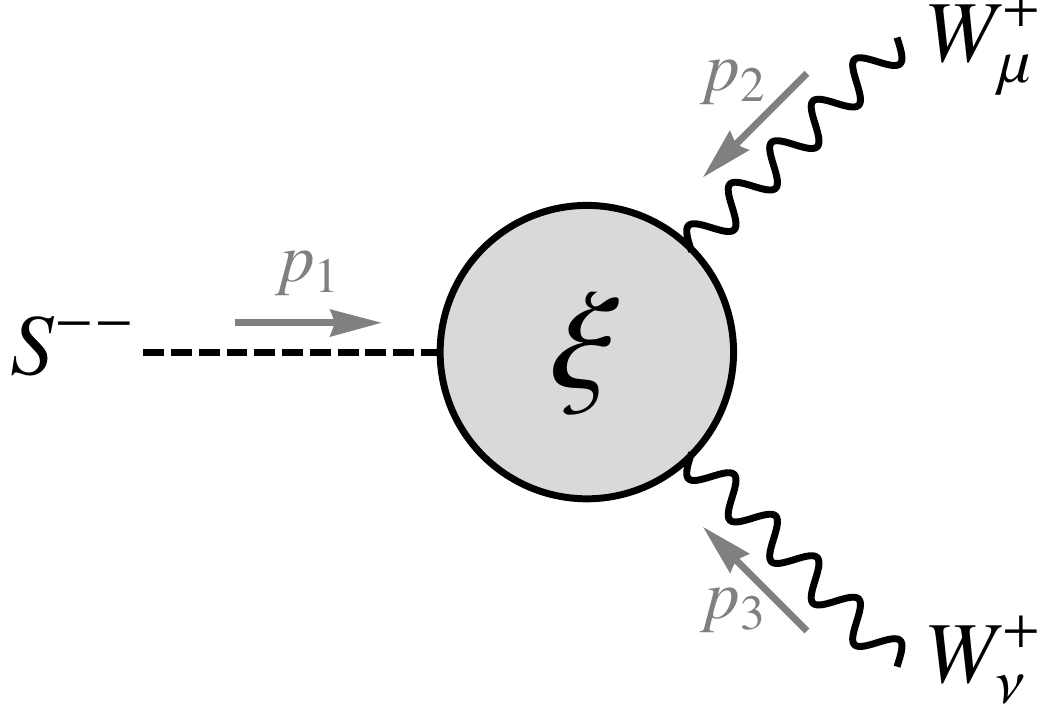} & $= -\frac{i g^2 \xi v^4}{2\Lambda^3} g_{\mu \nu}$
\end{tabular}
\end{center}
In general, all three vertices will contribute. However, for high energies as given in a collider experiment, in fact the very first vertex involving two would-be Goldstone bosons will by far dominate the cross section by virtue of the Goldstone boson equivalence theorem~\cite{Cornwall:1974km,Vayonakis:1976vz,Lee:1977eg,Chanowitz:1985hj}. Hence, e.g.\ for LHC-related studies, effectively the whole relevant part of the Lagrangian will in practice be $\mathcal{L}_{SGG}$ from Eq.~\eqref{eq:L_rel_expl}.

\section{\label{sec:nu-mass}Neutrino mass}

\begin{figure}
\centering
\begin{tabular}{lr}
\includegraphics[scale=0.45]{Mass_Diagram.pdf} &
\includegraphics[scale=0.45]{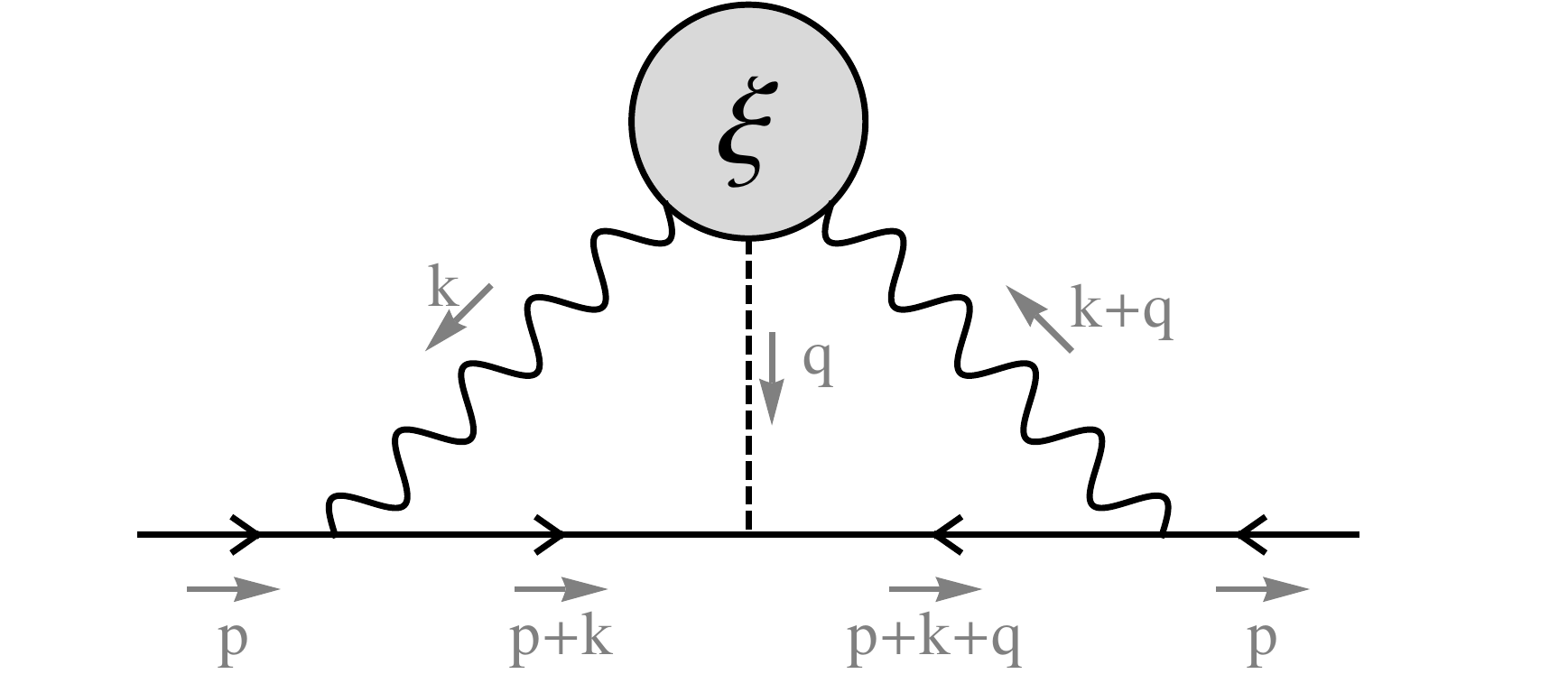}
\end{tabular}
\caption{\label{fig:neutrino_mass}
Two-loop diagram for the neutrino mass (left) and momentum-assignments for its computation (right).
}
\end{figure}

The vertex $SWW$, together with the $S$ coupling to pairs of like-sign right-handed charged leptons, $ f_{ab}S l_a l_b$, where $a,b=e,\mu,\tau$, leads to a neutrino mass at 2-loop level, as displayed in Fig.~\ref{fig:neutrino_mass}. This diagram has been estimated e.g.\ in Ref.~\cite{Chen:2006vn}, and it is intimately related to the Zee-Babu integral~\cite{Zee:1985id,Babu:1988ki,Babu:2002uu,McDonald:2003zj}.
A detailed computation of the neutrino mass can be found in Appendix~\hyperref[sec:app_A]{A}, where we go far beyond the approximations applied in Ref.~\cite{Chen:2006vn} and for the first time present a very detailed computation of the integral (note that a calculation beyond that in Ref.~\cite{Chen:2006vn} had already been performed in Ref.~\cite{delAguila:2011gr}, however, in this work we will go even further). From this appendix, we quote the final formula for the light neutrino mass matrix, see Eq.~\eqref{eq:Mnu_finite_APP}:
\begin{equation}
 \mathcal{M}_{\nu, ab}^\textrm{2-loop} = \frac{2 \xi m_a m_b M_S^2 f_{ab} (1 + \delta_{ab})}{\Lambda^3} \cdot \mathcal{\tilde I}(M_W, M_S,\mu).
 \label{eq:Mnu_finite}
\end{equation}
As shown in the appendix, the quasi exact evaluation of the integral results into a somewhat lengthy formula but, fortunately, it can be approximated by a relatively simple expression, see Eq.~\eqref{eq:int_approx}:
\begin{equation}
 \mathcal{\tilde I}(M_W, M_S,\mu) \simeq \frac{-1}{4 (16 \pi^2)^2} \left\{ \left( 1 - \frac{2}{\rho} \right) \left[ 2 (C_\gamma - L_W)^2 + \frac{\pi^2}{6} +1 \right] + (2 C_\gamma - L_W - L_S)^2 + \frac{\pi^2}{3} +2 \right\}, \nonumber
\end{equation}
where $\rho \equiv M_S^2 / M_W^2$, $L_W \equiv \log (M_W^2 / \mu^2)$, $L_S \equiv \log (M_S^2 / \mu^2)$, and $C_\gamma \equiv 1 - \gamma + \log (4\pi)$ with $\gamma = 0.5771...$ being the Euler-Mascheroni constant. We show in Appendix~\hyperref[sec:app_A]{A} that this is in fact a very good approximation, but more complete expressions could be found.

Note that Eq.~\eqref{eq:Mnu_finite} for the light neutrino mass matrix involves an approximate rescaling symmetry, which we will make use of later on. The trick is to realise that the flavour structure of the matrix is only set by the \emph{relative} magnitudes of the different Yukawa couplings $f_{ab}$ (all other ingredients which depend on flavour indices $a$ and $b$ cannot be varied). Thus, denoting $f_{\rm max} \equiv \text{max}~|f_{ab}|$, the light neutrino mass matrix stays approximately constant when changing the parameters $(\xi, M_S, f_{\rm max}, \Lambda)$, as long as
\begin{equation}
 \frac{\xi M_S^2 f_{\rm max}}{\Lambda^3} \equiv b \simeq {\rm const.}
 \label{eq:Mnu_rescaling_1}
\end{equation}
Of course, there is some variation of $\mathcal{\tilde I}(M_W, M_S,\mu)$ with $M_S$, but first this is only logarithmic\footnote{This is true except for the dependence on $1/\rho$, which is however only small number compared to the $\mathcal{O}(1)$ summand in front of it.} and second it can be at least partially compensated by a corresponding variation of $\mu$ when fitting to experimental data, see Appendix~\hyperref[sec:app_A]{A} and Sec.~\ref{sec:benchmarks} for details. Thus, once we have found a valid benchmark point which yields the correct light neutrino mass matrix, we can find another point with the same neutrino mass matrix prediction but, say, a smaller value of $M_S$ by rescaling the parameters $(\xi, M_S, f_{\rm max}, \Lambda)$ but keeping the ratio given in Eq.~\eqref{eq:Mnu_rescaling_1} constant.

We have to be a bit more careful though. In order to ensure that the EFT is sensible, we need to keep
\begin{equation}
 \xi \sim \mathcal{O}(1)\ \ {\rm and}\ \ M_S \ll \Lambda,
 \label{eq:Mnu_rescaling_2}
\end{equation}
which are two rough but nevertheless important conditions. Note that the latter condition is in practice already fulfilled for, e.g., $M_S \approx \Lambda/5$, since this would effectively mean that at most, even at colliders when the $SWW$ vertex is used to produce a doubly charged boson $S^{\pm \pm}$, the squared momenta in the propagators of the decisive particles are of $\mathcal{O} (M_S^2)$ which, when neglected compared to $\Lambda^2 \approx 25 M_S^2$, only introduces an error of a few percent.

We will furthermore see in Sec.~\ref{sec:0nbb} that the bound $b_{\rm LNV}$ arising from neutrinoless double beta decay yields the constraint
\begin{equation}
 \frac{\xi f_{\rm max}}{M_S^2 \Lambda^3} < b_{\rm LNV},
 \label{eq:Mnu_rescaling_3}
\end{equation}
and in Sec.~\ref{sec:LFV} that the various constraints arising from lepton flavour violating processes and anomalous leptonic magnetic moments are all of the same form, so that the strongest of them yields a final condition
\begin{equation}
 \frac{f_{\rm max}^2}{M_S^2} < b_{\rm LFV}.
 \label{eq:Mnu_rescaling_4}
\end{equation}
As long as the conditions from Eqs.~\eqref{eq:Mnu_rescaling_1} to~\eqref{eq:Mnu_rescaling_4} are fulfilled, any set of parameters $(\xi, M_S, f_{\rm max}, \Lambda)$ yields the same prediction for the light neutrino mass. This will turn out extremely useful later on when we aim to investigate how one can use LHC data to constrain or even exclude points with a different scalar mass $M_S$ leading to the same predictions for what concerns neutrinos, cf.\ Sec.~\ref{sec:collider}.

\section{\label{sec:0nbb}Non-standard contributions to neutrinoless double beta decay}

\begin{figure}
\centering
\includegraphics[scale=0.5]{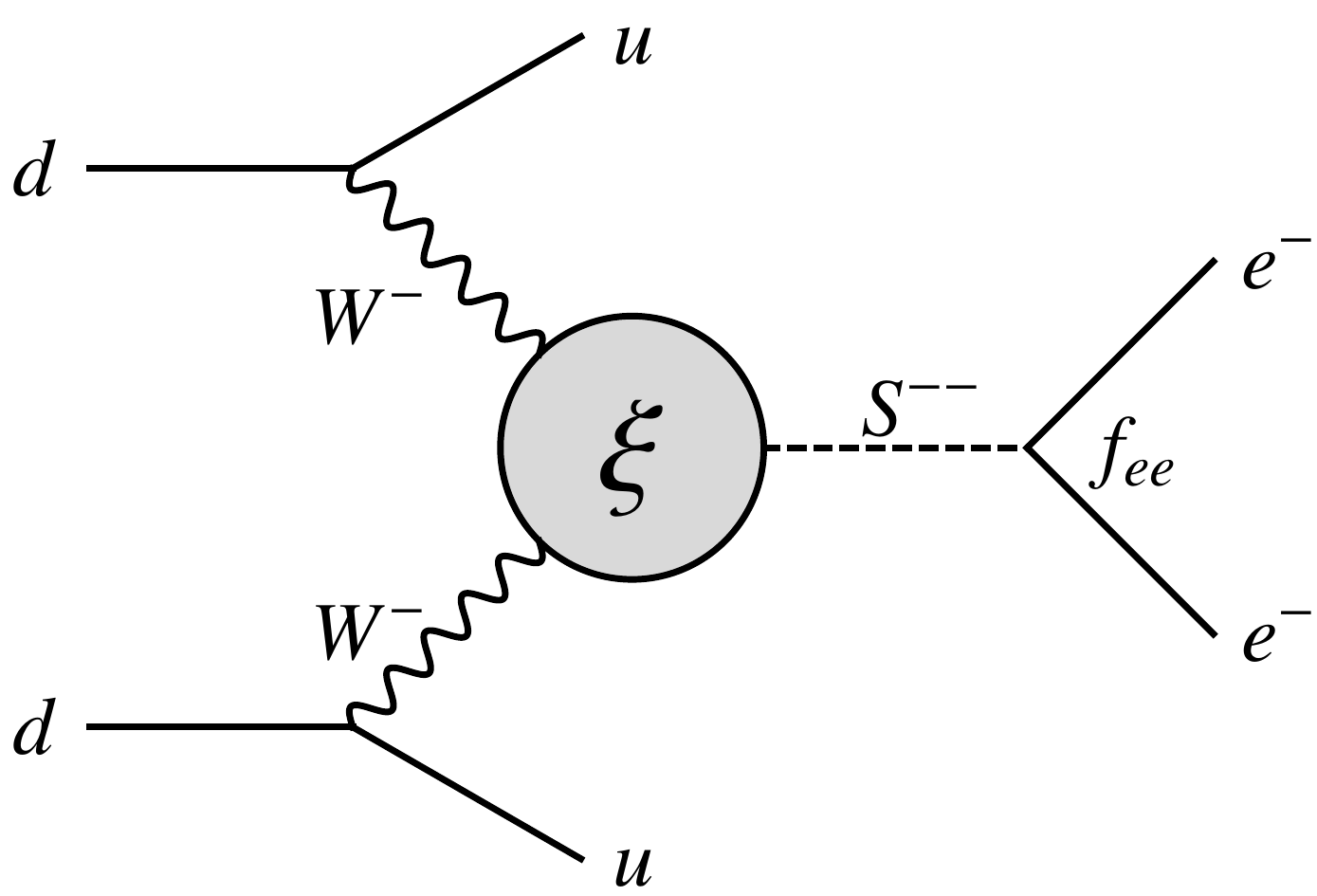}
\caption{\label{fig:0nbb}
The non-standard contribution to $0\nu\beta\beta$.
}
\end{figure}

Since the boson $S^{--}$ can couple to both, $W^- W^-$ and $e^- e^-$, our setting can lead to non-standard contributions to lepton number violating (LNV) processes, and in particular to neutrinoless double beta decay ($0\nu\beta\beta$), cf.\ Fig.~\ref{fig:0nbb}, which happens inside a nucleus. In order to get a bound from this contribution, it is decisive to understand that $0\nu\beta\beta$ is a low-energy process, to be precise, the typical momentum scale of nucleons inside the nucleus is of $\mathcal{O}(100~{\rm MeV})$ (see Ref.~\cite{Rodejohann:2011mu} and references therein). The masses $M_W$ and $M_S$ of the $W$-boson and the scalar $S$ are much larger than the characteristic energy scale of the process, which means that we can subsequently integrate out both of them to obtain a point-like non-standard effective operator contributing to neutrinoless double beta decay.

The bounds on such point-like non-standard contributions have first been calculated in Ref.~\cite{Pas:2000vn}, but they have been recently updated by Ref.~\cite{Bergstrom:2011dt}, using in particular improved values for the corresponding nuclear matrix elements associated with the short-range contributions~\cite{Simkovic:2010ka}. Integrating out $S$ and $W$ (in that order) from the Lagrangian from Eq.~\eqref{eq:L_rel_expl}, together with the contributions of the $S$-coupling to right-handed charged leptons, $\mathcal{L}_{Sll} = f_{ab} S^{--} \overline{l_a} P_L l_b^c + h.c.$, the ordinary weak interactions of quarks, and the mass terms for both bosons, we obtain the following effective Lagrangian,
\begin{equation}
 \mathcal{L}^{\rm effective}_{0\nu\beta\beta} = \frac{\xi f_{ee}}{4 M_S^2 \Lambda^3} J_{L \mu} J_L^\mu j_{L},
 \label{eq:L_0nbb}
\end{equation}
where $J_L^\mu \equiv \overline{u} \gamma^\mu (1- \gamma_5) d$ and $j_L \equiv \overline{e} (1- \gamma_5) e^c$. The coefficient has to be compared with the general coefficient $\frac{G_F^2}{2 m_p} \epsilon^{LLL}_3$, where $G_F = 1.166\cdot 10^{-5} {\rm GeV}^{-2}$ is the Fermi constant and $m_p$ is the proton mass. In Ref.~\cite{Bergstrom:2011dt}, a bound on the quantity $\epsilon^{LLL}_3$ has been found for the case (among others) where $0\nu\beta\beta$-searches only yield a limit, which corresponds to the situation today. Rescaling that bound to include the newest GERDA results of $T^{0\nu\beta\beta}_{1/2}({\rm Ge}) > 2.1\cdot 10^{25}$~y at $90\%$~C.L.~\cite{Agostini:2013mzu}, we obtain\footnote{Note that the upper bound on $\epsilon^{LLL}_3$ actually gets \emph{weaker} compared to the hypothetical scenario discussed in Ref.~\cite{Bergstrom:2011dt}, which assumes no detection of $0\nu\beta\beta$ even after phase~III of the GERDA experiment.}
\begin{equation}
 \epsilon^{LLL}_3 = \frac{\xi f_{ee} m_p}{2 G_F^2 M_S^2 \Lambda^3} < 1.4\cdot 10^{-8}\ \ \ \text{at $90\%$~C.L.}
 \label{eq:bound_0nbb}
\end{equation}
Getting rid of all known pieces, one obtains
\begin{equation}
 \frac{\xi f_{ee}}{M_S^2 \Lambda^3} < \frac{4.0\cdot 10^{-3}}{{\rm TeV}^5}\ \ \ \text{at $90\%$~C.L.}
 \label{eq:bound_0nbb_expl}
\end{equation}
Indeed, this imposes a strong bound on the model. However, since $0\nu\beta\beta$ involves only electrons, one can trivially fulfill it by choosing the LNV-couplings to two electrons to vanish, $f_{ee} \equiv 0$.

\section{\label{sec:LFV}Bounds from lepton flavour violation and dipole moments}

The doubly charged scalar can couple in a flavour violating way and thus lead to LFV processes such as $\mu \to e \gamma$ or $\mu \to 3 e$. Furthermore, these particles can contribute to the anomalous magnetic moment of leptons. This is very similar to what happens in the Zee-Babu model, with the exception that for that case, also a singly charged scalar exists, which can complicate things. An extensive collection of LFV-bounds in the Zee-Babu model has been presented in Ref.~\cite{Nebot:2007bc},\footnote{Note that very recently, an update~\cite{Herrero-Garcia:2014hfa} of Ref.~\cite{Nebot:2007bc} became available. We nevertheless decided to use the values that we ourselves had updated to ensure consistency in the process of updating the bounds. However, our numbers do not differ in any significant way from the ones used in~\cite{Herrero-Garcia:2014hfa}.} which we will update in the following. Due to the similarities in formalism, we can even take some of the bounds at face value and only need minor modifications.

\begin{table}
\hspace{-0.7cm}
\begin{tabular}{|l|l|l|}\hline 
Process (Tree) & Experimental limit ($90\%$~C.L.) & Resulting bound ($90\%$~C.L.)\\\hline\hline
$\mu^- \rightarrow e^+ e^- e^-$ & BR$< 1.0\cdot 10^{-12}$~\cite{Nebot:2007bc} & $|f_{e\mu} f_{ee}^*| < 2.3\times10^{-5} M_S^2 [{\rm TeV}]$\\\hline
$\tau^- \rightarrow e^+ e^- e^-$ & BR$< 2.7\cdot 10^{-8}$~\cite{Beringer:1900zz} & $|f_{e\tau} f_{ee}^*| < 0.0087 M_S^2 [{\rm TeV}]$\\\hline
$\tau^- \rightarrow e^+ e^- \mu^-$ & BR$< 1.8\cdot 10^{-8}$~\cite{Beringer:1900zz} & $|f_{e\tau} f_{e\mu}^*| < 0.005 M_S^2 [{\rm TeV}]$\\\hline
$\tau^- \rightarrow e^+ \mu^- \mu^-$ & BR$< 1.7\cdot 10^{-8}$~\cite{Beringer:1900zz} & $|f_{e\tau} f_{\mu\mu}^*| < 0.007 M_S^2 [{\rm TeV}]$\\\hline
$\tau^- \rightarrow \mu^+ e^- e^-$ & BR$< 1.5\cdot 10^{-8}$~\cite{Beringer:1900zz} & $|f_{\mu\tau} f_{ee}^*| < 0.007 M_S^2 [{\rm TeV}]$\\\hline
$\tau^- \rightarrow \mu^+ e^- \mu^-$ & BR$< 2.7\cdot 10^{-8}$~\cite{Beringer:1900zz} & $|f_{\mu\tau} f_{e\mu}^*| < 0.007 M_S^2 [{\rm TeV}]$\\\hline
$\tau^- \rightarrow \mu^+ \mu^- \mu^-$ & BR$< 2.1\cdot 10^{-8}$~\cite{Beringer:1900zz} & $|f_{\mu\tau} f_{\mu\mu}^*| < 0.0081 M_S^2 [{\rm TeV}]$\\\hline
$\mu^+ e^- \rightarrow \mu^- e^+$ & $G_{M \overline{M}} < 0.003 G_F$~\cite{Nebot:2007bc} & $|f_{ee} f_{\mu\mu}^*| < 0.2 M_S^2 [{\rm TeV}]$\\\hline\hline
Process (Loop) & Experimental limit ($90\%$~C.L.) & Resulting bound ($90\%$~C.L.)\\\hline\hline
$(g-2)_e$ & $\delta a_{e} = (1.2\pm 1.0)\cdot 10^{-11}$~\cite{Nebot:2007bc} & $|f_{ee}|^2 + |f_{e\mu}|^2 + |f_{e\tau}|^2 < 1.4\cdot 10^3 M_S^2 [{\rm TeV}]$\\\hline
$(g-2)_\mu$ & $\delta a_{\mu} = (2.1\pm 1.0)\cdot 10^{-9}$~\cite{Nebot:2007bc} & $|f_{e\mu}|^2 + |f_{\mu\mu}|^2 + |f_{\mu\tau}|^2 < 2.0 M_S^2 [{\rm TeV}]$\\\hline
$\mu \to e\gamma$ & BR$< 5.7 \cdot 10^{-13}$~\cite{MEG_EPS} & $|f_{ee}^* f_{e\mu} + f_{e\mu}^* f_{\mu\mu} + f_{e\tau}^* f_{\mu\tau}| < 3.2\cdot 10^{-4} M_S^2 [{\rm TeV}]$\\\hline
$\tau \to e \gamma$ & BR$< 3.3\cdot 10^{-8}$~\cite{Beringer:1900zz} & $|f_{ee}^* f_{e\tau} + f_{e\mu}^* f_{\mu\tau} + f_{e\tau}^* f_{\tau\tau}| < 0.18 M_S^2 [{\rm TeV}]$\\\hline
$\tau \to \mu \gamma$ & BR$< 4.4\cdot 10^{-8}$~\cite{Beringer:1900zz} & $|f_{e\mu}^* f_{e\tau} + f_{\mu\mu}^* f_{\mu\tau} + f_{\mu\tau}^* f_{\tau\tau}| < 0.21 M_S^2 [{\rm TeV}]$\\\hline
\end{tabular}
\caption{\label{tab:LFV}LFV-bounds resulting from various tree-level and 1-loop processes.}
\end{table}

In Tab.~\ref{tab:LFV} we have in particular updated the limit on ${\rm BR}(\mu \to e \gamma)$, which is bound to be smaller than $5.7\cdot 10^{-13}$ at $90\%$~C.L.\ according to the newest results from the MEG experiment~\cite{MEG_EPS}. Most of the other limits have been improved as well, but less dramatically~\cite{Beringer:1900zz}.

\section{\label{sec:benchmarks}Benchmark points}

The next step is to find benchmark scenarios where all bounds are fulfilled and a suitable light neutrino mass matrix is reproduced. In general, this is not an easy task since the model presented is in fact quite constrained for three reasons:
\begin{enumerate}

\item Due to the proportionality to products of charged lepton masses, the structure of the light neutrino mass matrix, cf.\ Eq.~\eqref{eq:Mnu_finite}, generically imposes a pattern that is not close to an ``anarchic'' mixing scenario~\cite{Haba:2000be} as resembled by the leptonic mixing parameters, which requires certain hierarchies in the couplings $f_{ab}$ to compensate for that. Furthermore, this structure (together with the power $\Lambda^{-3}$) suppresses the entries in whole light neutrino mass matrix. This makes it non-trivial to generate a mass scale of $\mathcal{O}(\sqrt{\Delta m^2_{31}})$, as required by phenomenology.

\item The bound from $0\nu\beta\beta$, cf.\ Eq.~\eqref{eq:bound_0nbb_expl}, poses a strong condition on the product $\xi f_{ee}$. In case $f_{ee}$ is not small enough, this may require a very small $\xi$ for this bound to be fulfilled, which however also suppresses the light neutrino masses. Alternatively, this bound may push $M_S$ to values so large that they are not desired if we want to have visible phenomenology at LHC, cf.\ Sec.~\ref{sec:collider}.

\item The bound from $\mu \to e \gamma$, cf.\ Tab.~\ref{tab:LFV}, poses a strong upper bound on a certain combination of couplings, $|f_{ee}^* f_{e\mu} + f_{e\mu}^* f_{\mu\mu} + f_{e\tau}^* f_{\mu\tau}|$. If this combination is not close to zero then again $M_S$ needs to be too large, thereby potentially destroying any interesting collider phenomenology.

\end{enumerate}

This problem has been investigated in a similar setting in Ref.~\cite{Gustafsson:2014vpa}. One could, for example, choose $f_{ee} \simeq 0$~\footnote{Note that this condition also leads to $\mathcal{M}_{\nu, ee} \simeq 0$, thereby simultaneously killing the light neutrino exchange and the scalar contributions to $0\nu\beta\beta$. It furthermore induces correlations between the neutrino mixing parameters~\cite{Lindner:2005kr} and it also implies that normal ordering must be present.} and $f_{e\mu} \simeq 0$ (or $f_{e\tau} \simeq 0$)\footnote{Also these conditions impose correlations $\mathcal{M}_{\nu, e\mu} \simeq 0$ or $\mathcal{M}_{\nu, e\tau} \simeq 0$ on the neutrino mixing parameters, which could also have observable consequences~\cite{Merle:2006du}.} in order to avoid the strong bounds mentioned in points~2 and~3. Indeed, if these conditions are at least approximately fulfilled, then the tension on the parameter space can be considerably reduced. Alternatively, one could in addition to $f_{ee} \simeq 0$ also require that
\begin{equation}
 f_{e\mu} \simeq - \frac{f_{\mu\tau}^*}{f_{\mu\mu}^*} f_{e\tau},
 \label{eq:condition}
\end{equation}
in which case $|f_{ee}^* f_{e\mu} + f_{e\mu}^* f_{\mu\mu} + f_{e\tau}^* f_{\mu\tau}| \simeq 0$ holds, too. Note that, by virtue of Eq.~\eqref{eq:Mnu_finite}, this also implies an interesting correlation between elements of the light neutrino mass matrix and two of the charged lepton masses, which is briefly discussed in Appendix~\hyperref[sec:app_B]{B}.

In order to find suitable benchmark points, we have departed from these conditions. To be precise, we have numerically found solutions based on the initial assumptions $(f_{ee}, f_{e\mu}) \simeq (0, - f_{\mu\tau}^* f_{e\tau}/f_{\mu\mu}^*)$, $(f_{ee}, f_{e\mu}) \simeq (0, 0)$, $(f_{ee}, f_{e\tau}) \simeq (0,0)$, $(f_{ee}, f_{\mu \mu}) \simeq (0, 0)$, and $(f_{ee}, f_{\mu \tau}) \simeq (0,0)$,\footnote{Note that $f_{\tau \tau} \simeq 0$ would \emph{not} be a good option, since it would have a strong tendency to yield too small neutrino masses.} which have then been partially relaxed while keeping all experimental constrains alive. By this procedure, we have been able to find 30 benchmark points which all agree with the $3\sigma$ neutrino oscillation parameters as well as with all bounds from low-energy charged lepton observables, neutrinoless double beta decay, and with Eq.~\eqref{eq:Mnu_rescaling_2}.

Note that the relations we started with do not necessarily need to be exactly fulfilled (and in fact they are not for some of the points found), but being somewhat close to them considerably reduces the tension in the parameter space. Of course one may criticise that these relations, even if fulfilled only approximately, are at this stage only phenomenological postulates. While this criticism is certainly justified, one could alternatively try to explain them by, e.g., a flavour symmetry~\cite{Altarelli:2010gt,Ishimori:2010au,King:2013eh,King:2014nza}. On the other hand, one could just take them for granted since after all they are enforced by the experimental bounds.

In the following, we will have a detailed look at the low-energy phenomenology of the 30 benchmark points we have found. First of all, we have found three categories of points:
\begin{itemize}

\item {\bf red} (light gray) points: two texture zeros $f_{ee} \simeq 0$ \& $f_{e\tau} \simeq 0$

This category of points features two couplings, $f_{ee}$ and $f_{e\tau}$, which are essentially zero. Due to the $m_e^2$-suppression in the $\mathcal{M}_{\nu, ee}$-element of the neutrino mass matrix from Eq.~\eqref{eq:Mnu_finite}, this yields, practically, $\mathcal{M}_{\nu, ee} = 0$. Note that this means nothing else than that the standard light neutrino exchange contribution to neutrinoless double beta decay, whose amplitude is proportional the so-called effective mass $m_{ee} = m_1 c_{12}^2 c_{13}^2 + \sqrt{m_1^2 + \Delta m^2_\odot} s_{12}^2 c_{13}^2 e^{i\alpha_{21}} + \sqrt{m_1^2 + \Delta m^2_A} s_{13}^2 e^{i(\alpha_{31}-2\delta)} \equiv \mathcal{M}_{\nu, ee}$, vanishes in our case.\footnote{Note that also the contribution comes from the non-standard diagram containing the doubly charged scalar, discussed in Sec.~\ref{sec:0nbb}, is also suppressed due to $f_{ee} \simeq 0$. Thus, this category of points yield Majorana neutrinos with, however, $0\nu\beta\beta$ unfortunately not being observable.} A vanishing effective mass implies \emph{normal ordering} (NO) of the light neutrino masses, $m_1 < m_2 < m_3$, and a lightest neutrino mass around $5$~meV~\cite{Lindner:2005kr} (see also Ref.~\cite{King:2013psa} for a more recent version of the corresponding plot), which are two clear predictions of this setting. Furthermore, $f_{e\tau} \simeq 0$ implies that $\mathcal{M}_{\nu, e\tau} \simeq 0$, too, which imposes further restrictions on the neutrino oscillation parameters and Majorana phases but not on the light neutrino mass scale~\cite{Merle:2006du}.

\item {\bf purple} (medium gray) points: one texture zero $f_{ee} \simeq 0$ \& correlation Eq.~\eqref{eq:condition}

This category of points also features $\mathcal{M}_{\nu, ee} = 0$ and thus NO and $m_1 \sim 5$~meV. However, the strong constraint from $\mu \to e\gamma$ is instead evaded by having $f_{e\mu} f_{\mu\mu}^* + f_{e\tau} f_{\mu\tau}^* \simeq 0$. As we will see, this category of points is much more predictive than the red points in what concerns lepton flavour phenomenology, due to the conditions imposed by the correlation. For example, a sizable coupling $f_{e\tau}$ (which is not so strongly constrained by experiments) enforces a sizable coupling $f_{e\mu}$ if the above correlation is to be fulfilled, and vice versa. It is non-trivial to see from the formulae how the neutrino oscillation parameters and phases are constrained, but our numerical analysis presented below will easily reveal these tendencies. Note further that, just as for the red points, $0\nu\beta\beta$ is not observable due to $f_{ee} \simeq 0$ suppressing both the standard and non-standard contributions.

\item {\bf blue} (dark gray) points: only the correlation Eq.~\eqref{eq:condition}

These points only obey the correlation $f_{e\mu} f_{\mu\mu}^* + f_{e\tau} f_{\mu\tau}^* \simeq 0$, but they do \emph{not} feature a tiny coupling $f_{ee}$. In fact, this coupling can be relatively large, of $\mathcal{O}(0.1)$, which makes these points most interesting from the LNV point of view. First of all, note that we can even for a sizable $f_{ee}$ assume that $\mathcal{M}_{\nu, ee} \simeq 0$, due to the $m_e^2$-suppression of that matrix element. This means that the standard light neutrino exchange contribution to $0\nu\beta\beta$ still vanishes and thus NO and a lightest neutrino mass of roughly $5$~meV is implied, similar to the scenario discussed in Ref.~\cite{Gustafsson:2014vpa}. However, this time, the non-standard contribution from the doubly charged scalar $S^{--}$, cf.\ Sec.~\ref{sec:0nbb}, is sizable. In fact, as Eq.~\eqref{eq:bound_0nbb_expl} reveals, the doubly charged scalar mass $M_S$ and the cutoff $\Lambda$ must be sizable in order to not be in conflict with the current experimental upper bound on $0\nu\beta\beta$. Thus, this category of points has a strong tendency to be falsifiable by a dedicated experiment. However, since at the same time a larger $M_S$ is required, these points are not so desirable from the point of view of collider phenomenology, cf.\ Sec.~\ref{sec:collider}.

\end{itemize}

\begin{table}[ht!]
\centering
\begin{tabular}{|l||l|l|l|}\hline 
 & Best-fit point (red) & Best purple point &  Best blue point\\\hline\hline
$\xi$ & $5.02$ & $6.38$ & $3.39$\\\hline
$M_S$ & $164.5$~GeV & $364.6$~GeV & $626.0$~GeV\\\hline
$\Lambda$ & $905.9$~GeV & $2505.1$~GeV & $5094.7$~GeV\\\hline
$\mu$ & $0.664$~keV & $1.36\cdot 10^3$~TeV & $33.5$~eV \\\hline
${\rm Re}f_{ee}$ & $8.32\cdot 10^{-17} \simeq 0$ & $7.21\cdot 10^{-16} \simeq 0$ & $-0.0467$\\\hline
${\rm Im}f_{ee}$ & $-4.62\cdot 10^{-17} \simeq 0$ & $0$ & $0.00881$\\\hline
${\rm Re}f_{e\mu}$ & $0.00804$ & $1.63\cdot 10^{-4}$ & $-9.25\cdot 10^{-6}$\\\hline
${\rm Im}f_{e\mu}$ & $-5.34\cdot 10^{-4}$ & $9.09\cdot 10^{-7}$ & $-1.64\cdot 10^{-5}$\\\hline
${\rm Re}f_{e\tau}$ & $4.25\cdot 10^{-20} \simeq 0$ & $-4.08\cdot 10^{-3}$ & $-0.00562$\\\hline
${\rm Im}f_{e\tau}$ & $4.79\cdot 10^{-20} \simeq 0$ & $3.55\cdot 10^{-4}$ & $-7.20\cdot 10^{-6}$\\\hline
${\rm Re}f_{\mu\mu}$ & $4.75\cdot 10^{-5}$ & $3.60\cdot 10^{-4}$ & $6.90\cdot 10^{-4}$\\\hline
${\rm Im}f_{\mu\mu}$ & $-2.09\cdot 10^{-5}$ & $-3.32\cdot 10^{-4}$ & $1.53\cdot 10^{-7}$\\\hline
${\rm Re}f_{\mu\tau}$ & $4.55\cdot 10^{-6}$ & $2.61\cdot 10^{-5}$ & $5.23\cdot 10^{-5}$\\\hline
${\rm Im}f_{\mu\tau}$ & $-2.44\cdot 10^{-6}$ & $-2.90\cdot 10^{-5}$ & $-1.18\cdot 10^{-8}$\\\hline
${\rm Re}f_{\tau\tau}$ & $1.84\cdot 10^{-7}$ & $8.50\cdot 10^{-7}$ & $1.55\cdot 10^{-6}$\\\hline
${\rm Im}f_{\tau\tau}$ & $-8.16\cdot 10^{-8}$ & $-7.17\cdot 10^{-7}$ & $7.86\cdot 10^{-10} \simeq 0$\\\hline\hline
$m_1$ & $5.32$~meV & $6.87$~meV & $6.84$~meV\\\hline
$m_2$ & $10.2$~meV & $10.9$~meV & $10.9$~meV\\\hline
$m_3$ & $49.9$~meV & $48.5$~meV & $48.2$~meV\\\hline
$\Delta m^2_\odot$ & $7.64\cdot 10^{-5}~{\rm eV}^2 \in 2\sigma$ & $7.17\cdot 10^{-5}~{\rm eV}^2 \in 2\sigma$ & $7.19\cdot 10^{-5}~{\rm eV}^2 \in 3\sigma$\\\hline
$\Delta m^2_A$ & $2.47\cdot 10^{-3}~{\rm eV}^2 \in 3\sigma$ & $2.31\cdot 10^{-3}~{\rm eV}^2 \in 2\sigma$ & $2.28\cdot 10^{-3}~{\rm eV}^2 \in 3\sigma$\\\hline
$\sin^2 \theta_{12}$ & $0.286 \in 2\sigma$ & $0.345 \in 3\sigma$ & $0.332 \in 3\sigma$\\\hline
$\sin^2 \theta_{13}$ & $0.0185 \in 3\sigma$ & $0.0188 \in 3\sigma$ & $0.0183 \in 3\sigma$\\\hline
$\sin^2 \theta_{23}$ & $0.493 \in 2\sigma$ & $0.604 \in 2\sigma$ & $0.612 \in 2\sigma$\\\hline
$\delta$ & $1.88\pi$ & $0.780\pi$ & $0.999\pi$\\\hline
$\alpha_{21}$ & $0.968\pi$ & $0.955\pi$ & $1.00\pi$\\\hline
$\alpha_{31}$ & $0.857\pi$ & $0.753\pi$ & $1.00\pi$\\\hline
\end{tabular}
\caption{\label{tab:points} Three example benchmark points, where from each category of points the best one has been chosen.}
\end{table}

\begin{table}[ht!]
\centering
\begin{tabular}{|l||l|l|l|}\hline 
 & Best-fit point (red) & Best purple point &  Best blue point\\\hline\hline
$0\nu\beta\beta:\ T_{1/2}^{-1}~[{\rm yrs}^{-1}]$ & $1.61\cdot 10^{-48}\simeq 0$ & $1.38\cdot 10^{-50}\simeq 0$ & $2.76\cdot 10^{-26}$\\\hline\hline
${\rm BR}(\mu^- \rightarrow e^+ e^- e^-)$ & $1.52\cdot 10^{-36}\simeq 0$ & $1.48\cdot 10^{-39}\simeq 0$ & $9.89\cdot 10^{-15}$\\\hline
${\rm BR}(\tau^- \rightarrow e^+ e^- e^-)$ & $1.81\cdot 10^{-71}\simeq 0$ & $1.76\cdot 10^{-37}\simeq 0$ & $1.66\cdot 10^{-10}$\\\hline
${\rm BR}(\tau^- \rightarrow e^+ e^- \mu^-)$ & $2.62\cdot 10^{-43}\simeq 0$ & $1.82\cdot 10^{-14}$ & $5.25\cdot 10^{-17}$\\\hline
${\rm BR}(\tau^- \rightarrow e^+ \mu^- \mu^-)$ & $5.23\cdot 10^{-48}\simeq 0$ & $7.93\cdot 10^{-14}$ & $3.40\cdot 10^{-14}$\\\hline
${\rm BR}(\tau^- \rightarrow \mu^+ e^- e^-)$ & $1.01\cdot 10^{-43}\simeq 0$ & $1.37\cdot 10^{-41}\simeq 0$ & $1.23\cdot 10^{-14}$\\\hline
${\rm BR}(\tau^- \rightarrow \mu^+ e^- \mu^-)$ & $1.30\cdot 10^{-15}$ & $1.26\cdot 10^{-18}$ & $3.48\cdot 10^{-21}\simeq 0$\\\hline
${\rm BR}(\tau^- \rightarrow \mu^+ \mu^- \mu^-)$ & $3.13\cdot 10^{-20}$ & $6.63\cdot 10^{-18}$ & $2.72\cdot 10^{-18}$\\\hline\hline
$\mu^+ e^- \rightarrow \mu^- e^+:\ G_{M \overline{M}}/G_F$ & $2.74\cdot 10^{-22}\simeq 0$ & $3.99\cdot 10^{-21}\simeq 0$ & $1.26\cdot 10^{-7}$\\\hline\hline
$(g-2)_e$ & $2.06\cdot 10^{-17}$ & $1.09\cdot 10^{-18}$ & $5.02\cdot 10^{-17}$\\\hline
$(g-2)_\mu$ & $2.52\cdot 10^{-12}$ & $2.12\cdot 10^{-15}$ & $1.29\cdot 10^{-15}$\\\hline\hline
${\rm BR}(\mu \to e\gamma)$ & $1.33\cdot 10^{-15}$ & $2.02\cdot 10^{-18}$ & $2.69\cdot 10^{-17}$\\\hline
${\rm BR}(\tau \to e \gamma)$ & $2.40\cdot 10^{-18}\simeq 0$ & $2.76\cdot 10^{-22}\simeq 0$ & $4.73\cdot 10^{-13}$\\\hline
${\rm BR}(\tau \to \mu \gamma)$ & $9.83\cdot 10^{-23}\simeq 0$ & $2.38\cdot 10^{-17}\simeq 0$ & $1.06\cdot 10^{-19}\simeq 0$\\\hline
\end{tabular}
\caption{\label{tab:pointsLFV} LNV and LFV predictions of the three example benchmark points detailed in Tab.~\ref{tab:points}. The values marked as ``$\simeq 0$'' are always viewed to not be reachable \emph{with respect to the current experimental bounds} for the particular process under consideration.}
\end{table}

After having explained the rough categories of points, we will now present a few example points (in each category the one which fits the neutrino data best), before analysing the benchmark points in more detail. The three example points are given in Tab.~\ref{tab:points}, along with their predictions for the different LNV and LFV observables in Tab.~\ref{tab:pointsLFV}. Some remarks should be given on the numerical values:
\begin{itemize}

\item As to be expected, $\xi$ is always of $\mathcal{O}(1)$, and we have ensured that all the benchmark points have couplings $\xi$ below the perturbativity limit of $4\pi$.

\item We also have found points with considerably larger scalar masses. However, we did not include them in our plots because we wanted to focus on the points which are at least potentially testable at LHC.

\item The dimensional regularisation scale $\mu$ is meaningless in the sense that its actual value does not decide in any way about the validity of a certain benchmark point~\cite{Georgi:1994qn}, and we only quote it in order to enable the inclined reader to reproduce our results. Any observable $\hat{\mathcal{O}}$ calculated at loop-level does in principle depend on some energy scale $p^2$, where $p$ is some 4-momentum, and on the unknown scale $\mu$:  $\hat{\mathcal{O}} = \hat{\mathcal{O}} (p^2, \mu)$. If the value of the observable is known at some energy scale $p_0$, $\hat{\mathcal{O}} (p_0^2, \mu) = \hat{\mathcal{O}}_{\rm obs}$, one can in principle use this relation to compute $\mu$ as a function of $p_0$, $\mu= \mu_0 \equiv \mu (p_0^2)$. Then, the value of $\hat{\mathcal{O}}$ at any other energy scale $p$ can be computed as $\hat{\mathcal{O}} (p^2, \mu_0)$.

In our case, there is no \emph{explicit} dependence of the neutrino mass matrix on the momentum $p^2$, cf.\ Eq.~\eqref{eq:Mnu_finite}. However, there is an \emph{implicit} dependence through the observables involved, e.g., by the $\Delta m^2$-values obtained from a global fit~\cite{GonzalezGarcia:2012sz}. Thus, the values of $\mu$ presented in the table are simply the $\mu_0$'s derived from matching the observables to their experimental values.

\item We purposely present the figures in the Tab.~\ref{tab:points} with a too good precision, in order to make it easier to reproduce our results. We are confident that any potential reader of this text will be able to round the numbers to a smaller precision if required.

\end{itemize}

From Tab.~\ref{tab:points}, the input parameters (upper part of the table) can be inserted into Eq.~\eqref{eq:Mnu_finite} to yield the light neutrino mass basis in the non-diagonal basis. As the charged lepton mass matrix is already diagonal (otherwise the formula for the neutrino mass matrix would not contain the charged lepton mass eigenvalues), all the leptonic mixing originates from the neutrino sector. The light neutrino mass matrix $\mathcal{M}_\nu$ can then be brought to a diagonal form, $\mathcal{D}_\nu = {\rm diag} (m_1, m_2, m_3)$, as usual for a Majorana neutrino matrix:
\begin{equation}
 \mathcal{D}_\nu = U_{\rm PMNS}^\dagger \mathcal{M}_\nu U_{\rm PMNS}^*,
 \label{eq:diag}
\end{equation}
where the Pontecorvo-Maki-Nagakawa-Sakata matrix is explicitly given by its standard form~\cite{Beringer:1900zz},
\begin{equation}
 U_{\rm PMNS} =
 \begin{pmatrix}
 c_{12} c_{13} & s_{12} c_{13} & s_{13} e^{-i\delta}\\
 - s_{12} c_{23} - c_{12} s_{13} s_{23} e^{i\delta} & c_{12} c_{23} - s_{12} s_{13} s_{23} e^{i\delta} & c_{13} s_{23}\\
s_{12} s_{23} - c_{12} s_{13} c_{23} e^{i\delta} & - c_{12} s_{23} - s_{12} s_{13} c_{23} e^{i\delta} & c_{13} c_{23}
 \end{pmatrix}
 \begin{pmatrix}
 1 & 0 & 0\\
 0 & e^{i \alpha_{21}/2} & 0\\
 0 & 0 & e^{i \alpha_{31}/2}
 \end{pmatrix},
 \label{eq:PMNS}
\end{equation}
with $s_{ij} \equiv \sin \theta_{ij}$ and $c_{ij} \equiv \cos \theta_{ij}$. In this formula, $\delta$ is the Dirac $CP$ phase and $\alpha_{21,31}$ are the two Majorana $CP$ phases.

Using the values for the mixing angles given in Tab.~\ref{tab:points}, one can use Eq.~\eqref{eq:diag} to diagonalise the (non-diagonal) light neutrino mass matrices obtained by the input parameters. The result will in each case be the diagonal matrix $\mathcal{D}_\nu$ for NO such that $\Delta m^2_\odot = m_2^2 - m_1^2$ and $\Delta m^2_A = m_2^3 - m_1^2$.

As already mentioned, the correlations discussed above can be translated into relations between the mixing angles, the phases, the lightest neutrino mass, and the two mass-square differences. For example, using Eqs.~\eqref{eq:diag} and~\eqref{eq:PMNS}, it is easy to show that $\mathcal{M}_{\nu, ee} = m_1 c_{12}^2 c_{13}^2 + \sqrt{m_1^2 + \Delta m^2_\odot} s_{12}^2 c_{13}^2 e^{i\alpha_{21}} + \sqrt{m_1^2 + \Delta m^2_A} s_{13}^2 e^{i(\alpha_{31}-2\delta)}$ is the effective neutrino mass measured in $0\nu\beta\beta$. Similarly, Eq.~\eqref{eq:correlation} results in a correlation, some details of which are given in Appendix~\hyperref[sec:app_B]{B}.

\begin{figure}[t]
\centering
\includegraphics[scale=0.6]{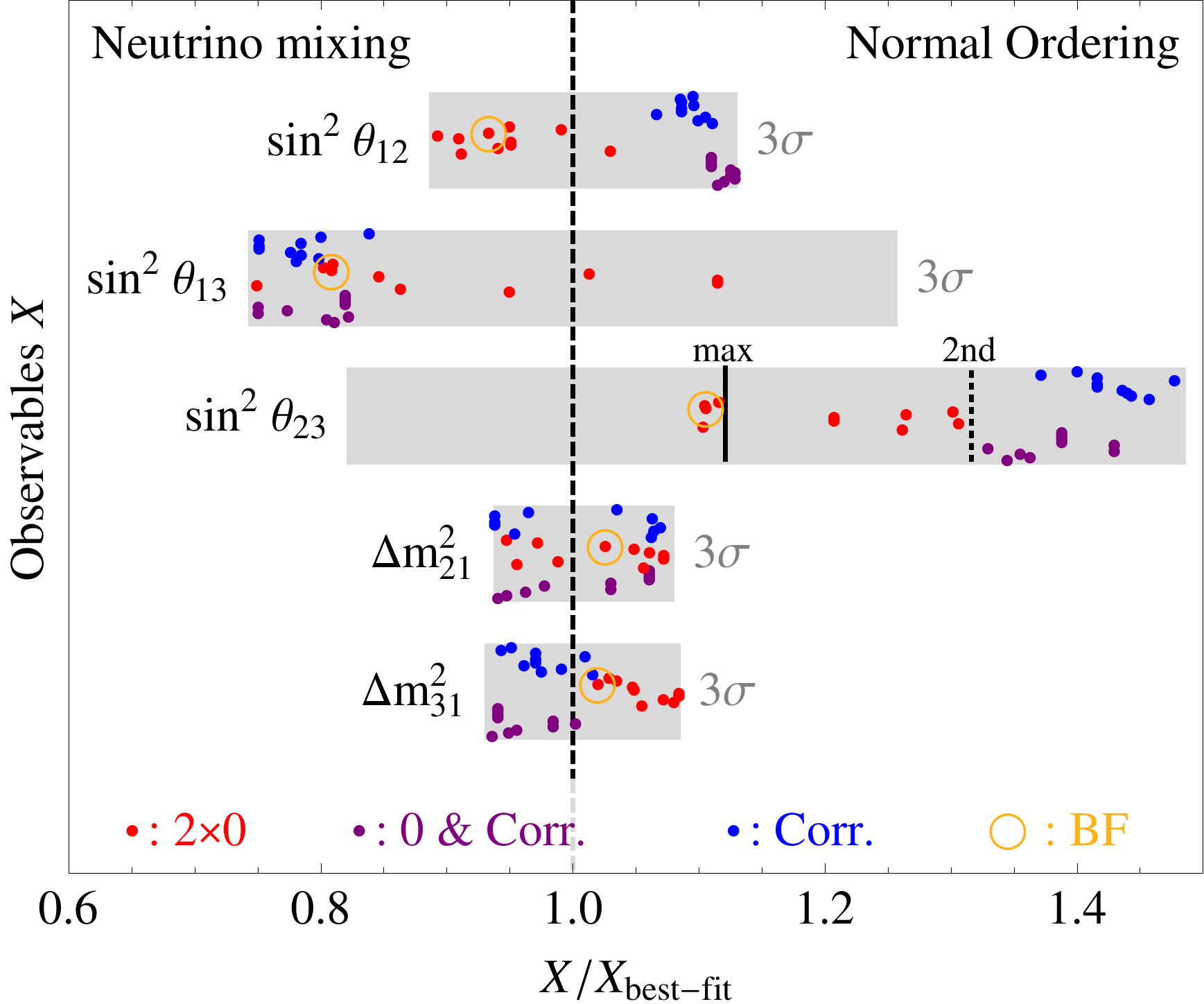}
\caption{\label{fig:NeutrinoMixing}
The predictions of the different benchmark points for the known neutrino oscillation parameters. As can be seen, all benchmark points lie within the $3\sigma$ ranges for the parameters.
}
\end{figure}

What do the benchmark points imply for phenomenology? Let us start with the predictions for the known neutrino oscillation parameters, which are depicted in Fig.~\ref{fig:NeutrinoMixing}. In this figure, we show the ratios of the different predictions to the best-fit values of the corresponding neutrino observables~\cite{GonzalezGarcia:2012sz}. As can be seen, all the points fit the neutrino oscillation parameters within their $3\sigma$ ranges indicated by the light gray thick bars, and the closer the points are to the vertical line at $1.0$, the better they are; the points representing the best-fit benchmark are indicated by the yellow (gray) circles. Note that, for $\theta_{23}$, we have also indicated the position of the (only slightly worse) second minimum of the $\chi^2$-function in the second octant, as well as the maximal value of $\theta_{23} = 45^\circ$ to simplify the orientation. Note that all the points drawn either yield maximal $\theta_{23}$ or a value of $\theta_{23}$ close to the second minimum. Thus, the resulting fits would be improved if at some point the second minimum was favoured by the data.

Starting with the points of the first category ($f_{ee, e\tau} \simeq 0$), which are indicated by the red (light gray) points which will from now on only be called ``red points'', one can see that the spread of these points is quite large. Indeed, these points are least constrained, which is also reflected by us having found many more points of this category than of any other. However, even for the red points, a tendency for smaller values of $\theta_{13}$ and larger values of $\theta_{23}$ can be seen. Furthermore, it seems that the atmospheric mass-square difference $\Delta m_{31}^2$ has a tendency to be larger than the best-fit value, while no such tendency can be seen for $\Delta m_{21}^2$.

The second category of points (``purple points''), which fulfill $f_{ee} \simeq 0$ and Eq.~\eqref{eq:condition}, are indicated by the purple (medium gray) points. Here, the picture is very different: the purple points seem to have very solid predictions for nearly all the observables, which allows to falsify these points with high precision neutrino oscillation experiments. The first mixing angle $\theta_{12}$ should be quite large, close to the upper $3\sigma$ bound, while $\theta_{13}$ should be rather small. Also $\theta_{23}$ should be large, but not quite at the upper $3\sigma$ edge. $\Delta m_{31}^2$ should be small, too, only for $\Delta m_{21}^2$ there is no clear tendency.

Finally, the third category of points (``blue points''), which only obey the correlation from Eq.~\eqref{eq:condition} and are indicated by the blue (dark gray) points yield equally solid predictions. While there is no clear prediction visible for $\Delta m_{21}^2$, one can see that $\theta_{12, 23}$ should be large while $\theta_{13}$ should be small. $\Delta m_{31}^2$, in turn, should be small or at least close to its best-fit value.

\begin{figure}
\centering
\begin{tabular}{lr}
\includegraphics[scale=0.61]{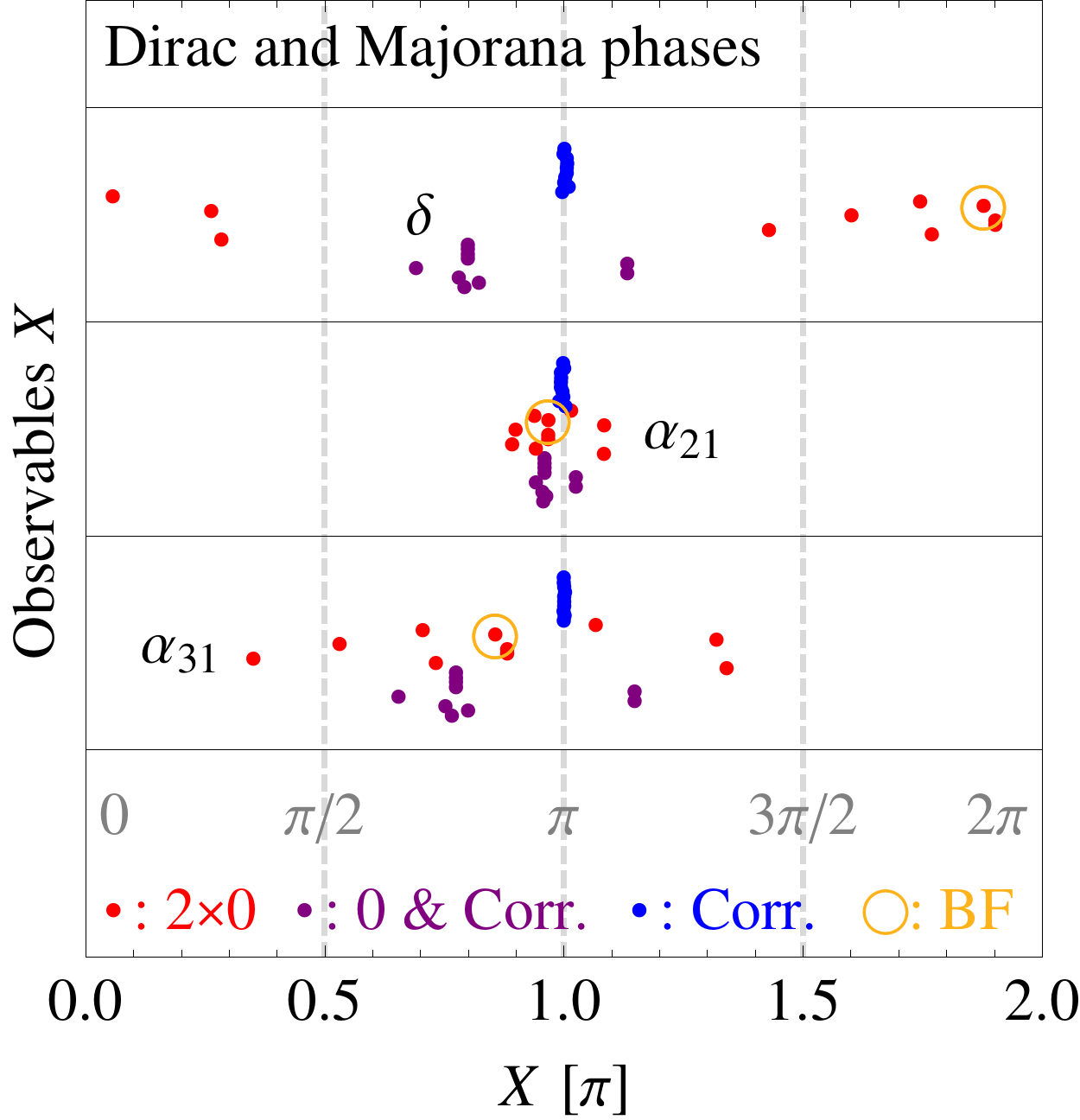} &
\includegraphics[scale=0.55]{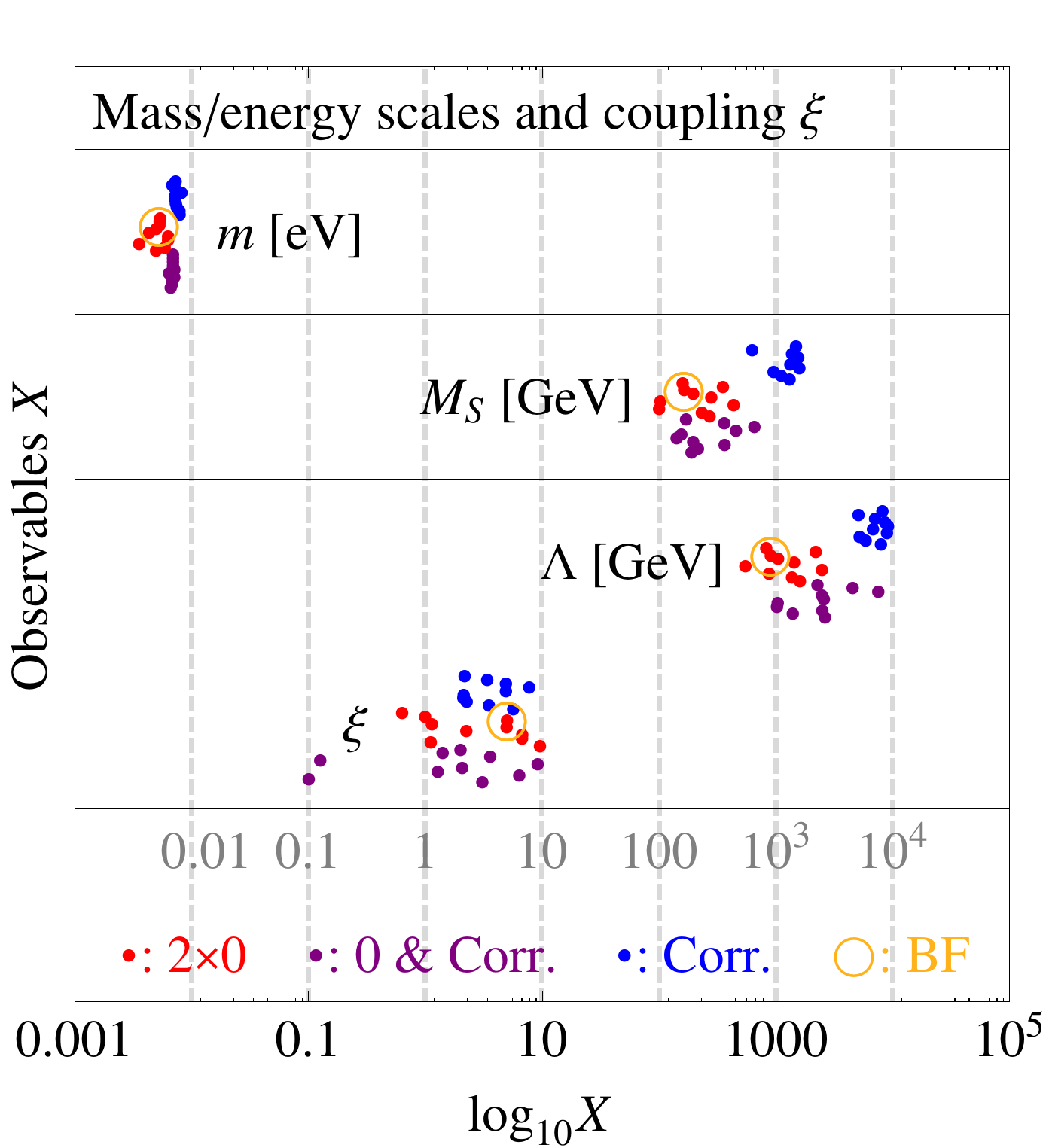}
\end{tabular}
\caption{\label{fig:predictions}
Predictions of the benchmark points for the neutrino-related phases (left) and for the mass/energy scales as well as the coupling $\xi$ (right).
}
\end{figure}

The next set of predictions is depicted in Fig.~\ref{fig:predictions}, where the Dirac and Majorana phases are shown on the left, whereas the lightest neutrino mass $m$, the scalar mass $M_S$, the cutoff scale $\Lambda$, and the effective coupling $\xi$ are presented on the right. Starting with the phases, it is immediately visible that in particular the Dirac phase $\delta$ can be used to distinguish the different types of points. While the red points prefer $\delta$ to be close to $0$ (or, equivalently, $2\pi$), the blue points clearly prefer $\delta \simeq \pi$ and the purple points cluster around $\delta \approx 0.8\pi$ or $1.2\pi$. Thus, with a future improved measurement of $\delta$ or if the hint by T2K~\cite{Abe:2013nka,T2K_NExT} persists, this would be very useful to distinguish the different categories of points. For the Majorana phases, interestingly, all the benchmark points we have found seem to prefer $\alpha_{21} \simeq \pi$, which is particularly notable since the red and blue points have, in fact, nothing in common in terms of the conditions leading to them. The explanation of this is that some bound must push $\alpha_{21}$ to being close to this value, and indeed this value is needed in order to make the required cancellation in $\mathcal{M}_{\nu, ee}$ possible~\cite{Lindner:2005kr}. For $\alpha_{31}$, in turn, there seems to be no clear tendency for the red points. The purple points, however, seem to enforce $\alpha_{31} \sim \delta$, while the blue points cluster strongly around $\alpha_{31} = \pi$, so that the blue points seem to forbid $CP$ violation (i.e., all phases are trivially equal to $-1$).

The mass/energy scales predicted as well as the predictions for the effective coupling $\xi$ are displayed on the right panel of Fig.~\ref{fig:predictions}. First of all, as we had already pointed out, all the points found predict NO with a lightest neutrino mass $m \approx 5$~meV, due to them being the cancellation region of $\mathcal{M}_{\nu, ee}$~\cite{Lindner:2005kr,King:2013psa}. The values predicted for $M_S$ and $\Lambda$ are spread over a slightly larger range, and we have also found points with considerably larger values of $M_S$ which we however do not display here due to their bad detection prospects for LHC. On the other hand, there is a clear tendency of the blue points to yield larger values of $M_S$ and $\Lambda$, which arises from the requirement of fulfilling the $0\nu\beta\beta$ bound, Eq.~\eqref{eq:bound_0nbb_expl}, as we had anticipated from observing that the coupling $f_{ee}$ is sizable for these points. The predictions for $\xi$ are all roughly around $1$ as they should be for the EFT to make sense~\cite{Georgi:1994qn} -- some are a bit larger, but always below the perturbativity limit. The only outliers are two single purple points which yield $\xi \sim 0.1$. The reason for $\xi$ not to be too small is that this would lead to an overall suppression of the light neutrino mass matrix, cf.\ Eq.~\eqref{eq:Mnu_finite}, which should not be too small in order to correctly reproduce the ``large'' mass-splitting $\Delta m_{31}^2$.

\begin{figure}
\centering
\begin{tabular}{lr}
\includegraphics[scale=0.53]{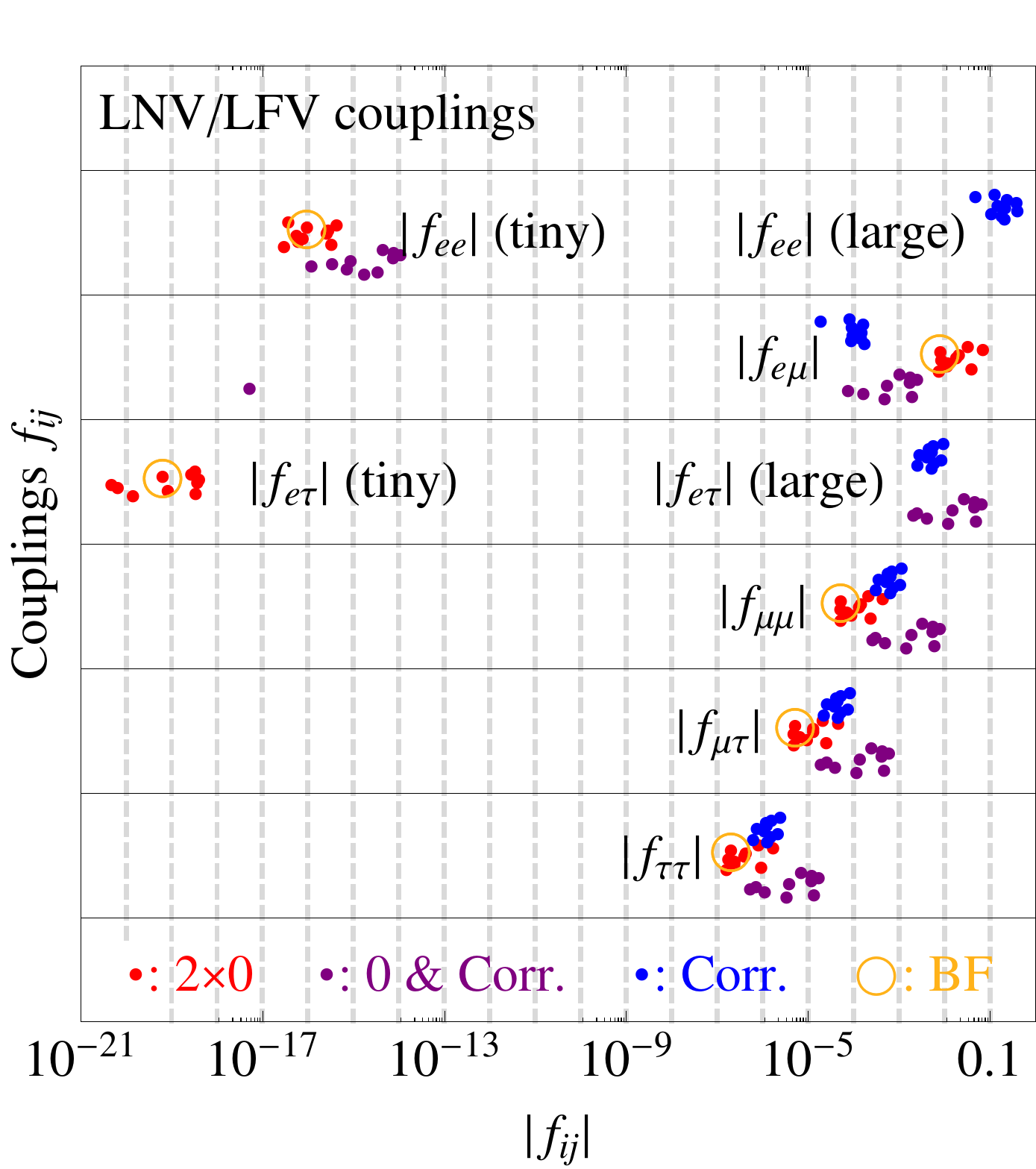} &
\includegraphics[scale=0.54]{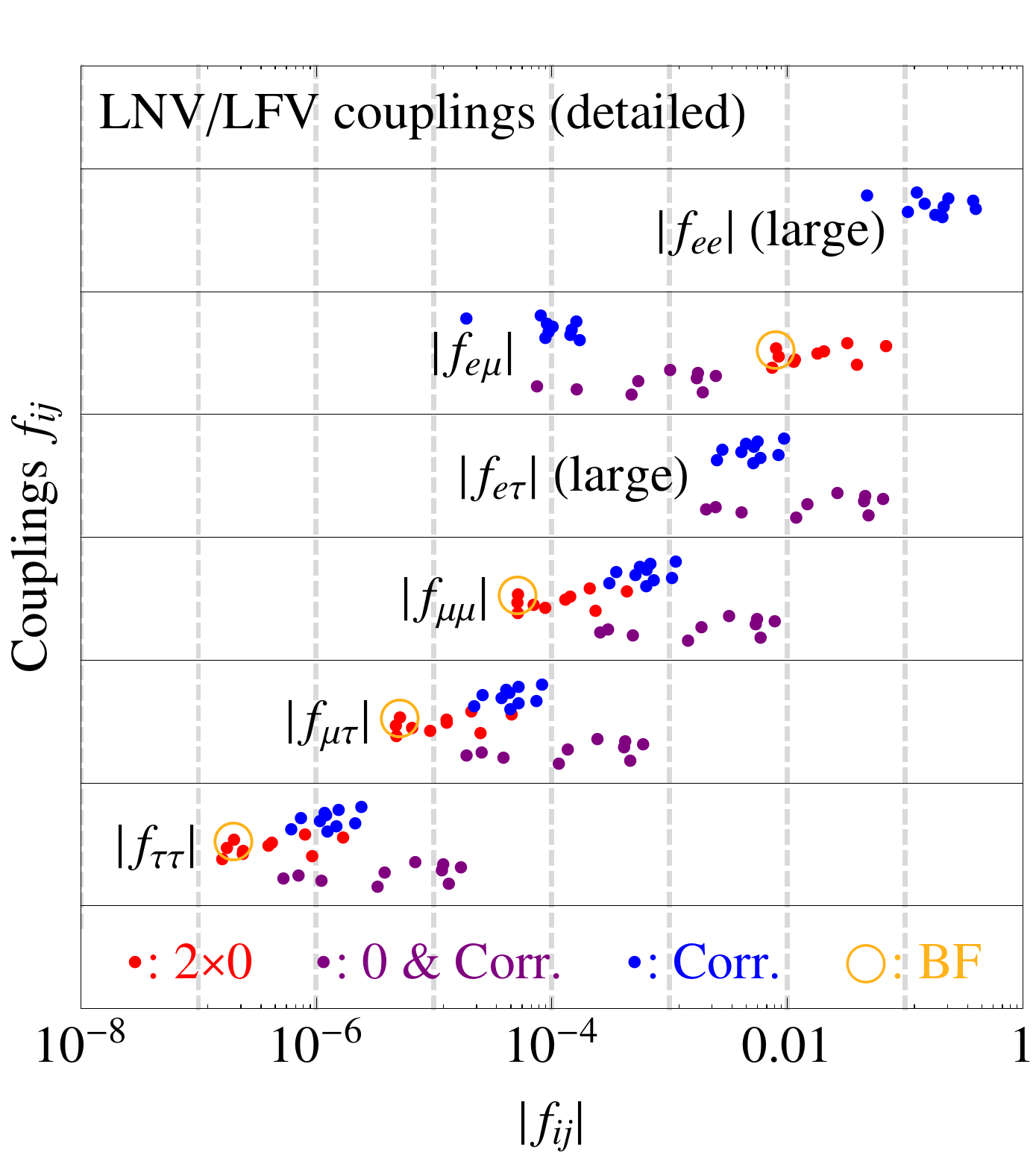}
\end{tabular}
\caption{\label{fig:couplings}
The Yukawa couplings of the doubly charged scalar to like-sign right-handed charged leptons. The full range is displayed on the left, while a more detailed view of the more interesting regions (i.e., larger couplings) is displayed on the right.
}
\end{figure}

The sizes of the Yukawa couplings of the doubly charged scalar to pairs of right-handed charged leptons are displayed in Fig.~\ref{fig:couplings}, left panel, with a blow-up of the region with the larger couplings shown on the right. As can be seen, $|f_{ee}|$ is trivially very tiny for the red and purple points, and so is $|f_{e\tau}|$ for the red points. However, the blue points admit quite large values of $|f_{ee}|$, which should translate into sizable rates for $0\nu\beta\beta$. For the red points, large values for $|f_{e\mu}|$ are required, as otherwise it would not be possible to obtain phenomenologically viable light neutrino masses, given that the mass matrices corresponding to these points inherit two texture zeros from Yukawa coupling matrix $f_{ij}$.\footnote{As we will see in a minute, it might lead to phenomenological differences whether the neutrino mass matrix inherits textures from the Yukawa coupling matrix or whether it only involves ``effective'' texture zeros due to the $m_i m_j$-suppressions being at work.} Naturally, because the corresponding Yukawa matrices only have one (no) texture zero(s), this requirement is not so strong for the purple (blue) points, such that smaller values of $|f_{e\mu}|$ are possible. They need a small $|f_{e\mu}|$, even though the correlation from Eq.~\eqref{eq:condition} does help considerably in order to avoid the strong bound from $\mu \to e \gamma$ as well as the tree-level decay $\mu \to 3 e$. That decay is also suppressed by a small $|f_{ee}|$, but this coupling is large for the blue points so that they need $|f_{e\mu}|$ to be small enough.

Apart from some exceptionally strong LFV bounds, the general tendency of the values for the couplings $f_{ij}$ is that they tend to compensate for the $m_i m_j$-suppressions present in the light neutrino mass matrix, cf.\ Eq.~\eqref{eq:Mnu_finite}. Put into a more intuitive form, $m_\tau^2 > m_\tau m_\mu > m_\mu^2 > m_\tau m_e > m_\mu m_e > m_e^2$ generically implies that $|f_{ee}| > |f_{e \mu}| > |f_{e \tau}| > |f_{\mu \mu}| > |f_{\mu \tau}| > |f_{\tau \tau}|$, in order to have large mixings as required in the neutrino sector, and this relation is only altered by certain strong bounds requiring some couplings to be particularly small. Indeed, as can even be seen in the plot, one could draw an imaginary straight diagonal line through the sets of points. The interesting region of the couplings is depicted in greater detail on the right panel of Fig.~\ref{fig:couplings}. From this figure one can in particular see that, for the blue and purple points, the correlation $f_{e\mu} f_{\mu\mu}^* + f_{e\tau} f_{\mu \tau}^* \simeq 0$ is another reason that $|f_{\mu\mu}| > |f_{\mu\tau}|$ is required, due to $|f_{e \mu}|$ being smaller than $|f_{e \tau}|$.

\begin{figure}
\centering
\begin{tabular}{lr}
\includegraphics[scale=0.54]{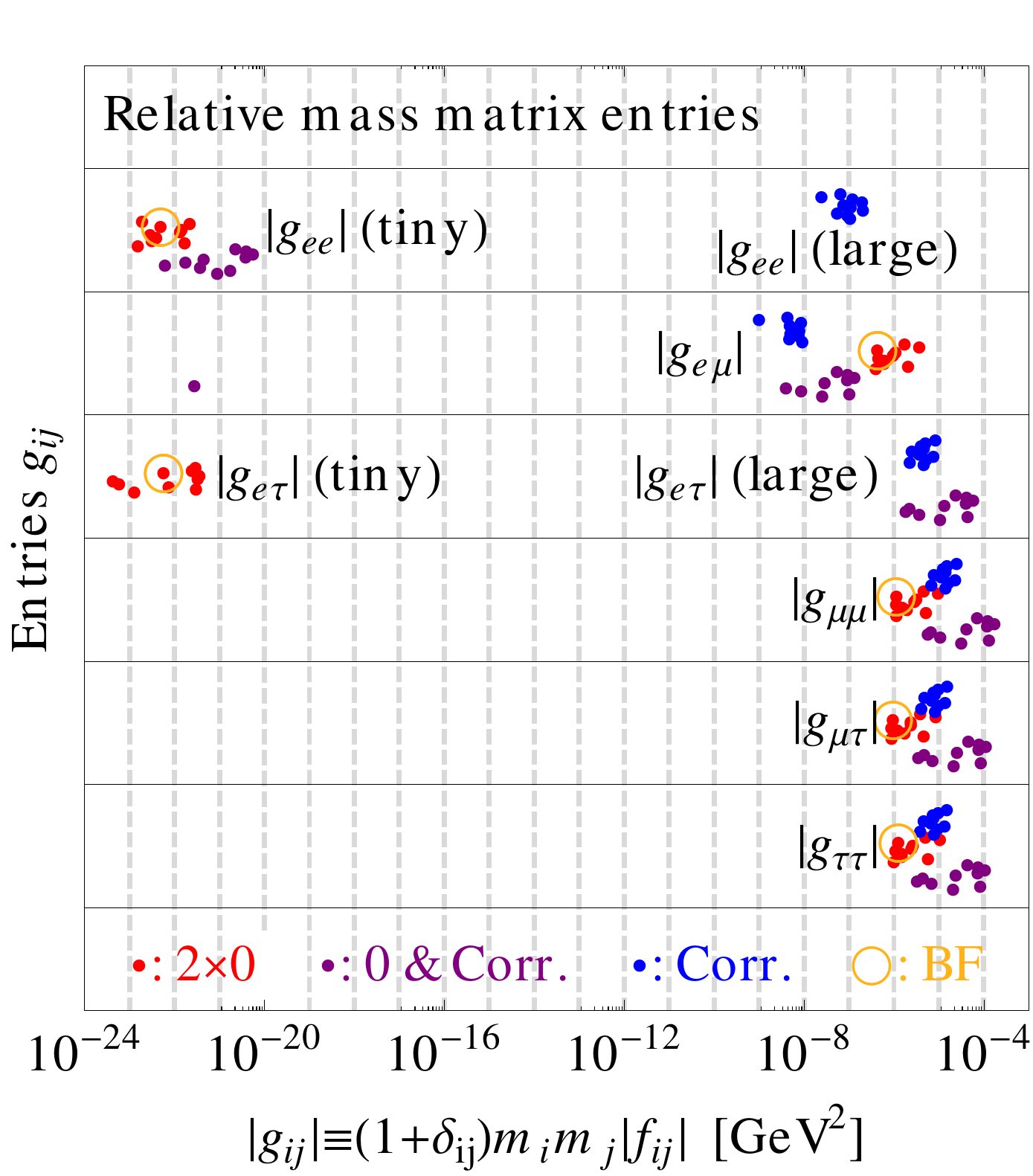} &
\includegraphics[scale=0.56]{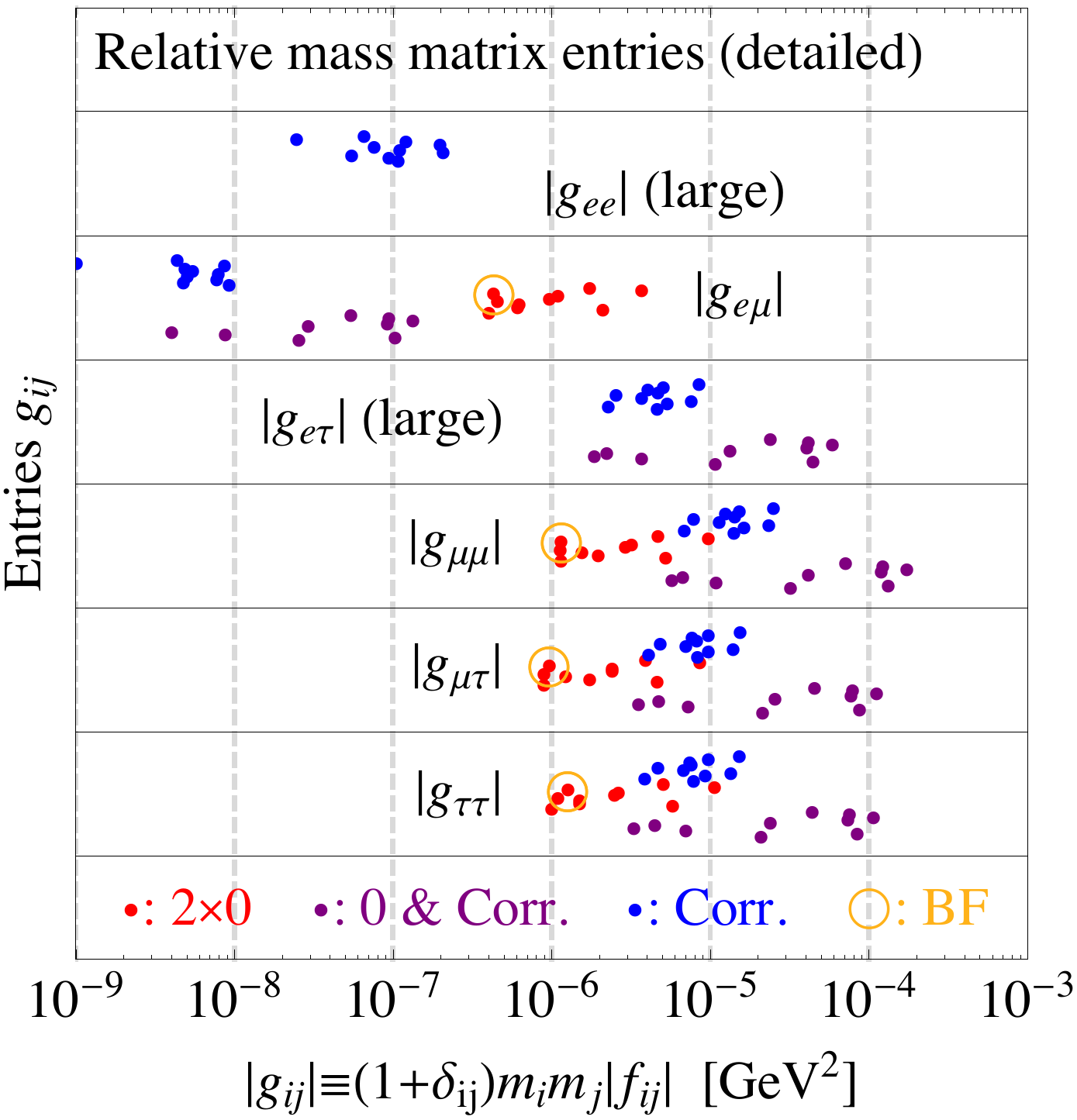}
\end{tabular}
\caption{\label{fig:entries}
Flavour-dependent parts $g_{ij}$ of the mass matrix entries in units of GeV$^2$ (left: whole range, right: only sizable entries). See text for details.
}
\end{figure}

For comparison, we have also displayed in Fig.~\ref{fig:entries} the sizes of the resulting mass matrix entries (left panel: whole range, right panel: sizable entries only), by calculating their flavour-dependent parts $g_{ij}\equiv (1 + \delta_{ij}) m_i m_j f_{ij}$. These reveal the actual physical texture zeros in what concerns the light neutrino mass matrix. As generic for phenomenologically valid light neutrino mass matrices, most of the matrix elements are nearly of the same size, with some additional structure imposed by certain texture zeros. This plot clearly reveals that the physical differences between our three categories of benchmark points: while at the level of the Yukawa matrices $f_{ij}$, the red [purple, blue] points were characterised by two texture zeros [by one texture zero and the correlation from Eq.~\eqref{eq:condition}, only by the correlation from Eq.~\eqref{eq:condition}], we can see from Fig.~\ref{fig:entries} that, effectively, all categories of points involve \emph{two} texture zeros on the level of the mass matrix. For the red points, these are trivially the elements $\mathcal{M}_{\nu, ee} \simeq 0$ and $\mathcal{M}_{\nu, e\tau} \simeq 0$, while both the purple and blue points are characterised by $\mathcal{M}_{\nu, ee} \simeq \mathcal{M}_{\nu, e\mu} \simeq 0$. Physically, this means that the purple and blue points should look very similar in what concerns neutrino phenomenology, and this is exactly confirmed by Figs.~\ref{fig:NeutrinoMixing} and~\ref{fig:predictions}, which illustrate that all predictions of neutrino mass and leptonic mixing parameters are essentially the same (the only small differences are visible in the phases $\delta$ and $\alpha_{31}$). Thus, the purple and blue points indeed would need to be distinguished by LNV/LFV data. This is well possible, as we will show in the next few paragraphs.

\begin{figure}
\centering
\begin{tabular}{lr}
\includegraphics[scale=0.33]{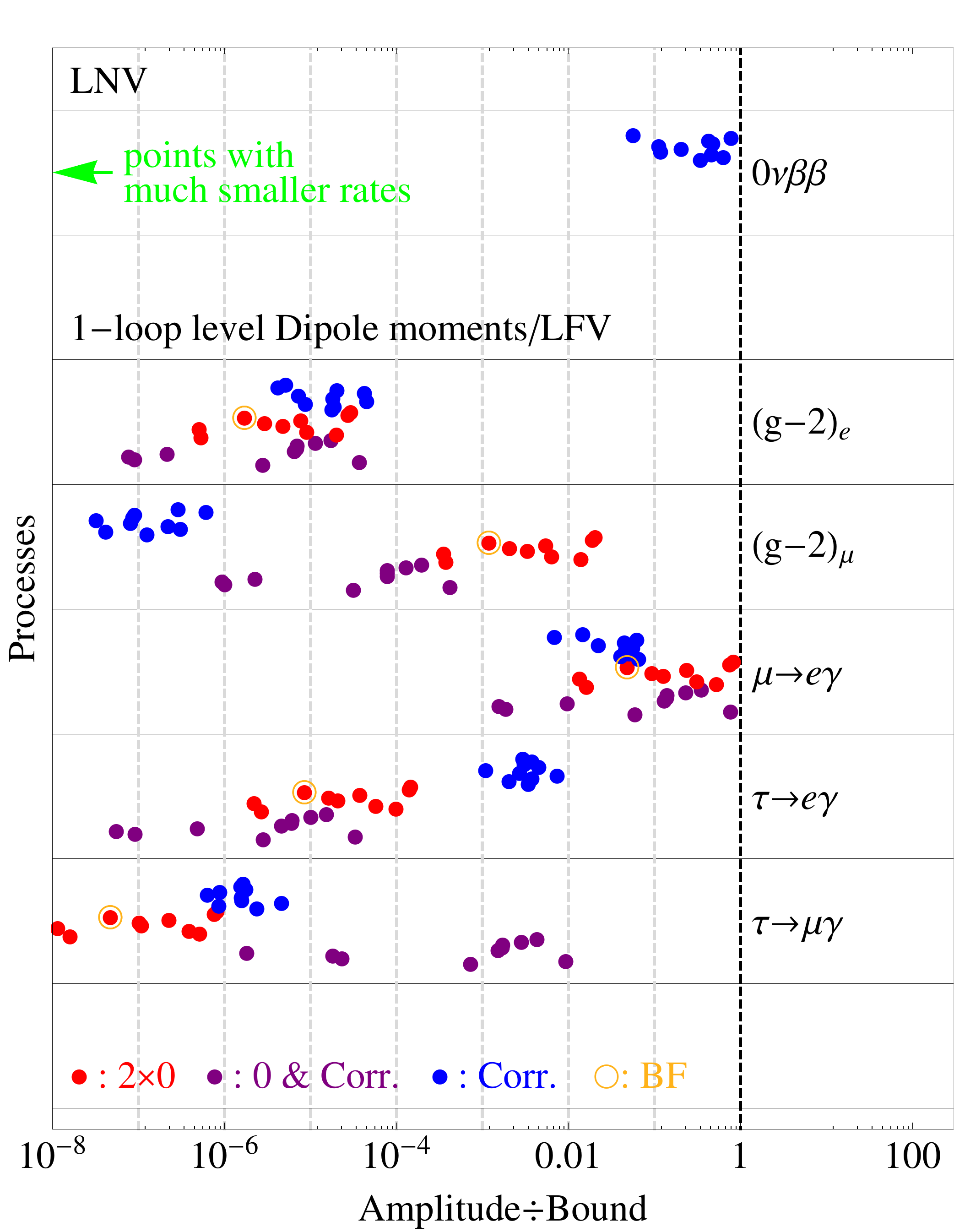} &
\includegraphics[scale=0.33]{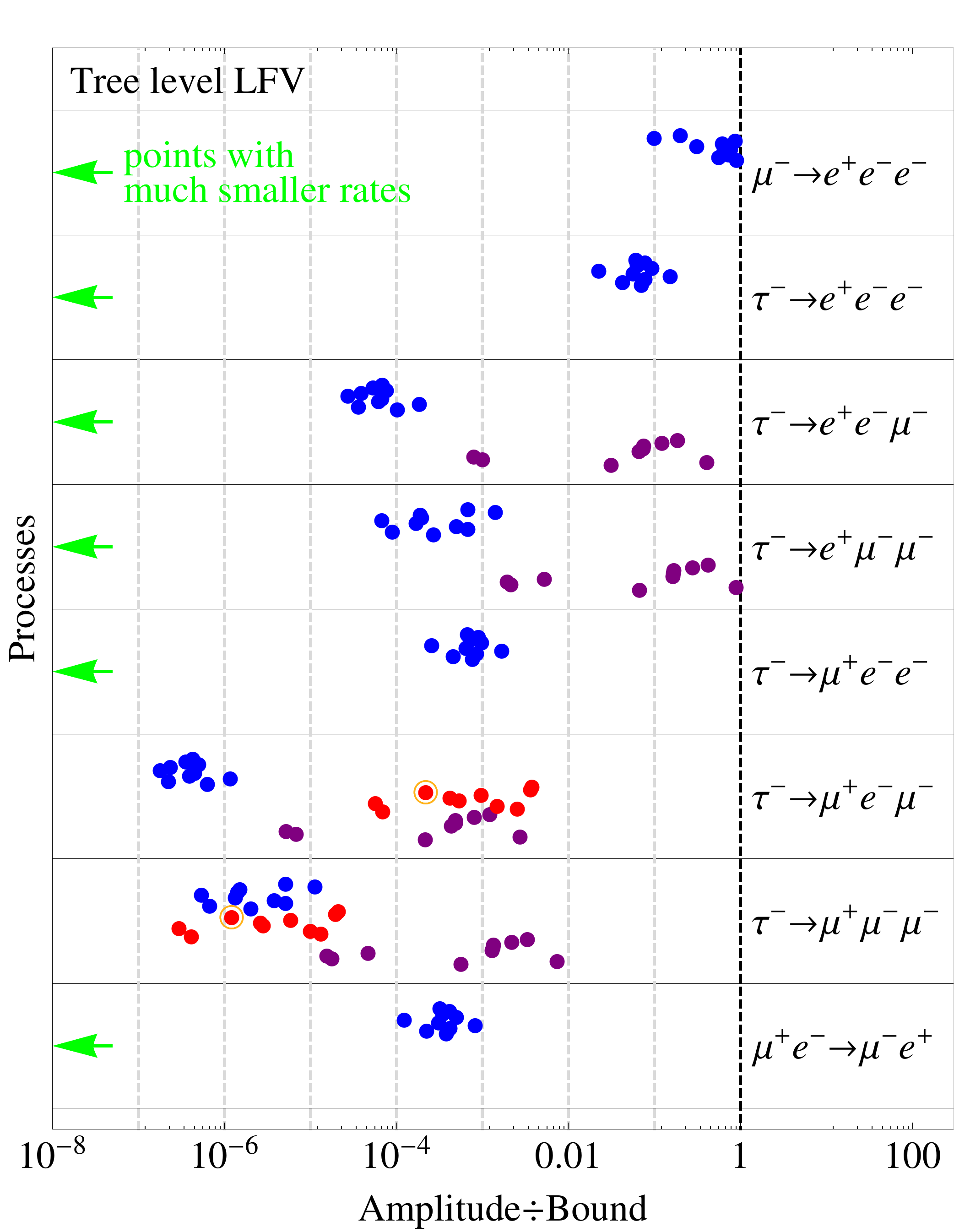}
\end{tabular}
\caption{\label{fig:LNVLFV}
The predictions for the different low-energy leptonic observables (lepton number/flavour violation and magnetic dipole moments).
}
\end{figure}

Finally, the predictions of the different points for the relevant observables in low-energy leptonic phenomenology are all depicted in Fig.~\ref{fig:LNVLFV}. Let us start with LNV processes, the only relevant of which being $0\nu\beta\beta$. The predictions for this process, i.e., the predicted amplitude divided by the experimental bound, cf.\ Sec.~\ref{sec:0nbb}, can be found on the left panel. Since too small predictions are of no interest, we have decided against plotting the whole range of points. Instead, we only show the region somewhat close to the bounds, while we indicate in the plots whenever there are some points left of that region and thus out of the plot. Not surprisingly, in the row corresponding to $0\nu\beta\beta$ no red or purple points are visible, which is a simple reflection of $f_{ee} \simeq 0$ holding for these types of points. However, the blue points have sizable couplings $f_{ee}$ and, although the standard light neutrino exchange contribution is nevertheless suppressed, they hence yield a potentially sizable signal. In fact, as can be seen from the plot, the blue points are very close to the current bound and could potentially be ruled out completely by the next generation of experiments on neutrinoless double beta decay.

The next category of bounds are those arising from the electron and muon anomalous magnetic moments. However, as can be seen from the plot, these bounds are not really competitive in our case, so that we will not discuss them further. The loop-level LFV processes show a more interesting picture: first of all, as anticipated, it is clearly visible from the plot that the bound coming from MEG's non-observation of $\mu \to e \gamma$~\cite{MEG_EPS} is by far the strongest one in the game: indeed, most of the points are close to the bound within one or two orders of magnitude in the amplitude, which shows that improved bounds on this observable could potentially rule out many of the valid points. A handful of points yields values that are smaller by about one more order, and only two purple points have been found which seem fairly safe even from improved measurements of the process. The radiative LFV decays of the $\tau$ do not seem so strong. This purely stems from the fact that these experiments are much more difficult to perform due to the short lifetime of the $\tau$. On the other hand, there exist blue (purple) points which are only a bit more than two orders of magnitude away from the current bound on $\tau \to e\gamma$ ($\tau \to \mu\gamma$) and, given that these bounds are relatively weak at the moment, it is not unthinkable that new experimental techniques might be devised in the future which could considerably improve the current values.

Finally, let us look at the tree-level LFV processes, right panel of Fig.~\ref{fig:LNVLFV}. Here, it is clear that the red and purple points, with $f_{ee} \simeq 0$, are safe from any bounds involving that coupling, so that only the blue points are of interest for $\mu \to 3e$, $\tau \to 3e$, and $\tau^- \to \mu^+ e^- e^-$, if so at all. However, in particular $\mu \to 3e$ imposes a strong bound on the blue points, so that all of them could probably be ruled out by an improved measurement. Going down in the figure, the red points are also entirely safe from $\tau^- \to e^+ \mu^- e^-$ and $\tau \to 3 e$, due to $f_{e\tau} \simeq 0$. However, these processes are potentially dangerous for the purple points, while the blue ones are fairly below the current bounds. The weakest bounds exist on decay modes like $\tau^- \to \mu^+ e^- \mu^-$ and $\tau \to 3 \mu$, which is why from the corresponding observables there originates no strong tension on the parameters. Thus, couplings like $f_{\mu \mu}$ or $f_{\mu \tau}$ can easily be comparatively large from an LFV point of view, and this is desired for the corresponding points to be interesting for phenomenology at LHC, which we will investigate in Sec.~\ref{sec:collider}. As sfinal remarks for this section note that, for completeness, muonium conversion does not impose a strong bound and except for the blue points all benchmark scenarios found are off the plot. However, what could yield an interesting bound \emph{in the future} would probably by $\mu$-$e$ conversion on nuclei~\cite{Chiang:1993xz,Raidal:1997hq}. Due to the difficulties involved with a detailed computation and because of the current limit not being competitive with the one on $\mu \to e \gamma$, we postpone an analysis of this particular process to later work~\cite{mueConAnalysis}.

\section{\label{sec:collider}Collider phenomenology and bounds}

The existence of doubly charged scalars being predicted in different scenarios of new physics beyond the SM, analyses at the LHC have been designed to identify signatures of these new particles. Searches by both ATLAS\cite{ATLAS:2012hi} and CMS\cite{Chatrchyan:2012ya} collaborations have been undertaken, though no signal has been found so far. From the phenomenological side, a large number of analyses have been performed to identify signatures of doubly charged scalars~\cite{Azuelos:2004mwa,Chen:2007dc,Han:2007bk,Chen:2008qb,Akeroyd:2011zza,Akeroyd:2011ir,Sugiyama:2012yw,delAguila:2013yaa,Babu:2013ega,Chaudhuri:2013xoa,delAguila:2013hla,Alloul:2013raa,Chun:2013vma,Dutta:2014dba}. A quite extensive treament of the production and decay channels of doubly charged scalars has recently been presented in~\cite{delAguila:2013mia}: in this paper the main production processes and decay channels of doubly charged scalars are analysed from a model-independent point of view; furthermore, a calculation of experimental efficiencies for different masses and decay channels has been performed, also providing predictions for future LHC energies. The results of this analysis will be extensively used in the following sections, to pose bounds on the classes of scenarios treated in this work.

\subsection{Production and decay channels}

\begin{figure}[t]
\centering
\subfigure[s-channel pair]{\includegraphics[scale=0.8]{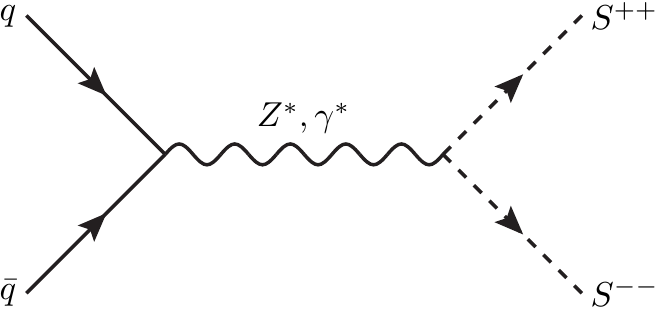}}
\hskip 20pt
\subfigure[VBF pair]{\includegraphics[scale=0.8]{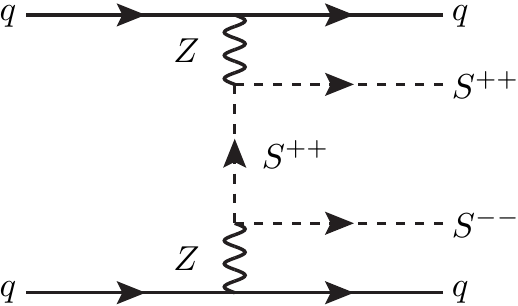}}
\hskip 20pt
\subfigure[VBF single]{\includegraphics[scale=0.8]{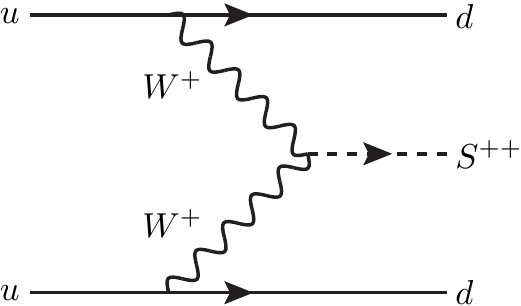}}
\caption{Examples of pair and single production processes of $\SPP$, partially in association with jets.}
\label{fig:productionprocesses}
\end{figure}

The doubly-charged scalars can be either produced in pairs or singly, as shown in Fig.~\ref{fig:productionprocesses}, with pair production being the dominant mechanism for a large range of $\SPP$ masses. 

While pair production in association with jets through vector boson fusion (VBF) is suppressed by the higher order of the process, single production is more model-dependent: the coupling with the $W$-boson depends on two parameters, the cutoff $\Lambda$ and the coupling parameter $\xi$. The cross sections for pair production as well as pair production in association with jets are provided in Tab.~\ref{tab:sigmapair}. However, to understand the relevance of single production with respect to pair production, the relevant parameters have been fixed to the values $\Lambda=5 M_S$ or $\Lambda=10 M_S$ and $\xi=4\pi$. These values are not related to any of the benchmark points used in this analysis; they have been chosen to maximise the contribution of single production, which increases for increasing $\xi$ and decreasing $\Lambda$: $\xi=4\pi$ is at the boundary of the perturbative region, while as the cutoff is concerned, in one case ($\Lambda=5 M_S$) the cutoff has been chosen to be in a region where the validity of the EFT approach may appear questionable, while in the other case ($\Lambda=10 M_S$) the EFT approach is safer. In any case, the typical scale of the process for single production of the doubly charged scalar is $p^2\simeq M_S^2$, therefore the error given by the EFT approximation scales as $p^2/\Lambda^2 \sim M_S^2/\Lambda^2$, which for $\Lambda=5M_S$ amounts to only about $4\%$. Smaller values of the cutoff/mass ratio would enter a region where the approximation may not be acceptable anymore, and where the details of the UV completion of the model are required to correctly describe the process. For this reason, in the following, the lower bound on the cutoff will always be 5$M_S$. The cross sections at different LHC energies for all production mechanisms are provided in Tab.~\ref{tab:sigmasingle}.\footnote{The values for pair production have been obtained using the model files provided in~\cite{delAguila:2013mia} and performing numerical simulations through {\tt MadGraph5}~\cite{Alwall:2011uj}. The cut on the minimum transverse momentum of jets has been fixed to be $10$~GeV. The new vertex for single production has been implemented in dedicated model files.}

\begin{table}[t]
\tiny
\centering
\begin{tabular}{c|cccc|cccc}
\toprule
\multirow{3}{*}{$M_S$ [GeV]} & \multicolumn{8}{c}{Cross section [fb]} \\
\cmidrule{2-9}
& \multicolumn{4}{c|}{pair production} & \multicolumn{4}{c}{pair production + jets (VBF)} \\
& 7 TeV & 8 TeV & 13 TeV & 14 TeV & 7 TeV & 8 TeV & 13 TeV & 14 TeV \\
\midrule
200  & 8.52                & 11.3                & 28.0                & 31.7                & 0.199               & 0.279               & 0.818               & 0.955               \\
300  & 1.36                & 1.93                & 5.74                & 6.64                & 3.65$\times10^{-2}$ & 5.44$\times10^{-2}$ & 0.196               & 0.229               \\
400  & 0.310               & 0.474               & 1.70                & 2.02                & 9.46$\times10^{-3}$ & 1.48$\times10^{-2}$ & 6.29$\times10^{-2}$ & 7.55$\times10^{-2}$ \\
500  & 8.59$\times10^{-2}$ & 0.142               & 0.620               & 0.749               & 2.90$\times10^{-3}$ & 4.87$\times10^{-3}$ & 2.41$\times10^{-2}$ & 2.97$\times10^{-2}$ \\
600  & 2.68$\times10^{-2}$ & 4.78$\times10^{-2}$ & 0.255               & 0.316               & 9.98$\times10^{-4}$ & 1.80$\times10^{-3}$ & 1.06$\times10^{-2}$ & 1.34$\times10^{-2}$ \\
700  & 9.04$\times10^{-3}$ & 1.75$\times10^{-2}$ & 0.115               & 0.145               & 3.74$\times10^{-4}$ & 7.23$\times10^{-4}$ & 5.00$\times10^{-3}$ & 6.47$\times10^{-3}$ \\ 
800  & 3.21$\times10^{-3}$ & 6.77$\times10^{-3}$ & 5.51$\times10^{-2}$ & 7.12$\times10^{-2}$ & 1.50$\times10^{-4}$ & 3.09$\times10^{-4}$ & 2.55$\times10^{-3}$ & 3.33$\times10^{-3}$ \\
900  & 1.18$\times10^{-3}$ & 2.73$\times10^{-3}$ & 2.77$\times10^{-2}$ & 3.66$\times10^{-2}$ & 6.06$\times10^{-5}$ & 1.34$\times10^{-4}$ & 1.34$\times10^{-3}$ & 1.81$\times10^{-3}$ \\
1000 & 4.43$\times10^{-4}$ & 1.13$\times10^{-3}$ & 1.44$\times10^{-2}$ & 1.95$\times10^{-2}$ & 2.52$\times10^{-5}$ & 6.15$\times10^{-5}$ & 7.41$\times10^{-4}$ & 1.02$\times10^{-3}$ \\
\bottomrule
\end{tabular}
\caption{Cross sections (in fb) for $\SPP$ pair production and for pair production in association with jets for different $\SPP$ masses and LHC energies.}
\label{tab:sigmapair}
\end{table}

\begin{table}[t]
\tiny
\centering
\begin{tabular}{c|cccc|cccc}
\toprule
\multirow{3}{*}{$M_S$ [GeV]} & \multicolumn{8}{c}{Single production cross section [fb]}\\
\cmidrule{2-9}
& \multicolumn{4}{c|}{$\Lambda=5 M_S$} & \multicolumn{4}{c}{$\Lambda=10 M_S$}\\
& 7 TeV & 8 TeV & 13 TeV & 14 TeV & 7 TeV & 8 TeV & 13 TeV & 14 TeV \\
\midrule
200  & 4.81                & 6.27                & 15.1                & 17.1                & 7.53$\times10^{-2}$ & 9.79$\times10^{-2}$ & 0.237               & 0.267               \\
300  & 0.201               & 0.270               & 0.717               & 0.823               & 3.14$\times10^{-3}$ & 4.22$\times10^{-3}$ & 1.12$\times10^{-2}$ & 1.29$\times10^{-2}$ \\
400  & 1.94$\times10^{-2}$ & 2.70$\times10^{-2}$ & 7.81$\times10^{-2}$ & 9.06$\times10^{-2}$ & 3.03$\times10^{-4}$ & 4.22$\times10^{-4}$ & 1.22$\times10^{-3}$ & 1.41$\times10^{-3}$ \\
500  & 2.98$\times10^{-3}$ & 4.26$\times10^{-3}$ & 1.34$\times10^{-2}$ & 1.57$\times10^{-2}$ & 4.65$\times10^{-5}$ & 6.65$\times10^{-5}$ & 2.10$\times10^{-4}$ & 2.46$\times10^{-4}$ \\
600  & 6.10$\times10^{-4}$ & 9.01$\times10^{-4}$ & 3.10$\times10^{-3}$ & 3.67$\times10^{-3}$ & 9.53$\times10^{-6}$ & 1.41$\times10^{-5}$ & 4.83$\times10^{-5}$ & 5.73$\times10^{-5}$ \\
700  & 1.53$\times10^{-4}$ & 2.34$\times10^{-4}$ & 8.71$\times10^{-4}$ & 1.04$\times10^{-3}$ & 2.40$\times10^{-6}$ & 3.65$\times10^{-6}$ & 1.36$\times10^{-5}$ & 1.63$\times10^{-5}$ \\ 
800  & 4.46$\times10^{-5}$ & 7.02$\times10^{-5}$ & 2.86$\times10^{-4}$ & 3.44$\times10^{-4}$ & 6.98$\times10^{-7}$ & 1.10$\times10^{-6}$ & 4.46$\times10^{-6}$ & 5.38$\times10^{-6}$ \\
900  & 1.46$\times10^{-5}$ & 2.37$\times10^{-5}$ & 1.05$\times10^{-4}$ & 1.27$\times10^{-4}$ & 2.27$\times10^{-7}$ & 3.70$\times10^{-7}$ & 1.64$\times10^{-6}$ & 1.99$\times10^{-6}$ \\
1000 & 5.19$\times10^{-6}$ & 8.72$\times10^{-6}$ & 4.20$\times10^{-5}$ & 5.16$\times10^{-5}$ & 8.10$\times10^{-8}$ & 1.36$\times10^{-7}$ & 6.56$\times10^{-7}$ & 8.07$\times10^{-7}$ \\
\bottomrule
\end{tabular}
\caption{Cross sections (in fb) for $\SPP$ single production, given for different $\SPP$ masses and LHC energies. The cross sections have been computed for different values of the parameters that determine the $S W W$ coupling: $\xi=4\pi$ and $\Lambda=5M_S$ or $\Lambda=10M_S$.}
\label{tab:sigmasingle}
\end{table}

\begin{figure}[t]
\centering
\includegraphics[scale=0.5]{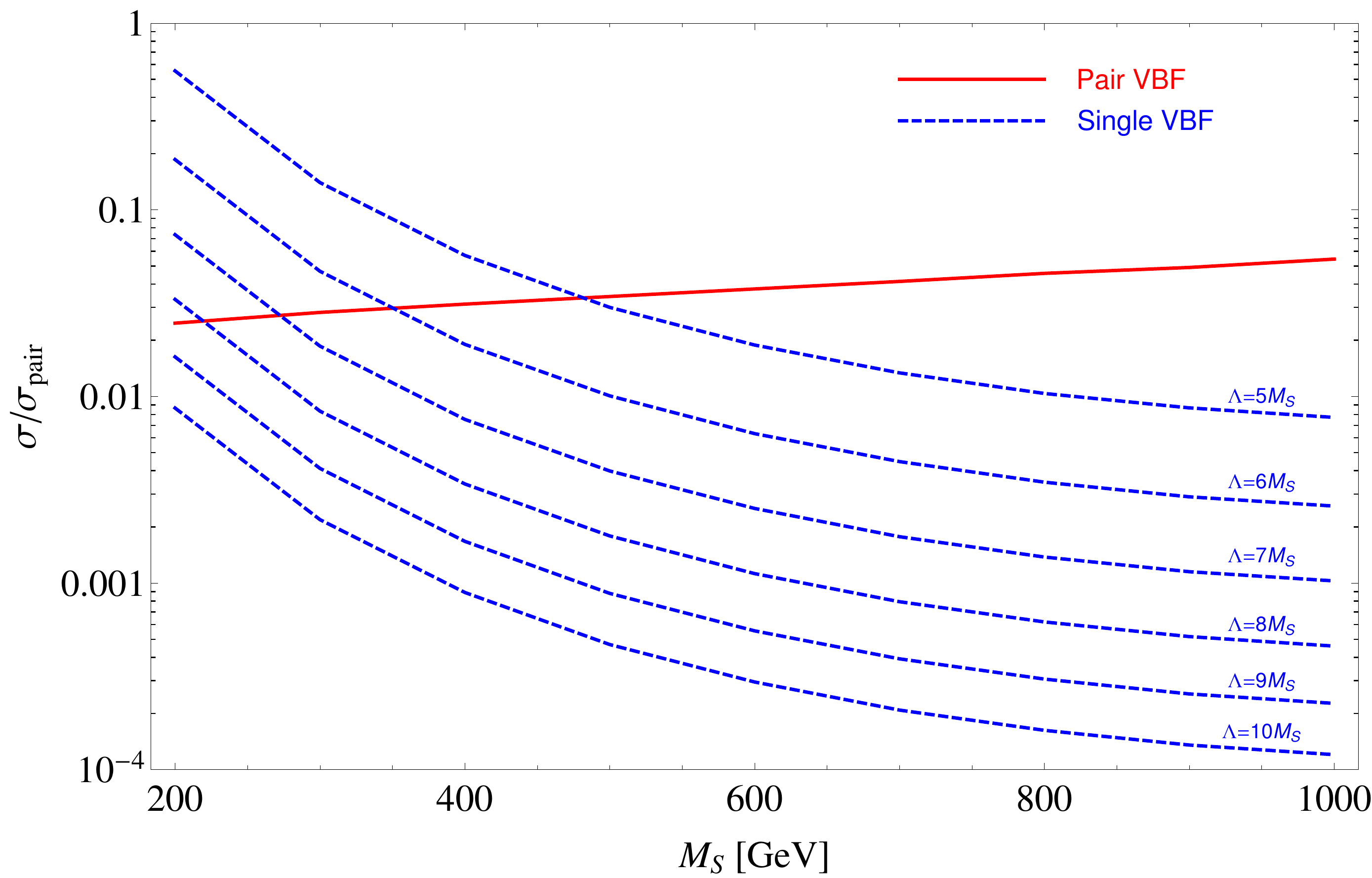}
\caption{\label{fig:sigmaratio}Ratios of cross sections at 8 TeV for single and pair production of $\SPP$ via VBF with respect to $s$-channel pair production ($\sigma_{\mbox{\scriptsize VBF single}}/\sigma_{\mbox{\scriptsize $s$-channel pair}}$ and $\sigma_{\mbox{\scriptsize VBF pair}}/\sigma_{\mbox{\scriptsize $s$-channel pair}}$). Single production cross sections have been computed for different values of the $\Lambda/M_S$ ratio.}
\end{figure}

\begin{figure}[t]
\centering
\includegraphics[scale=0.5]{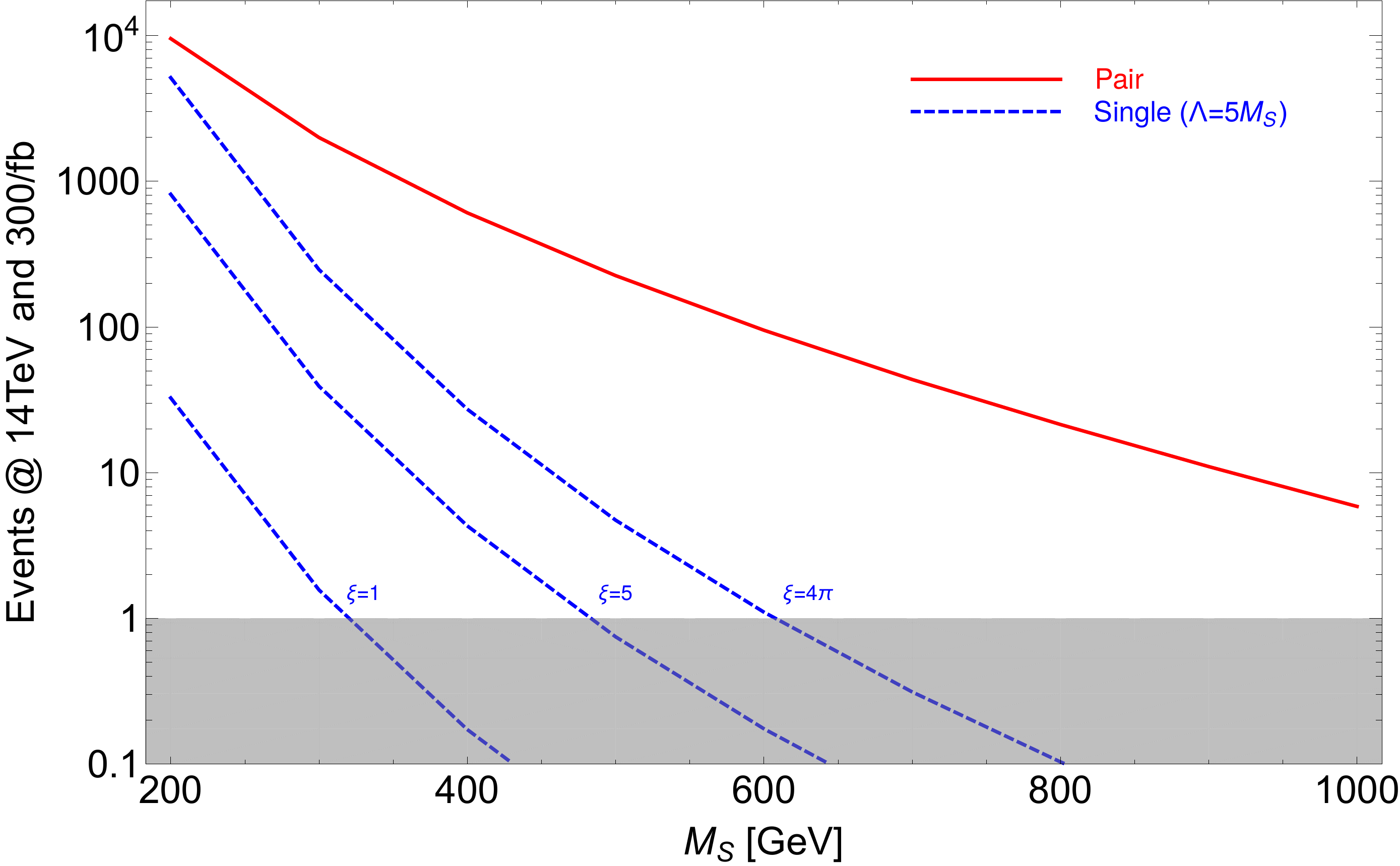}
\caption{\label{fig:events}Number of signal events for pair production and single production at 14~TeV, with an integrated luminosity of 300 fb$^{-1}$. The shaded region corresponds to less than one event and therefore an undetectable signal. The curves for single production have been drawn for different values of $\xi$ and can be derived from Tab.~\ref{tab:sigmasingle} by taking into account that $\sigma_{\text{single}} (\xi) = \sigma_{\text{single}}(\xi=4\pi)\times [\xi /(4\pi)]^2$.}
\end{figure}

An interesting feature of single production, which can be immediately noticed in Fig.~\ref{fig:sigmaratio}, is its peculiar scaling property: comparing the ratios between the cross sections of single production with pair production ($\sigma_{\mbox{\scriptsize VBF single}}/\sigma_{\mbox{\scriptsize $s$-channel pair}}$) and of pair production via VBF with pair production ($\sigma_{\mbox{\scriptsize VBF pair}}/\sigma_{\mbox{\scriptsize $s$-channel pair}}$), it can be noticed that single production is more relevant at low $\SPP$ masses and becomes negliglible at high $\SPP$ masses. The behaviour of pair production via VBF is complementary: the cross section grows relatively to pair production for increasing $\SPP$ masses. In the production processes of different states (such as quarks) at the LHC, single production usually becomes dominant at high masses. But, due to the peculiar scaling property of the $W W S$ vertex, this is \emph{not} the case for the classes of models described here. However, this behaviour opens up the possibility to develop dedicated strategies for searches of singly-produced light doubly charged scalars. The number of signal events in the final state is not huge, of course, and it strongly depends on the values of the $S W W$ coupling parameters $\Lambda$ and $\xi$. However, as shown in Fig.~\ref{fig:events}, the numbers of events for single production can be large enough to be detectable, especially at low masses. A dedicated analysis is in progress~\cite{PhenoAnalysis}.

The search strategies for doubly-charged scalars depend on their decay channels, which can be divided into $WW$ and $l_al_b$, also accounting for LFV. Considering specific selections and kinematic cuts, the number of expected signal events can be obtained by multiplying the cross section by the integrated luminosity $L$, the branching ratios (BRs) of the $\SPP$ in all its possible decay channels, and the efficiencies $\epsilon$ corresponding to all channels as:
\begin{equation}
 N^{\rm events}_{ij}=\sigma\ L \ {\rm BR}(\SPP)_i \ {\rm BR}(\SMM)_j \ \epsilon_{ij},
\end{equation}
where $i$ and $j$ run over all possible decay channels. The three categories of points identified in the previous sections are characterised by different patterns of branching ratios into these channels. Rescaling (cf.\ Sec.~\ref{sec:nu-mass}) the benchmark points that comply with all the bounds discussed in the previous sections, it is possible to obtain branching ratios and total widths for different $\SPP$ masses. A sample of the results obtained for various $\SPP$ masses between 200~GeV and 1000~GeV and for each category of points is given in Appendix~\hyperref[sec:app_C]{C}. The BR values have been obtained using a MC generator (BRIDGE~\cite{Meade:2007js}) on our MadGraph model files, and therefore the samples shown in the tables are just representative and they have the purpose to identify general features of the different categories.

The results shown in the tables allow us to derive some consequences for the different search strategies:

\begin{itemize}

\item For {\bf all points} the balance between decays into leptons and $WW$ is crucially determined by the values of $\xi$ and $\Lambda$. However, while $\xi$ is constrained to be in the perturbative region $0<\xi<4\pi$, the value of $\Lambda$ is only bounded from below by the (rough) $5M_S$ limit, which allows to describe the process of production of the doubly charged scalar process with an accuracy of the order of $(M_S/\Lambda)^2 \approx 4\%$. Due to the scaling of the $SWW$ coupling ($\xi/\Lambda^3$), the decays into $WW$ will be more suppressed as the mass of the scalar increases and indeed, for large masses $M_S \sim 1000$~GeV, the decay into $WW$ is almost always suppressed.

\item For {\bf purple points}, in all the range of masses considered, the doubly-charged scalar decays either to $WW$ or $e\tau$, with a small fraction of events possibly decaying to $\mu\mu$ when the leptonic decays dominate. This means that, to test purple points, different channels can be explored, regardless of the value of $M_S$ in the explored range (200~GeV -- 1~TeV):
\begin{enumerate}

 \item ${\rm BR}(WW)=100\%$, ${\rm BR}(e\tau)=0\%$, ${\rm BR}(\mu\mu)=0\%$: this scenario leads to the only possible final state $WWWW$.
 
 \item ${\rm BR}(WW)=50\%$, ${\rm BR}(e\tau)=50\%$, ${\rm BR}(\mu\mu)=0\%$: this scenario allows for three different final states, namely $WWWW$, $WWe\tau$, and $e\tau e\tau$.
 
 \item ${\rm BR}(WW)=0\%$, ${\rm BR}(e\tau)=98\%$, ${\rm BR}(\mu\mu)=2\%$: this scenario can be explored in two final states, $e\tau e\tau$ and $e\tau \mu\mu$, while the branching ratios are too small to make the $\mu\mu\mu\mu$ final state relevant enough.
 
\end{enumerate}

\item For {\bf red points} the only allowed decays are into $WW$ or $e\mu$. However, the leptonic decay is always dominant for large masses, while the $WW$ decay can dominate at low masses depending on the parameters of the $SWW$ coupling. Therefore, to test red points, different channels can be explored at different masses:
\begin{enumerate}

 \item ${\rm BR}(WW)=100\%$, ${\rm BR}(e\mu)=0\%$: this scenario leads to the only possible final state $WWWW$ and can be explored in the small mass region ($M_S\simeq200$~GeV).
 
 \item ${\rm BR}(WW)=50\%$, ${\rm BR}(e\mu)=50\%$: this scenario allows for three different final states: $WWWW$, $WWe\mu$, and $e\mu e\mu$, and they can be explored up to the medium-sized mass region ($M_S\lesssim 600$~GeV).
 
 \item ${\rm BR}(WW)=0\%$, ${\rm BR}(e\mu)=100\%$: this scenario leads to the only possible final state $e\mu e\mu$ and can be explored in the full range of $M_S$ masses.
 
\end{enumerate}

\item For {\bf blue points} it is only possible to define strategies for large masses, as it is not feasible to determine any benchmark point that satisfies all the flavour constraints in the low mass region. The preferred decay of the doubly charged scalar in this scenario is almost always into $ee$, with a small fraction of events possibly decaying to $e\tau$. However, depending on the $SWW$ coupling parameters, it is possible to have sizable $WW$ decays in the $600$~GeV mass region. Therefore, the most interesting channels for blue points are:
\begin{enumerate}

 \item BR($WW$)=40\%, BR($ee$)=50\%, BR($e\tau$)=10\%: this scenario leads to various final states: $WWWW$, $WWee$, $eeee$, and $eee\tau$, while the $WWe\tau$ and $e\tau e\tau$ final states are not very interesting due to the small branching ratios. However, this scenario can only be explored in the medium mass region, $M_S\simeq 600$~GeV.
 
 \item BR($WW$)=0\%, BR($ee$)=83\%, BR($e\tau$)=17\%: this scenario leads to three final states: $eeee$, $eee\tau$, and $e\tau e\tau$. It can be explored for all masses where blue points can be defined (i.e., $M_S\gtrsim 600$~GeV).
 
 \item BR($WW$)=0\%, BR($ee$)=100\%, BR($e\tau$)=0\%: this scenario only leads to the final state $eeee$ and can only be explored in the large mass region ($M_S\gtrsim 1000$~GeV).
 
\end{enumerate}

\end{itemize}
Due to the smallness of couplings, it could also be useful to understand if the $\SPP$ can be long-lived enough to produce displaced vertices or different signatures, but the size of the total widths in all the benchmarks analysed largely rules out this possibility: considering a displaced vertex resolution of $\sim10~\mu$m~\cite{Chatrchyan:2014fea}, corresponding to a lifetime $\tau\sim10^{-13}$s, the width of the $\SPP$ has to be $\Gamma\lesssim 10^{-12}$GeV. However, the scenarios explored with widths larger than $10^{-5}$~GeV are quite far from this regime.

The efficiencies associated to the decay channels can only be obtained running a full simulation that also reproduces detector effects. This simulation has been performed in~\cite{delAguila:2013mia} for different values of the LHC energy, and the values provided in this analysis will be used in the following. The numbers of events for representative values of the branching ratios and the exclusion confidence levels for the expected signal events in the different channels are shown in Tab.~\ref{tab:eventseCL7}. According to~\cite{delAguila:2013mia}, no expected background and no observed events have been assumed. The uncertainty on the signal events has been taken to be as large as $20\%$. Exclusion confidence levels above 95\% correspond to a 2$\sigma$ exclusion; it is already possible to see that some of the scenarios considered in our study, even if complying with all the flavour bounds, are already excluded by LHC data, and they have been highlited in the table. However, most of the points considered are well within the allowed region, and we report all results for transparency and reproductibility. 

\begin{table}[t!]
\tiny
\centering
\begin{tabular}{c|ccc||cc|cc|cc|cc}
\toprule
\midrule
\multicolumn{12}{c}{\bf\scriptsize Purple Points} \\
\midrule\midrule
\multirow{3}{*}{$M_S$ [GeV]} & \multicolumn{3}{c||}{BR (\%)} & \multicolumn{8}{c}{Channels} \\
& \multirow{2}{*}{$WW$} & \multirow{2}{*}{$e\tau$} & \multirow{2}{*}{$\mu\mu$} & \multicolumn{2}{c|}{$WWWW$} & \multicolumn{2}{c|}{$WW e\tau$} & \multicolumn{2}{c|}{$e\tau e\tau$} & \multicolumn{2}{c}{$e\tau \mu\mu$} \\
& & & & Events & eCL (\%) & Events & eCL (\%) & Events & eCL (\%) & Events & eCL (\%) \\
\midrule
\multirow{3}{*}{200}  & 100 & 0  & 0 & 0.04 & 3.9 & -    & -   & -    & -    & -    & -    \\
                      & 50  & 50 & 0 & 0.01 & 1.0 & 0.02 & 2.0 & 0.94 & 60.2 & -    & -    \\
                      & 0   & 98 & 2 & -    & -   & -    & -   & 3.61 & {\bf\scriptsize 96.5} & 0.19 & 17.2 \\
\midrule
\multirow{3}{*}{600}  & 100 & 0  & 0 & 0.00 & 0.0 & -    & -   & -    & -   & -    & -   \\
                      & 50  & 50 & 0 & 0.00 & 0.0 & 0.00 & 0.0 & 0.00 & 0.0 & -    & -   \\
                      & 0   & 98 & 2 & -    & -   & -    & -   & 0.02 & 2.0 & 0.00 & 0.0 \\
\bottomrule
\end{tabular}

\vskip 10pt

\begin{tabular}{c|cc||cc|cc|cc}
\toprule
\midrule
\multicolumn{9}{c}{\bf\scriptsize Red Points} \\
\midrule\midrule
\multirow{3}{*}{$M_S$ [GeV]} & \multicolumn{2}{c||}{BR (\%)} & \multicolumn{6}{c}{Channels} \\
& \multirow{2}{*}{$WW$} & \multirow{2}{*}{$e\mu$} & \multicolumn{2}{c|}{$WWWW$} & \multicolumn{2}{c|}{$WW e\mu$} & \multicolumn{2}{c}{$e\mu e\mu$}  \\
& & & Events & eCL (\%) & Events & eCL (\%) & Events & eCL (\%) \\
\midrule
\multirow{3}{*}{200}  & 100 & 0   & 0.04 & 3.9 & -    & -   & -    & -     \\
                      & 50  & 50  & 0.01 & 1.0 & 0.06 & 5.8 & 5.53 & {\bf\scriptsize 99.3}  \\
                      & 0   & 100 & -    & -   & -    & -   & 22.1 & {\bf\scriptsize 100.0} \\
\midrule
\multirow{2}{*}{600}  & 50  & 50  & 0.00 & 0.0 & 0.00 & 0.0 & 0.02 & 2.0  \\
                      & 0   & 100 & -    & -   & -    & -   & 0.09 & 85.9 \\
\bottomrule
\end{tabular}

\vskip 10pt

\begin{tabular}{c|ccc||cc|cc|cc|cc}
\toprule
\midrule
\multicolumn{12}{c}{\bf\scriptsize Blue Points} \\
\midrule\midrule
\multirow{3}{*}{$M_S$ [GeV]} & \multicolumn{3}{c||}{BR (\%)} & \multicolumn{8}{c}{Channels} \\
& \multirow{2}{*}{$WW$} & \multirow{2}{*}{$ee$} & \multirow{2}{*}{$e\tau$} & \multicolumn{2}{c|}{$WWWW$} & \multicolumn{2}{c|}{$WW ee$} & \multicolumn{2}{c|}{$eeee$} & \multicolumn{2}{c}{$eee\tau$} \\
& & & & Events & eCL (\%) & Events & eCL (\%) & Events & eCL (\%) & Events & eCL (\%) \\
\midrule
\multirow{2}{*}{600}  & 40 & 50 & 10 & 0.00 & 0.0 & 0.00 & 0.0 & 0.02 & 2.0 & 0.00 & 0.0 \\
                      & 0  & 83 & 17 & -    & -   & -    & -   & 0.06 & 5.8 & 0.00 & 0.0 \\
\bottomrule
\end{tabular}

\caption{\label{tab:eventseCL7}Numbers of signal events and exclusion confidence levels at 7~TeV and an integrated luminosity of $4.9$~fb$^{-1}$, for different $\SPP$ masses and for the most relevant channels. The efficiencies used to determine the exclusion confidence levels are those computed in~\cite{delAguila:2013mia}. If a specific final state is not allowed with the given BRs, it has been labelled by the symbol ``-''; on the other hand, if the number of events is numerically smaller than $0.01$, the number of events has been set to $0.00$ and the corresponding exclusion CL to $0.0$. Scenarios which can be excluded at more than 95\% CL have been highlighted.}
\end{table}

Current experimental searches do not consider $\SPP$ decays into $WW$, and the selection and kinematic cuts are obsviously optimised for decays into leptons. It would be interesting to understand which kind of strategies could be useful to explore the channels fed by the decays into $W$'s (see e.g.~\cite{Han:2007bk}). In our case, an example can be provided for red points at low masses, for which we can consider the following process:
\begin{equation}
PP\to \SPP \SMM \to (W^\pm W^\pm)(e^\mp\mu^\mp).
\end{equation}
The branching ratios of the $W$-boson are ${\rm BR}(jj)\sim68\%$ and $\sum_i {\rm BR}(l_i\nu_i)\sim32\%$, so, considering a scenario in which the doubly-charged scalar decays in $50\%$ of all cases into $W$-bosons, we obtain the following possible channels: 
\begin{eqnarray}
\setlength{\arraycolsep}{2pt}
\begin{array}{llcccl}
1)&{\rm BR}(e^-\mu^-+4j)&\lesssim& 0.5\times0.5\times0.68^2 &\sim& 0.116\ , \\
2)&{\rm BR}(e^-\mu^-+2j+l^++\nu)&\lesssim& 0.5\times0.5\times0.68\times0.32 &\sim& 0.054\ , \\
3)&{\rm BR}(e^-\mu^-+2l^++2\nu)&\lesssim& 0.5\times0.5\times0.32^2 &\sim& 0.026\ .
\end{array}
\end{eqnarray}
The expected numbers of events for the three scenarios before any kinematics cut are shown in Tab.~\ref{tab:EventsllWW}.

\begin{table}[!t]
\centering\begin{tabular}{ccccccc}
Mass [GeV] & \multicolumn{6}{c}{Expected number of events} \\
\midrule
& \multicolumn{2}{c}{Scenario 1} & \multicolumn{2}{c}{Scenario 2} & \multicolumn{2}{c}{Scenario 3} \\
& 8~TeV & 14~TeV & 8~TeV & 14~TeV & 8~TeV & 14~TeV\\
\midrule
200  & 26.22 & 367.72 & 12.20 & 171.18 & 5.88 & 82.42 \\ 
300  & 4.48  & 77.02  & 2.08  & 35.86  & 1.00 & 17.26 \\
400  & 1.10  & 23.43  & 0.51  & 10.91  & 0.25 & 5.25  \\
500  & 0.33  & 8.69   & 0.15  & 4.04   & 0.07 & 1.95  \\
600  & 0.11  & 3.67   & 0.05  & 1.71   & 0.02 & 0.82
\end{tabular}
\caption{Expected number of events for the $llWW$ channel considering $W$ decays and before any selection or kinematics cuts. The luminosity associated to the energy of 8 TeV (14 TeV) is 20/fb (100/fb).}
\label{tab:EventsllWW}
\end{table}

The strategies we can propose to test the various final states are the following:
\begin{itemize}

\item[1)] Require a final state with same-sign electron and muon and 4 jets. Furthermore, impose that the invariant mass of same-sign $e\mu$ pair is in the $\SPP$ window and that the invariant mass of the two jet pairs corresponds to two $W$-bosons.

\item[2)] Require a final state with same-sign electron and muon, 2 jets, and missing transverse momentum. Furthermore, impose that the invariant mass of same-sign $e\mu$ pair is in the $\SPP$ window and that the invariant mass of the jet pair corresponds to the $W$-boson.

\item[3)] The identification of final states is quite challenging due to both the reduction in the cross section by the branching ratios and the difficulty in reconstructing the invariant mass of the $W$ because of the presence of 2 neutrinos, but it can be worth exploring this final state for very low masses of the doubly-charged scalar.

\end{itemize}

In this analysis we have just provided an overview of the possible phenomenology of this class of models; a more specific analysis will be performed in a forthcoming dedicated study~\cite{PhenoAnalysis}.

\section{\label{sec:conc}Conclusions and Outlook}

We have considered the implications of a rather minimal extension of the Standard Model involving just one extra particle, namely a single $SU(2)_L$ singlet scalar $S^{++}$ and its antiparticle $S^{--}$. In Sec.~\ref{sec:EFT} we proposed a model independent effective operator, which yields an effective coupling of $S^{\pm \pm}$ to pairs of same sign weak gauge bosons, $W^{\pm} W^{\pm}$. We also allowed tree-level couplings of $S^{\pm \pm}$ to pairs of same sign right-handed charged leptons $l^{\pm}_Rl'^{\pm}_R$ of the same or different flavour. In Sec.~\ref{sec:nu-mass} and Appendix~A we calculated explicitly the resulting two-loop diagrams in the effective theory responsible for neutrino mass and mixing. We discussed lepton number violation and lepton flavour violation in Secs.~\ref{sec:0nbb} and~\ref{sec:LFV}, then in Sec.~\ref{sec:benchmarks}, presented thirty example benchmark points for various $S^{\pm \pm}$ masses and couplings which can yield successful neutrino masses and mixing, consistent with limits on charged lepton flavour violation (LFV) and neutrinoless double beta decay. 

We then turned to the high energy phenomenology of the effective vertex in Sec.~\ref{sec:collider}, where we showed that the particle $S$ can also lead to very interesting phenomenology at colliders. In particular, we showed that LHC searches at 7 TeV are already able to exclude certain configurations otherwise allowed by flavour bounds. Furthermore, due to the structure of the $SWW$ vertex in this class of models, the yet experimentally unexplored decays into the $WW$ channel may be largely dominant, and the peculiar scaling behaviour of single production mechanism makes it especially relevant at low $S$ masses. In Appendix~\hyperref[sec:app_B]{B} we presented a new correlation for elements of the light neutrino mass matrix, which arises for a certain category of benchmark points and which is, in principle, also testable. In Appendix~\hyperref[sec:app_C]{C}, we gave the numerical values of the branching ratios of the doubly charged scalar for all classes of benchmark points found. A more detailed phenomenological study of the experimental reach for the most interesting channels of this class of models is under way.

In conclusion, we have studied the different phenomenological aspects of a class of models involving a single doubly charged scalar $S^{\pm\pm}$, with an effective coupling to two $W$-bosons and flavour violating couplings to right-handed charged leptons. This very basic setting leads to a huge variety of phenomena, starting from a viable neutrino mass matrix generated at 2-loop level, and extending over a variety of lepton number/flavour changing processes, through to the $S^{\pm\pm}$ discovery prospects in various channels at the LHC. We have shown that it is possible to find benchmark points which are consistent with all the low and high energy data available, demonstrating the complementarity between neutrino and LFV experiments and the LHC.

\section*{\label{sec:ack}Acknowledgements}

We would like to thank D.~A.~Ross and M.~A.~Schmidt for useful discussions. SFK acknowledges support from the STFC Consolidated ST/J000396/1 grant. During his time in Southampton, AM has been supported by a Marie Curie Intra-European Fellowship within the 7th European Community Framework Programme FP7-PEOPLE-2011-IEF, contract PIEF-GA-2011-297557. LP is financed in part through the NExT Institute. Finally, SFK and AM both acknowledge partial support from the European Union FP7 ITN-INVISIBLES (Marie Curie Actions, PITN-GA-2011-289442).

\section*{\label{sec:app_A}Appendix~A: The neutrino mass in detail}
\renewcommand{\theequation}{A-\arabic{equation}}
\setcounter{equation}{0}  

Even though it is based on a 2-loop diagram, the neutrino mass generated by the vertex displayed can be calculated analytically and the resulting expression are still somewhat economic in the simple approximation of neglecting the squared charged lepton masses in the denominators of the propagators inside the loop. This allows to write down a fully analytic expression for the neutrino mass, if some tricks are used. We will detail the calculation in this Appendix~A.

\begin{figure}
\centering
\begin{tabular}{lr}
\includegraphics[scale=0.45]{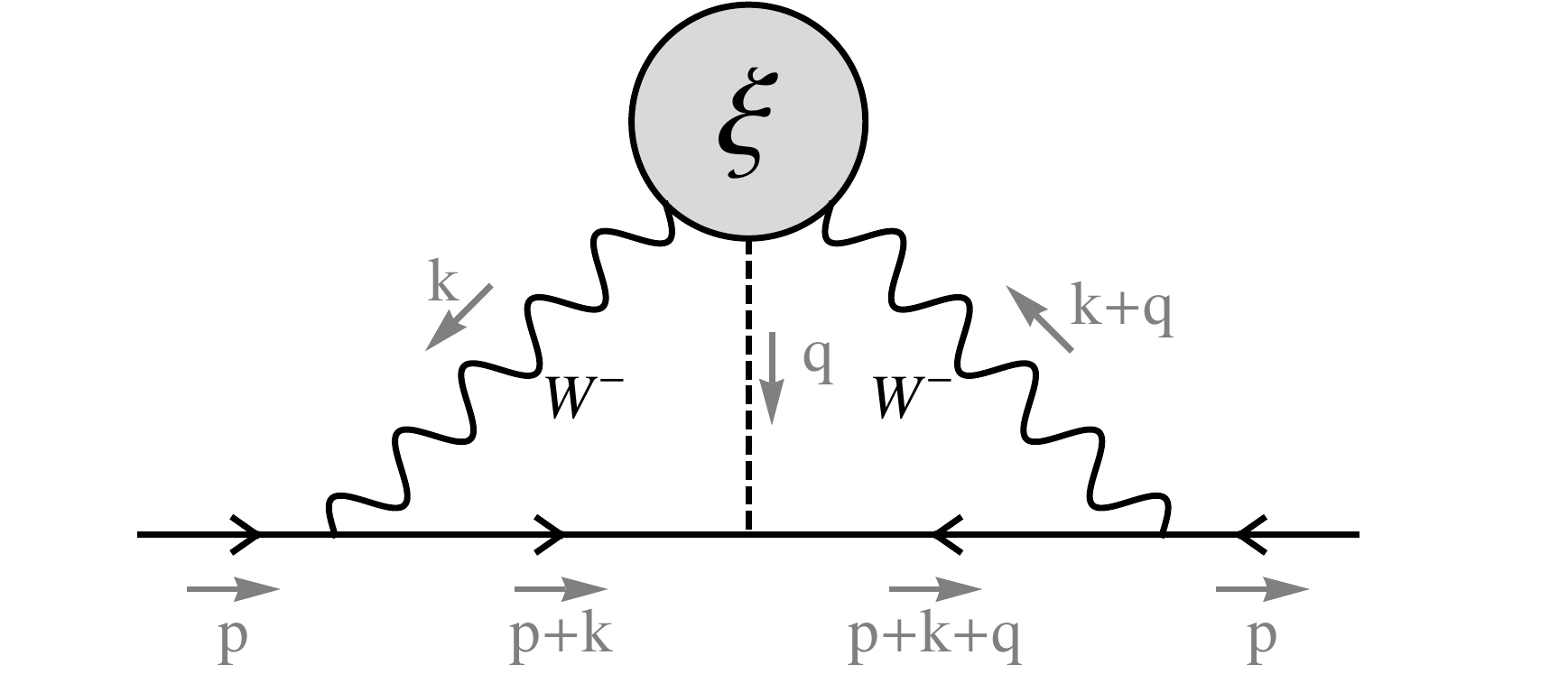} &
\includegraphics[scale=0.45]{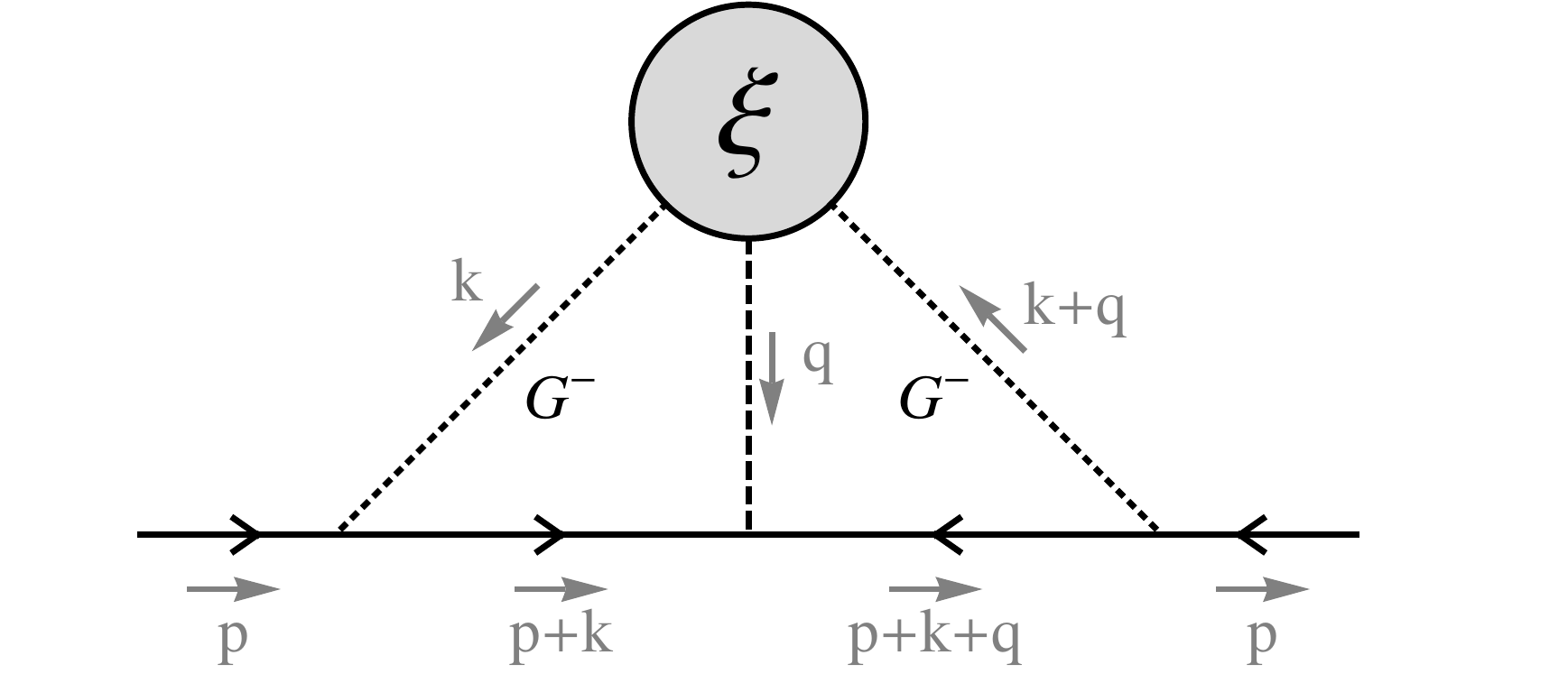}
\end{tabular}
\caption{\label{fig:neutrino_mass_calculation}
Momentum-assignments for the two relevant Feynman diagrams.
}
\end{figure}

First if all, depending on the $R_\xi$ gauge used, there can be more diagrams than displayed in Fig.~\ref{fig:neutrino_mass}. In particular, one can replace one of the $W$-bosons (or both of them) by their longitudinal modes, i.e., the corresponding would-be Goldstone bosons. It is only in unitary gauge ($\xi_W = \infty$) that only the diagram with $W$-bosons contributes, and this gauge is known to suffer from complicated and unfortunate forms of the gauge boson propagators and resulting divergences. We have instead decided to use Feynman-'t~Hooft gauge, $\xi_W = 1$, where it is easy to show that the two diagrams involving exactly one $W$-boson and one Goldstone boson cancel.\footnote{More generally, these diagrams must cancel in any gauge because of the form of the second vertex in Eq.~\eqref{eq:L_rel_expl}. The corresponding coupling is proportional to the momentum of the Goldstone boson, which is exactly opposite for the two diagrams involving exactly one $W$-boson and one Goldstone boson, while all other factors are identical. Thus, these diagrams have to cancel in any gauge, which also includes the special case of both being exactly zero in unitary gauge.} The remaining two diagrams are displayed in Fig.~\ref{fig:neutrino_mass_calculation} with the corresponding momentum assignments.

Using the Feynman rule $i f_{ab} (1 + \delta_{ab}) C P_R$ for the $S l_a l_b$ vertex,\footnote{The origin of the $\delta_{ab}$ is the additional factor 2 arising in the Feynman rule from the 2nd derivative of the path integral, in case the two charged leptons are identical in flavour. Note that this detail was not explicitly displayed in Ref.~\cite{Gustafsson:2012vj} which, however, does not play any role since it can be compensated by a simple redefinition of the corresponding couplings $f_{ab}$.} one can easily compute the resulting self-energy corrections:
\begin{eqnarray}
-i \Sigma_{ab}^W &=& \frac{i g^4 \xi v^4 m_a m_b f_{ab} (1 + \delta_{ab})}{4 \Lambda^3} C P_L \int \frac{d^4 k}{(2 \pi)^4} \frac{d^4 q}{(2 \pi)^4} \gamma_\mu \gamma_\nu \frac{1}{(p+k)^2-m_a^2} \frac{1}{(p+k+q)^2-m_b^2}\nonumber\\
&& \cdot \frac{1}{q^2-M_S^2} \frac{1}{k^2-M_W^2} \frac{1}{(k+q)^2-M_W^2} \left[ g^{\mu \alpha} - \frac{(1- \xi_W) k^\mu k^\alpha}{k^2 - \xi_W M_W^2} \right] \left[ g_\alpha^\nu - \frac{(1- \xi_W) k_\alpha k^\nu}{k^2 - \xi_W M_W^2} \right],\nonumber\\
-i \Sigma_{ab}^G &=& \frac{-i g^2 \xi v^2 m_a m_b f_{ab} (1 + \delta_{ab})}{2 \Lambda^3 M_W^2} C P_L \int \frac{d^4 k}{(2 \pi)^4} \frac{d^4 q}{(2 \pi)^4} (\slashed{p} + \slashed{k}) (\slashed{p} + \slashed{k} + \slashed{q}) \ k\cdot (k+q)\nonumber\\
 &&  \cdot \frac{1}{(p+k)^2-m_a^2} \frac{1}{(p+k+q)^2-m_b^2} \frac{1}{q^2-M_S^2} \frac{1}{k^2-\xi_W M_W^2} \frac{1}{(k+q)^2-\xi_W M_W^2},
\end{eqnarray}
where $a$ and $b$ are flavour indices, $\xi$ is the coupling of the effective vertex from Eq.~\eqref{eq:L_rel_expl}, and $\Lambda$ is the high energy cutoff of the EFT. Obviously, $\Sigma_{ab}^W$ denotes the self-energy corrections coming from the diagram with two $W$-bosons and $\Sigma_{ab}^G$ does the same for the one with two Goldstone bosons, while the mixed diagrams cancel.

As already explained, we can greatly simplify the calculation if we use Feynman-'t~Hooft gauge ($\xi_W = 1$) and neglect $m_{a,b}^2$ in the denominators. We can also use the fact that the external momentum can be set to zero, $p = 0$, if we are only interested in a mass correction. Furthermore, substituting the $q$-integration by an integration over $r \equiv k + q$, it is easy to see that the denominators are symmetric under the exchange $k \leftrightarrow r$, which also forces the numerators to have the same symmetry property. This symmetrisation gets rid of most of the $\gamma$-matrices. Using the fact that $M_W = \frac{1}2 g v$, the 2-loop neutrino mass matrix, $\mathcal{M}_{\nu, ab}^\textrm{2-loop} C P_L = \Sigma_{ab}^W + \Sigma_{ab}^G$, can be written as follows:
\begin{equation}
 \mathcal{M}_{\nu, ab}^\textrm{2-loop} = \frac{2 \xi m_a m_b f_{ab} (1 + \delta_{ab})}{\Lambda^3} \cdot \mathcal{I},
 \label{eq:M_nu}
\end{equation}
where the remaining integral looks comparatively simple and is given by
\begin{equation}
 \mathcal{I} \equiv \int \frac{d^4 k}{(2 \pi)^4} \frac{d^4 r}{(2 \pi)^4} \frac{(k r)^2 - 8 M_W^2}{k^2 (k^2 - M_W^2) r^2 (r^2 - M_W^2) [(k-r)^2 - M_S^2]}.
 \label{eq:I_def}
\end{equation}
Note that the basic structure of Eq.~\eqref{eq:M_nu} is in perfect agreement with e.g.\ the result obtained in Ref.~\cite{Gustafsson:2012vj}, which is no surprise as that model contains an explicit realisation of the effective vertex displayed in Fig.~\ref{fig:vertex}. Our remaining task in this appendix is to explicitly calculate the integral $\mathcal{I}$.

To tackle the integration, it is best to make use of dimensional regularisation~\cite{tHooft:1972fi},\footnote{See, e.g., Ref.~\cite{Collins:1984xc} for a very detailed treatment.} $d^4 p \to \mu^\epsilon d^d p$ and $(2 \pi)^4 \to (2 \pi)^d$ where $d = 4 - \epsilon$ is the number of dimensions and $\mu$ is an arbitrary energy scale. This is the most appropriate regularisation scheme for EFTs~\cite{Pich:1998xt}, since it does not lead to any problems with summing up the lowest order contributions correctly. Then, the basic trick is to rewrite the numerator in Eq.~\eqref{eq:I_def} in order to obtain a series of simpler integrals:
\begin{eqnarray}
 (k r)^2 &=& \frac{1}{4} [(k-r)^2 - M_S^2]^2 - \frac{1}2 [(k-r)^2 - M_S^2] (k^2 + r^2) + \frac{M_S^2}2 [(k-r)^2 - M_S^2]\nonumber \\
 && + \frac{1}{4} [k^4 + r^4 + 2 k^2 r^2 - 2 M_S^2 (k^2 + r^2) + M_S^4].
 \label{eq:numerator}
\end{eqnarray}
The next point is to restore the charged lepton mass scale. This is not strictly important for the final result, but it will make it easier to do the integral decomposition, as it avoids artificial infrared divergences on the way to the final result which could make it very difficult to perform a clean computation of the limit $m \to 0$. Thus, the denominator of the integral from Eq.~\eqref{eq:I_def} is changed to:
\begin{equation}
 k^2 (k^2 - M_W^2) r^2 (r^2 - M_W^2) [(k-r)^2 - M_S^2] \to (k^2-m^2) (k^2 - M_W^2) (r^2-m^2) (r^2 - M_W^2) [(k-r)^2 - M_S^2].
 \label{eq:I_denom}
\end{equation}
We will finally take the limit $m \to 0$, but if the reader would like to keep the dependence on the charged lepton mass, one could for example take $m = \sqrt{m_e^2 + m_\mu^2 + m_\tau^2} \simeq m_\tau$ or $m = (m_e + m_\mu + m_\tau)/3$. The exact value does not matter as long as the charged lepton masses are indeed negligible compared to the other masses involved, since the result only depends logarithmically on it. In the calculation presented, however, we have to be careful to take $m$ to be the same in both factors in the denominator, in order not to spoil the symmetry $k \leftrightarrow r$ of the integrand. 

Adopting the method of dimensional regularisation, the integral to solve is then:
\begin{equation}
 \mathcal{I} \equiv \mu^{2 \epsilon} \int \frac{d^d k}{(2 \pi)^d} \frac{d^d r}{(2 \pi)^d} \frac{(k r)^2 - 8 M_W^2}{(k^2-m^2) (k^2 - M_W^2) (r^2-m^2) (r^2 - M_W^2) [(k-r)^2 - M_S^2]}.
 \label{eq:I_DimReg}
\end{equation}
The next step is to make use of the following decomposition:
\begin{equation}
 \frac{1}{(p^2 - M^2) (p^2 - m^2)} = \frac{1}{M^2-m^2} \left( \frac{1}{p^2 - M^2} - \frac{1}{p^2 - m^2}\right),
 \label{eq:fractions}
\end{equation}
which allows to rewrite the denominator $D \equiv (k^2-m^2) (k^2 - M_W^2) (r^2-m^2) (r^2 - M_W^2) [(k-r)^2 - M_S^2]$ as
\begin{equation}
 \frac{1}{D} = \frac{1}{(M_W^2-m^2)^2} \sum_{(4)} \frac{1}{(k^2-m_1^2) (r^2-m_2^2) [(k-r)^2 - M_S^2]},
 \label{eq:denom}
\end{equation}
with the sum $\sum_{(4)}$ being defined as
\begin{equation}
 \sum_{(4)} F(m_1, m_2) \equiv F(M_W,M_W) - F(M_W,m) - F(m,M_W) + F(m,m).
 \label{eq:sum_4}
\end{equation}
Using Eq.~\eqref{eq:numerator}, one can rewrite the numerator $N \equiv (k r)^2 - 8 M_W^2$ as
\begin{eqnarray}
 N &=& \frac{N_1 + N_2 + N_3 + N_4 + N_5}{4},\ \ {\rm where}\ \ N_1 = [(k - r)^2 - M_S^2]^2 - 2 [(k - r)^2 - M_S^2] (k^2 + r^2 - M_S^2),\nonumber\\
 && N_2 = k^4 + r^4,\ \ N_3 = 2 k^2 r^2,\ \ N_4 = -2 M_S^2 (k^2 + r^2),\ \ N_5 = M_S^4 - 32 M_W^4.
 \label{eq:expl}
\end{eqnarray}
Thus, we have to compute five different integrals which add up to the total integral,
\begin{equation}
 \mathcal{I} = \mathcal{I}_1 + \mathcal{I}_2 + \mathcal{I}_3 + \mathcal{I}_4 + \mathcal{I}_5,\ \ \ {\rm where}\ \ \ \mathcal{I}_i \equiv \frac{\mu^{2\epsilon}}{4} \int \frac{d^d k}{(2 \pi)^d} \frac{d^d r}{(2 \pi)^d} \frac{N_i}{D},
 \label{eq:total_int}
\end{equation}
with the integrands explicitly given by
\begin{equation}
 \frac{N_i}{D} = \frac{1}{(M_W^2-m^2)^2} \sum_{(4)} \frac{N_i}{(k^2-m_1^2) (r^2-m_2^2) [(k-r)^2 - M_S^2]}.
 \label{eq:integrands}
\end{equation}
One can determine the different contributions to the integral to be
\begin{eqnarray}
 (M_W^2-m^2)^2\ \mathcal{I}_1 &=& \sum_{(4)} \big[\underbrace{(M_S^2 - m_1^2 - m_2^2) \mathcal{J}_1(m_1) \mathcal{J}_1(m_2)}_{=(1)} \underbrace{- \mathcal{J}_0(0) [\mathcal{J}_1(m_1) + \mathcal{J}_1(m_2)]}_{=(2)} \big],\nonumber\\
 (M_W^2-m^2)^2\ \mathcal{I}_2 &=& \sum_{(4)} \big[ \underbrace{[\mathcal{J}_1(m_1) + \mathcal{J}_1(m_2)] \mathcal{J}'_1(M_S)}_{=(3)} + \underbrace{[\mathcal{J}'_1(m_1) + \mathcal{J}'_1(m_2)] \mathcal{J}_1(M_S)}_{=(4)} \nonumber\\
 && + \underbrace{[m_1^2 \mathcal{J}_1(m_1) + m_2^2 \mathcal{J}_1(m_2)]}_{=(5)} + \underbrace{(m_1^4 + m_2^4) \mathcal{K}(m_1, m_2, M_S)}_{=(6)} \big],\nonumber\\
 (M_W^2-m^2)^2\ \mathcal{I}_3 &=& \sum_{(4)} \underbrace{2 \mathcal{J}_0(0) \mathcal{J}_1(M_S)}_{=(7)} + \underbrace{2[m_1^2 \mathcal{J}_1(m_1) + m_2^2 \mathcal{J}_1(m_2)] \mathcal{J}_1(M_S)}_{=(8)} + \underbrace{2 m_1^2  m_2^2\ \mathcal{K}(m_1, m_2, M_S)}_{=(9)},\nonumber\\
 (M_W^2-m^2)^2\ \mathcal{I}_4 &=& \sum_{(4)} \big[ \underbrace{-2 M_S^2 [\mathcal{J}_1(m_1) + \mathcal{J}_1(m_2)] \mathcal{J}_1(M_S)}_{=(10)} \underbrace{-2 M_S^2 (m_1^2 +  m_2^2) \mathcal{K}(m_1, m_2, M_S)}_{=(11)} \big],\nonumber\\
 (M_W^2-m^2)^2\ \mathcal{I}_5 &=& \sum_{(4)} \underbrace{(M_S^4 - 32 M_W^4) \mathcal{K}(m_1, m_2, M_S)}_{=(12)},
 \label{eq:integral_definitions}
\end{eqnarray}
where the integral functions introduced are given by
\begin{eqnarray}
 && \mathcal{J}_s (M) \equiv \frac{\mu^\epsilon}2 \int \frac{d^d p}{(2 \pi)^d} \frac{1}{(p^2-M^2)^s},\ \ \ \mathcal{J}'_s (M) \equiv \frac{\mu^\epsilon}2 \int \frac{d^d p}{(2 \pi)^d} \frac{p^2}{(p^2-M^2)^s},\nonumber\\
 && \mathcal{K}( M_1, M_2, M_3 ) \equiv \frac{\mu^{2\epsilon}}{4} \int \frac{d^d k}{(2 \pi)^d} \frac{d^d r}{(2 \pi)^d} \frac{1}{(k^2-M_1^2) (r^2-M_2^2) [(k-r)^2 - M_3^2]}.
 \label{eq:integral_pieces}
\end{eqnarray}
Performing the summation explicitly, cf.\ Eq.~\eqref{eq:sum_4}, one can see that the contributions $(2)$, $(3)$, $(4)$, $(7)$, $(8)$, and $(10)$ all vanish identically. For example, the contribution $(7)$ is independent of $m_{1,2}$, such that the summation leads to
\begin{equation}
 \sum_{(4)} (7) = \sum_{(4)} 2 \mathcal{J}_0(0) \mathcal{J}_1(M_S) = 2 \mathcal{J}_0(0) \mathcal{J}_1(M_S) \cdot (1-1-1+1) = 0.
 \label{eq:sum_vanish}
\end{equation}
Thus, only the contributions $(1)$, $(5)$, $(6)$, $(9)$, $(11)$, and $(12)$ survive, and we only need the two integrals $\mathcal{J}_1$ and $\mathcal{K}$ in Eq.~\eqref{eq:integral_pieces}, where the former is an effective 1-loop integral that appears always in products with other 1-loop integrals in the final 2-loop expression. Putting all the pieces together, it is easy to show that the final result will be of the form
\begin{eqnarray}
 \mathcal{I} &\equiv& \frac{1}{(M_W^2-m^2)^2}\sum_{(4)} [(1) + (5) + (6) + (9) + (11) + (12)] = \nonumber\\
 && = \Big[ (M_S^2 - 2 M_W^2)\ \mathcal{J}_1^2(M_W) + (M_S^2 - 2 m^2)\ \mathcal{J}_1^2(m) - (M_S^2 - M_W^2 - m^2)\ \mathcal{J}_1(M_W) \mathcal{J}_1(m)\nonumber\\
 && + 2 (M_W^2 - m^2) [\mathcal{J}_1(M_W) - \mathcal{J}_1(m)] \mathcal{J}_1(M_S) + [(M_S^2 - 2 M_W^2)^2 - 32 M_W^4]\ \mathcal{K}(M_W, M_W, M_S)\nonumber\\
 && - 2 [(M_S^2 - M_W^2 - m^2)^2 - 32 M_W^4]\ \mathcal{K}(m, M_W, M_S) + [(M_S^2 - 2 m^2)^2 - 32 M_W^4]\ \mathcal{K}(m, m, M_S)\Big]\nonumber\\
 && \times \frac{1}{(M_W^2-m^2)^2},
 \label{eq:integral_solution}
\end{eqnarray}
since $\mathcal{K}( M_1, M_2, M_3 ) = \mathcal{K}( M_2, M_1, M_3 )$.

We remain with the calculation of the two integrals. Starting with $\mathcal{J}_1$, one can make use of Wick rotation and integration in $d$ dimensions to show
\begin{equation}
 \int \frac{d^d p}{(2 \pi)^d} \frac{1}{(p^2 - \Delta)^s} = \frac{i \Gamma (s- \frac{d}2)}{2^d \pi^{d/2} (-1)^s} \left( \frac{1}{\Delta} \right)^{s - d/2},
 \label{eq:J_int_1}
\end{equation}
where $\Gamma(x)$ is the Gamma function. Thus, the full integral is given by
\begin{equation}
 \mathcal{J}_1 (M) = \frac{i}{16 \pi^2} \frac{M^2}2 \Big\{ \frac2{\epsilon} + [C_\gamma - L_M] + \frac{\epsilon}{4} \Big[ (C_\gamma - L_M)^2 + \frac{\pi^2}{6} + 1 \Big] \Big\},
 \label{eq:J_int_2}
\end{equation}
where we have used the abbreviations
\begin{equation}
 C_\gamma = 1 - \gamma + \log (4\pi)\ \ \ {\rm and}\ \ \ L_M \equiv \log \left( \frac{M^2}{\mu^2} \right),
 \label{eq:J_int_3}
\end{equation}
and $\gamma = 0.5771...$ is the Euler-Mascheroni constant. What we actually need are products of $\mathcal{J}_1$'s, which is to be expected since these integrals arise from contributions to the 2-loop integrations which can be factorised into products of two 1-loop integrals. The relevant combinations are
\begin{eqnarray}
 \mathcal{J}_1^2 (M) &=& \frac{-1}{(16 \pi^2)^2} \frac{M^4}{4} \Big[ \frac{4}{\epsilon^2} + \frac{4}{\epsilon} (C_\gamma - L_M) + 2 (C_\gamma - L_M)^2 + \frac{\pi^2}{6} + 1 + \mathcal{O}(\epsilon) \Big], \nonumber\\
 \mathcal{J}_1 (M) \mathcal{J}_1 (m) &=& \frac{-1}{(16 \pi^2)^2} \frac{M^2 m^2}{4} \Big[ \frac{4}{\epsilon^2} + \frac2{\epsilon} (2C_\gamma - L_M - L_m) + \frac{1}2 (2 C_\gamma - L_M - L_m)^2 + \frac{\pi^2}{6} + 1 + \mathcal{O}(\epsilon) \Big],\nonumber\\
 && {\rm where}\ \ \ L_m \equiv \log \left( \frac{m^2}{\mu^2} \right).
 \label{eq:J_int_4}
\end{eqnarray}
Now, it is very easy to perform the limit $m\to 0$:
\begin{equation}
 \lim_{m\to 0} \mathcal{J}_1^2 (m) \to 0\ \ \ {\rm and}\ \ \ \lim_{m\to 0} \left[ \mathcal{J}_1 (M) \mathcal{J}_1 (m) \right] \to 0,
 \label{eq:J_int_5}
\end{equation}
such that the final resulting contributions are given by
\begin{equation}
 \sum_{(4)} \left. [(1) + (5)]\right|_{m \to 0} = (M_S^2 - 2 M_W^2)\ \mathcal{J}_1^2(M_W) + 2 M_W^2 \mathcal{J}_1(M_W) \mathcal{J}_1(M_S).
 \label{eq:J_int_6}
\end{equation}
Note that this contribution vanishes in the limit $M_W \ll M_S$.

The integral $\mathcal{K}( M_1, M_2, M_3 )$, in turn, is discussed in great detail in Ref.~\cite{Davydychev:1992mt}. In terms of the expressions used there, our integral is given by\footnote{In order to translate the conventions, one has to perform the replacement $\epsilon \to \epsilon/2$ in the expressions from Ref.~\cite{Davydychev:1992mt}.}
\begin{equation}
 \mathcal{K}( M_1, M_2, M_3 ) = \frac{\mu^{2 \epsilon}}{4 (2 \pi)^{2d}} I(1, 1, 1; M_1, M_2, M_3) \equiv P \times C.
 \label{eq:K_int_1}
\end{equation}
The easiest way to obtain a reasonable expression for $\mathcal{K}( M_1, M_2, M_3 )$ is to expand the prefactor $P \equiv \frac{\mu^{2 \epsilon}}{4 (2 \pi)^{2d}} \frac{\pi^{4-\epsilon} M_3^{2(1-\epsilon)}}{\Gamma (2-\epsilon/2)}$ and the remaining body $C$ of the expressions separately in $\epsilon$. We obtain
\begin{equation}
 P = P_0 + P_1 \epsilon + P_2 \epsilon^2 + \mathcal{O}(\epsilon^3)\ \ \ {\rm and}\ \ \ C = \frac{C_{-2}}{\epsilon^2} + \frac{C_{-1}}{\epsilon} + C_0 + \mathcal{O}(\epsilon),
 \label{eq:K_int_2}
\end{equation}
such that the final expression for the integral reads
\begin{equation}
 \mathcal{K}( M_1, M_2, M_3 ) = \frac{\Delta_{-2}}{\epsilon^2} + \frac{\Delta_{-1}}{\epsilon} + \Delta_0 + \mathcal{O}(\epsilon),
 \label{eq:K_int_3}
\end{equation}
where
\begin{eqnarray}
 \Delta_{-2} &=& P_0 C_{-2},\nonumber\\
 \Delta_{-1} &=& P_0 C_{-1} + P_1 C_{-2},\nonumber\\
     \Delta_0 &=& P_0 C_0 + P_1 C_{-1} + P_2 C_{-2}.
 \label{eq:K_int_4}
\end{eqnarray}
In full generality, the different pieces $P_i$ can be written as
\begin{eqnarray}
 P_0 &=& \frac{M_3^2}{4 (16 \pi^2)^2},\nonumber\\
 P_1 &=& \frac{M_3^2}{4 (16 \pi^2)^2} (C_\gamma - L_3),\nonumber\\
 P_2 &=& \frac{-M_3^2}{8 (16 \pi^2)^2} \left[ \frac{\pi^2}{6} - [C_\gamma - 2 (1-\gamma) - L_3]^2 \right],
 \label{eq:K_int_5}
\end{eqnarray}
where $L_3 \equiv \log (M_3^2/\mu^2)$. Note that we can set $M_3 = M_S$ in the final result. The pieces $C_i$ turn out to be
\begin{eqnarray}
 C_0 &=& \frac{-2 w (z-2)+4 z-2}{(1-w) (1-z)},\nonumber\\
 C_{-1} &=& \frac{1}{(1-w) (1-z)} \Big\{ -2 + 8 (w + z) - 6 w z + \gamma (1- 2 w - 2 z + w z) - 2 w \log \left( \frac{-w}{(1-w) (1-z)} \right)\nonumber\\
 && - 2 z \log \left( \frac{-z}{(1-w) (1-z)} \right) + 2 (1-w z) \log(1 - w z) + 2 (1-w z) \log \left( \frac{1 - w z}{(1-w) (1-z)} \right)\nonumber\\
 && - 2 (1- w z) \log \left( \frac{(1- w z)^2}{(1-w) (1-z)} \right) \Big],\nonumber
 \end{eqnarray}
 \begin{eqnarray}
 C_{-2} &=& \frac{1}{(1-w) (1-z)} \Big\{ -2 - \frac{7 \pi^2}{24} + 16 w + \frac{\pi^2}{4} w +  \gamma - 4 w \gamma - \frac{\gamma^2}{4} + \frac{w}2 \gamma^2 \nonumber\\
 && + z \left[ 2 (8 - 7 w) + \frac{\pi^2}{24} (6 + w) - (4 - 3 w) \gamma + (2 - w) \frac{\gamma^2}{4} \right] \nonumber\\
 && + \frac{w}2 \log^2 \left( \frac{-w}{(1-w) (1-z)} \right) +  \frac{z}2 \log^2 \left( \frac{-z}{(1-w) (1-z)} \right)+ 2 \log \left( \frac{1- w z}{(1-w) (1-z)} \right)\nonumber\\
 && +  \log \left( \frac{-w}{(1-w) (1-z)} \right) \Big[ w (-4 + \gamma) + w z \log \left( \frac{-z}{(1-w) (1-z)} \right) \nonumber\\
 && - (1 - w z) \log (1 - w z) + (1 - w z) \log \left( \frac{1 - w z}{1-w} \right) \Big] - 4 \log \left( \frac{(1- w z)^2}{(1-w)(1-z)} \right) \nonumber\\
 && +  \log \left( \frac{-z}{(1-w) (1-z)} \right) \left(z (-4 + \gamma) - (1 - w z) \log(1 - w z) + (1 - w z) \log \left( \frac{1- w z}{1-z} \right) \right)\nonumber\\
 && +  (1 - w z) (4 - \gamma) \log(1 - w z) + (-1 + w z) \log^2(1 - w z) \nonumber\\
 && -[w z (2 - \gamma) + \gamma] \log \left( \frac{1 - w z}{(1-w) (1-z)} \right) - (1 - w z) \log^2 \left( \frac{1 - w z}{(1-w) (1-z)} \right)\nonumber\\
 && + \log \left( \frac{1 - w z}{1-w} \right) \Big[-w z (-4 + \gamma) + \gamma + (1 - w z) \log \left( \frac{1 - w z}{1-w} \right) \Big]\nonumber\\
 && + (-w z (-4 + \gamma) + \gamma) \log \left( \frac{1 - w z}{1-z} \right) + (1 - w z) \log^2 \left( \frac{1 - w z}{1-z} \right) \nonumber\\
 && +  (1 - w z) \left[ {\rm Li}_2 \left( -w\frac{1-z}{1-w} \right)  + {\rm Li}_2 \left( -z\frac{1-w}{1-z} \right)  -  {\rm Li}_2 (w z) \right] \Big\},
 \label{eq:K_int_6}
\end{eqnarray}
where $w = (-1 + x + y + \lambda)/(2y)$, $z = (-1 + x + y + \lambda)/(2x)$, $\lambda = \sqrt{(1-x-y)^2 - 4 x y}$, $x = M_1^2 / M_3^2$, and $y = M_2^2 / M_3^2$, and the dilogarithm is given by
\begin{equation}
 {\rm Li}_2 (y) = - \int\limits_0^1 \frac{\log (1 - y t)}{t} dt.
 \label{eq:K_int_7}
\end{equation}
Glancing at Eq.~\eqref{eq:integral_solution}, we need to compute the three limits $M_1 = M_2 (=0)$ and $M_1 = 0 \neq M_2$ [since we need $\mathcal{K}(M_W, M_W, M_S)$, $\mathcal{K}(m, M_W, M_S)$, and $\mathcal{K}(m, m, M_S)$ and take the limit $m \to 0$]. The limit $M_1 = M_2  \neq 0$ implies $x = y = M_W^2 / M_S^2$, $w = z = (-1 + 2 x + \lambda)/(2x)$, and $\lambda = \sqrt{1-4 x}$, which leads to
\begin{eqnarray}
 C_0 (M_W, M_W, M_S) &=& -\frac{2 (1 - 4 z + z^2)}{(1 - z)^2},\nonumber\\
 C_{-1} (M_W, M_W, M_S) &=& \frac{1}{(1 - z)^2} \Big[ -2 + \gamma + z (16 - 6 z - 4 \gamma + z \gamma) - 2 (1 - z^2) \log(1 - z)\nonumber\\
 && + 2 (1 - z^2) \log(1 - z) - 4 z \log \left( \frac{-z}{(1 - z)^2)} \right) \Big],\nonumber\\
 C_{-2} (M_W, M_W, M_S) &=& \frac{1}{(1 - z)^2} \Big[ \frac{\pi^2}{24} (-7 + 12 z + z^2) - 2 (1 -16 z + 7 z^2) + (1 -8 z + 3 z^2) \gamma \nonumber\\
 && - (1 -4 z + z^2) \frac{\gamma^2}{4} - (1 - z^2) \log^2(1 - z) - 2 \log(1 + z)\nonumber\\
 && + (1 - z^2) \log(1 - z) [-2 + \gamma + 2 \log(1 + z)]  -(1 - z^2) \log^2(1 - z) \nonumber\\
 && - (1 - z^2) \log(1 - z) \left\{ -4 + \gamma + 2 \log\left( \frac{-z}{(1 - z)^2} \right) + 2 \log(1 + z) \right\} \nonumber\\
 && + z (2 (-4 + \gamma) \log\left( \frac{-z}{(1 - z)^2} \right) + (1 + z) \log^2\left( \frac{-z}{(1 - z)^2} \right)\nonumber\\
 && + 2 z \log(1 + z) ) - 2 (1 - z^2) {\rm Li}_2(z)\Big].
 \label{eq:K_int_8}
 \end{eqnarray}
Things get even easier when $M_1 = 0$ (i.e., $M_1 = m \to 0$). Carefully taking the limits allows to derive that $x=0$, $y = M_W^2 / M_S^2$, $w = 0$, $z = -\frac{y}{\lambda}$, and $\lambda = 1-y$, which implies
\begin{eqnarray}
 C_0 (0, M_W, M_S) &=& -2 (1 + y),\nonumber\\
 C_{-1} (0, M_W, M_S) &=& -2 - y (6 - \gamma ) + \gamma + 2 y \log y,\nonumber\\
 C_{-2} (0, M_W, M_S) &=& \frac{\pi^2}{24} (-7 + y) - \frac{1}{4} [8 + (-4 + \gamma) \gamma + 56 y - (12 - \gamma) \gamma y ] \label{eq:K_int_9}\\
 && - (1 - y) \log(1 - y) (2 - \log y) - \frac{y}2 \log y (-8 + 2 \gamma + \log y ) + (1 - y) {\rm Li}_2(y).\nonumber
 \end{eqnarray}
Finally, the limit $M_1 = M_2 = 0$, where $x = y = 0$, $w = z = 0$, and $\lambda = 1$, yields the simplest expressions:
\begin{eqnarray}
 C_0 (0, 0, M_S) &=& -2,\nonumber\\
 C_{-1} (0, 0, M_S) &=& -2 + \gamma,\nonumber\\
 C_{-2} (0, 0, M_S) &=& -2 + \gamma - \frac{7 \pi^2}{24} -\frac{\gamma^2}{4}.
 \label{eq:K_int_10}
 \end{eqnarray}
 Thus, the final resulting contributions are given by
 \begin{eqnarray}
 \mathcal{I} &\equiv& \sum_{(4)} [(6) + (9) + (11) + (12)] =  [(M_S^2 - 2 M_W^2)^2 - 32 M_W^4]\ \mathcal{K}(M_W, M_W, M_S)\nonumber\\
 && - 2 [(M_S^2 - M_W^2)^2 - 32 M_W^4]\ \mathcal{K}(0, M_W, M_S) + [M_S^4 - 32 M_W^4]\ \mathcal{K}(0, 0, M_S).
 \label{eq:K_int_11}
\end{eqnarray}
 
After having extracted the divergences, the next question is how to renormalise the 2-loop neutrino mass. The simple answer to this question is that, after having calculated the $\epsilon$-expansion of the total integral $\mathcal{I}$ from Eq.~\eqref{eq:I_DimReg}, we can simply drop all the divergent terms and only keep the ones which are finite in the limit $\epsilon \to 0$. This is easy to understand when taking into account that we can split all the divergent (denoted by $\Delta$) and finite (denoted by $F$) pieces into their high-energy (``UV'') and low-energy (``EFT'') contributions, such that any observable quantity $Q$ can be written as follows:
\begin{equation}
 Q = F_{\rm EFT} + F_{\rm UV} + \sum_{n=1}^N \frac{\Delta_n^{\rm EFT}}{\epsilon^n} + \sum_{m=1}^N \frac{\Delta_m^{\rm UV}}{\epsilon^m},
 \label{eq:renormalisation_1}
\end{equation}
where the divergences go up to $\mathcal{O}(1/\epsilon^N)$. The decisive point is that the low-energy EFT can only give us the low-energy finite ($F_{\rm EFT}$) and divergent ($\Delta_n^{\rm EFT}$) pieces. However, as long as we know that a renormalisable UV-completion exists, we know that the divergent pieces have to cancel at each order in $1/\epsilon$ [in the so-called minimal subtraction (MS) scheme]:
\begin{equation}
 \Delta_n^{\rm EFT} + \Delta_n^{\rm UV} = 0,\ \ \ \forall n.
 \label{eq:renormalisation_2}
\end{equation}
For our effective model, we are aware of at least two renormalisable UV-completions~\cite{Chen:2006vn,Gustafsson:2012vj,Gustafsson:2014vpa}, so that we can savely assume Eq.~\eqref{eq:renormalisation_2} to hold and thus drop the divergent pieces only (MS scheme). Furthermore, at low energies, practically all the relevant physics are covered by the EFT, so that one can safely neglect the suppressed finite terms from the high-energy part of the theory:
\begin{equation}
 F_{\rm UV} \simeq 0.
 \label{eq:renormalisation_3}
\end{equation}
Thus, one finally remains with the EFT finite piece $Q_{\rm physical} \simeq F_{\rm EFT}$, which will be a good approximation of the true result for energies below the UV-cutoff $\Lambda$ of the EFT. This is the simplest way to obtain sensible results.

\begin{figure}
\centering
\includegraphics[scale=0.47]{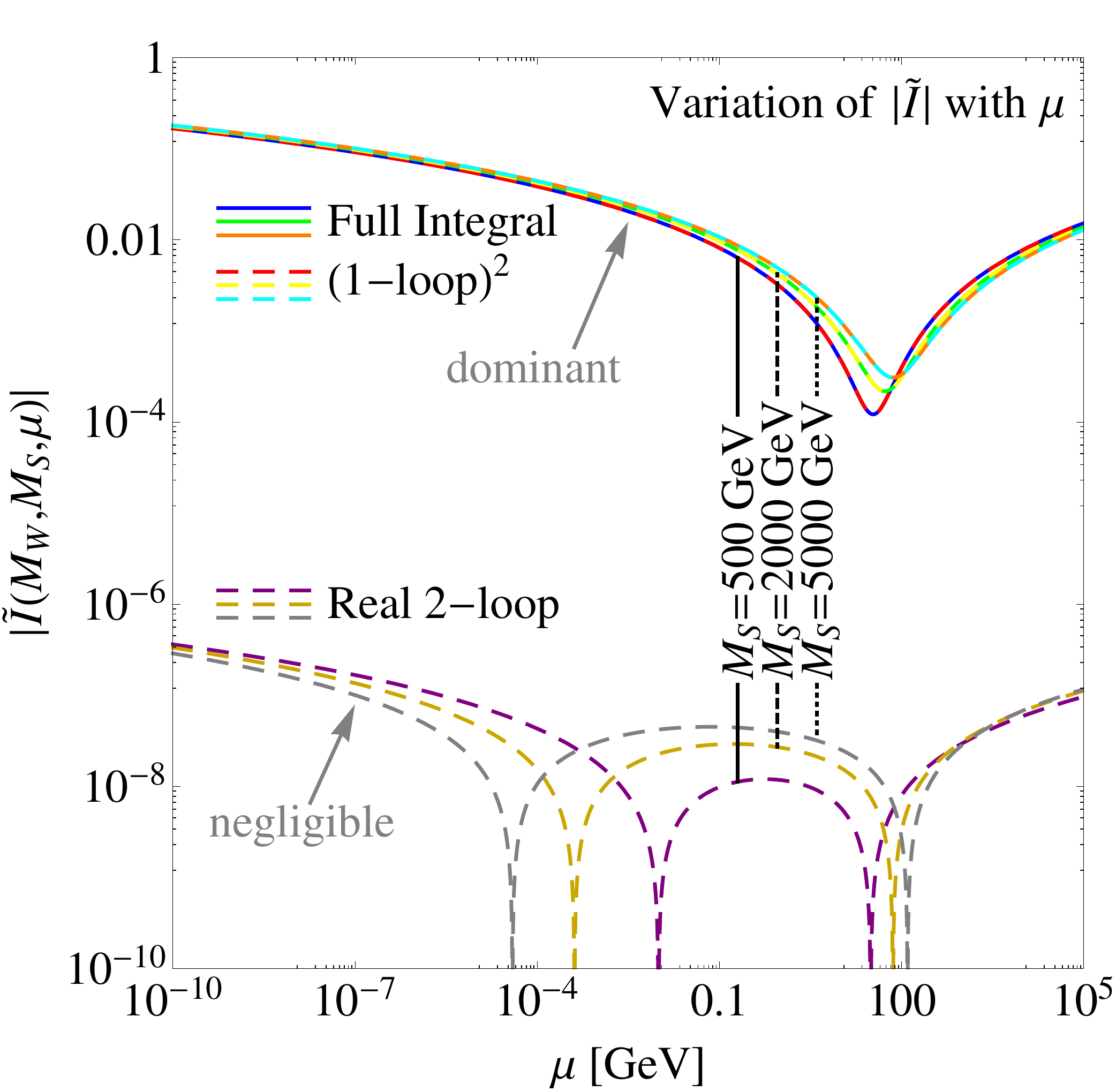}
\caption{\label{fig:Integral-Size}
The size of the integral $\mathcal{\tilde I}(M_W, M_S,\mu)$. Apparently, the decisive contribution comes from the products of effective 1-loop integrals. As can be seen from the plot, the variation with the dimensional regularisation scale $\mu$ is extremely mild.
}
\end{figure}

Taking only the finite pieces, the renormalised result for the integral $\mathcal{I}$, which only depends on the masses $M_W$ and $M_S$ as well as the unknown energy scale $\mu$, is given by
\begin{eqnarray}
 -4 (16 \pi^2)^2 \mathcal{I}_{\rm finite}^{\rm MS}(M_W, M_S,\mu) &=& (M_S^2 - 2 M_W^2) \left[ 2 (C_\gamma - L_W)^2 + \frac{\pi^2}{6} +1 \right]\nonumber\\
 && + 2 M_S^2 \left[ \frac{1}{2} (2 C_\gamma - L_W - L_S)^2 + \frac{\pi^2}{6} +1 \right]\nonumber\\
 && + [(\rho - 2)^2 - 32] \left.\left( P_0 C_0 + P_1 C_{-1} + P_2 C_{-2} \right)\right|_{\text{\eqref{eq:K_int_5},\eqref{eq:K_int_8}}}\nonumber\\
 && - 2 [(\rho - 1)^2 - 32] \left.\left( P_0 C_0 + P_1 C_{-1} + P_2 C_{-2} \right)\right|_{\text{\eqref{eq:K_int_5},\eqref{eq:K_int_9}}}\nonumber\\
 && + [\rho^2 - 32] \left.\left( P_0 C_0 + P_1 C_{-1} + P_2 C_{-2} \right)\right|_{\text{\eqref{eq:K_int_5},\eqref{eq:K_int_10}}},
 \label{eq:renormalisation_4}
\end{eqnarray}
where $\rho \equiv M_S^2 / M_W^2$, $L_W \equiv \log (M_W^2 / \mu^2)$, and $L_S \equiv \log (M_S^2 / \mu^2)$. The first (second) subscript refers to the equation from which the expressions for the $P_i$ ($C_i$) should be taken. Extracting the overall dependence on $M_S^2$, we finally arrive at the following decisive integral
\begin{equation}
 \mathcal{\tilde I}(M_W, M_S,\mu) = \mathcal{\tilde I}_1 + \mathcal{\tilde I}_2 + \mathcal{\tilde I}_3 + \mathcal{\tilde I}_4 + \mathcal{\tilde I}_5,
 \label{eq:renormalisation_5}
\end{equation}
with the different pieces given by
\begin{eqnarray}
 \mathcal{\tilde I}_1 &\equiv& \frac{-1}{4 (16 \pi^2)^2} \left( 1 - \frac{2}{\rho} \right) \left[ 2 (C_\gamma - L_W)^2 + \frac{\pi^2}{6} +1 \right], \nonumber\\
 \mathcal{\tilde I}_2 &\equiv& \frac{-1}{4 (16 \pi^2)^2} \left[ (2 C_\gamma - L_W - L_S)^2 + \frac{\pi^2}{3} +2 \right],\nonumber\\
 \mathcal{\tilde I}_3 &\equiv& \frac{-1}{4 (16 \pi^2)^2} [(\rho - 2)^2 - 32] \left.\left( \tilde P_0 C_0 + \tilde P_1 C_{-1} + \tilde P_2 C_{-2} \right)\right|_{\text{\eqref{eq:K_int_5},\eqref{eq:K_int_8}}},\nonumber\\
 \mathcal{\tilde I}_4 &\equiv& \frac{+1}{4 (16 \pi^2)^2} 2 [(\rho - 1)^2 - 32] \left.\left( \tilde P_0 C_0 + \tilde P_1 C_{-1} + \tilde P_2 C_{-2} \right)\right|_{\text{\eqref{eq:K_int_5},\eqref{eq:K_int_9}}},\nonumber\\
 \mathcal{\tilde I}_5 &\equiv& \frac{-1}{4 (16 \pi^2)^2} [\rho^2 - 32] \left.\left( \tilde P_0 C_0 + \tilde P_1 C_{-1} + \tilde P_2 C_{-2} \right)\right|_{\text{\eqref{eq:K_int_5},\eqref{eq:K_int_10}}},
 \label{eq:renormalisation_6}
\end{eqnarray}
where $\tilde P_i \equiv P_i/M_S^2$. The variation of $\mathcal{\tilde I}(M_W, M_S,\mu)$ with the energy scale $\mu$ is displayed in Fig.~\ref{fig:Integral-Size}. As can be seen, the contributions arising from products of effective 1-loop integrals ($\mathcal{\tilde I}_{1,2}$) by far dominate the contributions from irreducible 2-loop integrals ($\mathcal{\tilde I}_{3,4,5}$).\footnote{This is physically motivated, since on the one hand the large logarithms of the two individual effective 1-loop diagrams have a tendency to enhance each other and on the other hand there is an enhancement of these contributions stemming from the fact that the constant term $\frac{\pi^2}{6} + 1$ [cf.\ Eq.~\eqref{eq:J_int_2}], which is of $\mathcal{O(\epsilon)}$ and thus usually negligible in generic 1-loop integrals, can in the product lead to a finite contribution when multiplied with the divergent contribution $2/\epsilon$.} This allows to find an easy analytical approximation to the full integral,
\begin{eqnarray}
 \mathcal{\tilde I}(M_W, M_S,\mu) &\simeq& \mathcal{\tilde I}_1(M_W, M_S,\mu) + \mathcal{\tilde I}_2(M_W, M_S,\mu) \label{eq:int_approx}\\
 &=&  \frac{-1}{4 (16 \pi^2)^2} \left\{ \left( 1 - \frac{2}{\rho} \right) \left[ 2 (C_\gamma - L_W)^2 + \frac{\pi^2}{6} +1 \right] + (2 C_\gamma - L_W - L_S)^2 + \frac{\pi^2}{3} +2 \right\}. \nonumber
\end{eqnarray}

Then, the final result for the 2-loop neutrino mass matrix, as obtained in the MS scheme, is given by
\begin{equation}
 \mathcal{M}_{\nu, ab}^\textrm{2-loop} = \frac{2 \xi m_a m_b M_S^2 f_{ab} (1 + \delta_{ab})}{\Lambda^3} \cdot \mathcal{\tilde I}(M_W, M_S,\mu).
 \label{eq:Mnu_finite_APP}
\end{equation}
Note that the final result still carries a dependence on the unknown energy scale $\mu$, which is typically taken to be the energy scale of the problem. This comes from the fact that an EFT necessarily neglects some finite but small contributions which would allow to cancel the $\mu$-dependence, which is a typical phenomenon in effective pictures (see, e.g., Refs.~\cite{Pich:1998xt,Binoth:2012rjo}). However, this dependence is only logarithmic and, furthermore, it can be used to estimate the running of the neutrino mass.

This is, to our knowledge, the first time that a fully analytic expression for the 2-loop neutrino mass has been obtained for the class of models under consideration.

\section*{\label{sec:app_B}Appendix~B: Correlations for the purple and blue points}
\renewcommand{\theequation}{B-\arabic{equation}}
\setcounter{equation}{0}  

As shown in Tab.~\ref{tab:LFV}, the LFV transition $\mu \to e\gamma$ depends on a very specific combination of Yukawa couplings, $|f_{ee}^* f_{e\mu} + f_{e\mu}^* f_{\mu\mu} + f_{e\tau}^* f_{\mu\tau}|$. The red and purple points presented in this paper, cf.\ Sec.~\ref{sec:benchmarks}, fulfill $f_{ee} \simeq 0$, so that for  $\mu \to e\gamma$ the strong bound only pushes $|f_{e\mu}^* f_{\mu\mu} + f_{e\tau}^* f_{\mu\tau}|$ to have small values. If this combination of parameters is zero or close to zero, cf.\ Eq.~\eqref{eq:condition}, then the form of the light neutrino mass matrix as detailed in Eqs.~\eqref{eq:Mnu_finite} and~\eqref{eq:Mnu_finite_APP}, allows to translate the above condition into a correlation of light neutrino mass matrix elements,
\begin{equation}
 \mathcal{M}_{\nu, e\mu} \simeq - \frac{m_\mu^2}{m_\tau^2} \frac{\mathcal{M}_{\nu, \mu\tau}^*}{\mathcal{M}_{\nu, \mu\mu}^*} \mathcal{M}_{\nu, e\tau},
 \label{eq:correlation}
\end{equation}
which induces a connection between neutrino oscillation parameters and phases.

Expressing the light neutrino mass matrix in terms of mass eigenvalues $m_{1,2,3}$ and the complete set of mixing parameters, cf.\ Eqs.~\eqref{eq:diag} to~\eqref{eq:PMNS}, allows to express the mass matrix elements appearing in Eq.~\eqref{eq:correlation} in terms of physical parameters:
\begin{eqnarray}
 \mathcal{M}_{\nu, e\mu} &=& c_{13} \Bigg(\sqrt{m_1^2 + \Delta m^2_\odot} s_{12} e^{i \alpha_{21}} \left(c_{12} c_{23}-s_{12} s_{13} s_{23} e^{i \delta} \right) - m_1 c_{12} \left(c_{23} s_{12}+c_{12} s_{13} s_{23} e^{i \delta} \right)\nonumber\\
 && + \sqrt{m_1^2 + \Delta m^2_A} s_{13} s_{23} e^{i (\alpha_{31}-\delta )}\Bigg),\nonumber\\
 \mathcal{M}_{\nu, e\tau} &=& c_{13} \Bigg(- \sqrt{m_1^2 + \Delta m^2_\odot} s_{12} e^{i \alpha_{21}} \left(c_{12} s_{23}+c_{23} s_{12} s_{13} e^{i \delta}\right)+ \sqrt{m_1^2 + \Delta m^2_A} c_{23} s_{13} e^{i (\alpha_{31}-\delta )}\nonumber\\
 && + m_1 c_{12} \left(s_{12} s_{23}-c_{12} c_{23} s_{13} e^{i \delta}\right)\Bigg),\nonumber\\
 \mathcal{M}_{\nu, \mu\mu}^* &=& \sqrt{m_1^2 + \Delta m^2_\odot} e^{-i \alpha_{21}} \left(c_{12} c_{23}-s_{12} s_{13} s_{23} e^{-i \delta }\right)^2+ \sqrt{m_1^2 + \Delta m^2_A} c_{13}^2 s_{23}^2 e^{-i \alpha_{31}} \nonumber\\
 && + m_1 \left(c_{23} s_{12}+c_{12} s_{13} s_{23} e^{-i\delta}\right)^2,\nonumber\\
 \mathcal{M}_{\nu, \mu\tau}^* &=& c_{13} \Bigg(- \sqrt{m_1^2 + \Delta m^2_\odot} s_{12} e^{-i \alpha_{21}} \left(c_{12} s_{23}+c_{23} s_{12} s_{13} e^{-i \delta}\right)+ \sqrt{m_1^2 + \Delta m^2_A} c_{23} s_{13} e^{-i (\alpha_{31}-\delta)}\nonumber\\
 && + m_1 c_{12} \left(s_{12} s_{23}-c_{12} c_{23} s_{13} e^{-i \delta}\right)\Bigg).
 \label{eq:elements}
\end{eqnarray}
Apparently, the complex Eq.~\eqref{eq:correlation} translates into two real equations which allow to constrain two of the unknown neutrino-related parameters, e.g., the lightest neutrino mass and one of the Majorana phases.

\newpage
\section*{\label{sec:app_C}Appendix~C: Branching ratios for the different classes of benchmarks}

\begin{table}[!ht]
\tiny
\centering\begin{tabular}{c|cc|c|ccccccc|c}
\toprule
\midrule
\multicolumn{12}{c}{\bf\scriptsize Purple Points}\\
\midrule\midrule
\multirow{2}{*}{$M_S$ [GeV]} & \multicolumn{2}{c|}{\multirow{2}{*}{$\xi$}} & \multirow{2}{*}{$\Lambda/M_S$} & \multicolumn{7}{c|}{\bf Branching ratios} & \multirow{2}{*}{width [GeV]} \\
& \multicolumn{2}{c|}{} & & $WW$ & $ee$ & $e\mu$ & $e\tau$ & $\mu\mu$ & $\mu\tau$ & $\tau\tau$ & \\
\midrule
\multirow{20}{*}{200} & \multirow{6}{*}{$\begin{array}{c} \mbox{small} \\ (\xi\gtrsim0) \end{array}$} 
& 0.1 & 5 & 0 & 0 & 0.002 & 0.984 & 0.014 & 0 & 0 & 0.00132847 \\
&& 0.1 & 6 & 0 & 0 & 0.002 & 0.984 & 0.014 & 0 & 0 & 0.0039821 \\
&& 0.2 & 7 & 0 & 0 & 0.002 & 0.984 & 0.014 & 0 & 0 & 0.00249862 \\
&& 0.2 & 8 & 0 & 0 & 0.002 & 0.984 & 0.014 & 0 & 0 & 0.0056342 \\
&& 0.3 & 9 & 0 & 0 & 0.002 & 0.984 & 0.014 & 0 & 0 & 0.00507809 \\
&& 0.4 & 10 & 0 & 0 & 0.002 & 0.984 & 0.014 & 0 & 0 & 0.00534624 \\
\cmidrule{2-12}
&\multirow{6}{*}{$\begin{array}{c} \mbox{medium} \\ (\xi \simeq 2\pi) \end{array}$}
& 6 & 5 & 1 & 0 & 0 & 0 & 0 & 0 & 0 & 0.00105002 \\
&& 6 & 6 & 0.997 & 0 & 0 & 0.003 & 0 & 0 & 0 & 0.000350831 \\
&& 6 & 7 & 0.98 & 0 & 0 & 0.019 & 0 & 0 & 0 & 0.000142106 \\
&& 6 & 8 & 0.908 & 0 & 0 & 0.09 & 0.001 & 0 & 0 & 0.0000682291 \\
&& 6 & 9 & 0.712 & 0 & 0 & 0.283 & 0.004 & 0 & 0 & 0.000043683 \\
&& 6 & 10 & 0.412 & 0 & 0 & 0.579 & 0.008 & 0 & 0 & 0.000039957 \\
\cmidrule{2-12}
&\multirow{6}{*}{$\begin{array}{c} \mbox{large} \\ (\xi\lesssim 4\pi) \end{array}$}
& 12 & 5 & 1 & 0 & 0 & 0 & 0 & 0 & 0 & 0.00418131 \\
&& 12 & 6 & 1 & 0 & 0 & 0 & 0 & 0 & 0 & 0.00138856 \\
&& 12 & 7 & 0.999 & 0 & 0 & 0.001 & 0 & 0 & 0 & 0.000555644 \\
&& 12 & 8 & 0.994 & 0 & 0 & 0.006 & 0 & 0 & 0 & 0.000251118 \\
&& 12 & 9 & 0.975 & 0 & 0 & 0.024 & 0 & 0 & 0 & 0.000127217 \\
&& 12 & 10 & 0.893 & 0 & 0 & 0.105 & 0.002 & 0 & 0 & 0.0000735477 \\
\midrule
$\begin{array}{c} 364.6 \\ \mbox{(best-fit)} \end{array}$ & \multicolumn{2}{c|}{6.38} & 6.87 & 0.310 & 0 & 0.001 & 0.679 & 0.010 & 0 & 0 & 0.000178094 \\
\midrule
\multirow{20}{*}{600} & \multirow{6}{*}{$\begin{array}{c} \mbox{small} \\ (\xi\gtrsim0) \end{array}$} 
& 0.1 & 5 & 0 & 0 & 0.002 & 0.984 & 0.014 & 0 & 0 & 0.0479347 \\
&& 0.1 & 6 & 0 & 0 & 0.002 & 0.984 & 0.014 & 0 & 0 & 0.106938 \\
&& 0.2 & 7 & 0 & 0 & 0.002 & 0.984 & 0.014 & 0 & 0 & 0.0679208 \\
&& 0.2 & 8 & 0 & 0 & 0.002 & 0.984 & 0.014 & 0 & 0 & 0.151055 \\
&& 0.3 & 9 & 0 & 0 & 0.002 & 0.984 & 0.014 & 0 & 0 & 0.136088 \\
&& 0.4 & 10 & 0 & 0 & 0.002 & 0.984 & 0.014 & 0 & 0 & 0.144131 \\
\cmidrule{2-12}
&\multirow{6}{*}{$\begin{array}{c} \mbox{medium} \\ (\xi \simeq 2\pi) \end{array}$}
& 6 & 5 & 0.899 & 0 & 0 & 0.1 & 0.001 & 0 & 0 & 0.0000975578 \\
&& 6 & 6 & 0.5 & 0 & 0 & 0.492 & 0.007 & 0 & 0 & 0.0000595187 \\
&& 6 & 7 & 0.104 & 0 & 0.001 & 0.882 & 0.013 & 0 & 0 & 0.000112445 \\
&& 6 & 8 & 0.03 & 0 & 0.002 & 0.954 & 0.014 & 0 & 0 & 0.000173764 \\
&& 6 & 9 & 0.007 & 0 & 0.002 & 0.977 & 0.014 & 0 & 0 & 0.000344195 \\
&& 6 & 10 & 0.002 & 0 & 0.002 & 0.982 & 0.015 & 0 & 0 & 0.000854768 \\
\cmidrule{2-12}
&\multirow{6}{*}{$\begin{array}{c} \mbox{large} \\ (\xi\lesssim 4\pi) \end{array}$}
& 12 & 5 & 0.993 & 0 & 0 & 0.007 & 0 & 0 & 0 & 0.00035147 \\
&& 12 & 6 & 0.628 & 0 & 0 & 0.366 & 0.005 & 0 & 0 & 0.000185544 \\
&& 12 & 7 & 0.711 & 0 & 0 & 0.285 & 0.004 & 0 & 0 & 0.0000650648 \\
&& 12 & 8 & 0.335 & 0 & 0 & 0.654 & 0.01 & 0 & 0 & 0.0000626024 \\
&& 12 & 9 & 0.108 & 0 & 0.002 & 0.877 & 0.013 & 0 & 0 & 0.0000955724 \\
&& 12 & 10 & 0.034 & 0 & 0.002 & 0.951 & 0.014 & 0 & 0 & 0.000164169 \\
\midrule
\multirow{20}{*}{1000} & \multirow{6}{*}{$\begin{array}{c} \mbox{small} \\ (\xi\gtrsim0) \end{array}$} 
& 0.1 & 5 & 0 & 0 & 0.002 & 0.984 & 0.014 & 0 & 0 & 0.165629 \\
&& 0.1 & 6 & 0 & 0 & 0.002 & 0.984 & 0.014 & 0 & 0 & 0.500951 \\
&& 0.2 & 7 & 0 & 0 & 0.002 & 0.984 & 0.014 & 0 & 0 & 0.312644 \\
&& 0.2 & 8 & 0 & 0 & 0.002 & 0.984 & 0.014 & 0 & 0 & 0.70179 \\
&& 0.3 & 9 & 0 & 0 & 0.002 & 0.984 & 0.014 & 0 & 0 & 0.835827 \\
&& 0.4 & 10 & 0 & 0 & 0.002 & 0.984 & 0.014 & 0 & 0 & 0.667518 \\
\cmidrule{2-12}
&\multirow{6}{*}{$\begin{array}{c} \mbox{medium} \\ (\xi \simeq 2\pi) \end{array}$}
& 6 & 5 & 0.302 & 0 & 0.001 & 0.687 & 0.01 & 0 & 0 & 0.0000667239 \\
&& 6 & 6 & 0.047 & 0 & 0.001 & 0.938 & 0.013 & 0 & 0 & 0.000145385 \\
&& 6 & 7 & 0.008 & 0 & 0.002 & 0.977 & 0.014 & 0 & 0 & 0.000351838 \\
&& 6 & 8 & 0.002 & 0 & 0.002 & 0.983 & 0.014 & 0 & 0 & 0.000779377 \\
&& 6 & 9 & 0 & 0 & 0.002 & 0.984 & 0.014 & 0 & 0 & 0.00156814 \\
&& 6 & 10 & 0 & 0 & 0.002 & 0.984 & 0.014 & 0 & 0 & 0.00296203 \\
\cmidrule{2-12}
&\multirow{6}{*}{$\begin{array}{c} \mbox{large} \\ (\xi\lesssim 4\pi) \end{array}$}
& 12 & 5 & 0.876 & 0 & 0 & 0.122 & 0.002 & 0 & 0 & 0.0000925977 \\
&& 12 & 6 & 0.366 & 0 & 0 & 0.625 & 0.009 & 0 & 0 & 0.0000733874 \\
&& 12 & 7 & 0.11 & 0 & 0.001 & 0.876 & 0.013 & 0 & 0 & 0.00009764 \\
&& 12 & 8 & 0.018 & 0 & 0.001 & 0.965 & 0.015 & 0 & 0 & 0.000262922 \\
&& 12 & 9 & 0.006 & 0 & 0.002 & 0.978 & 0.014 & 0 & 0 & 0.000396615 \\
&& 12 & 10 & 0.001 & 0 & 0.001 & 0.983 & 0.014 & 0 & 0 & 0.000992832 \\
\bottomrule
\end{tabular}
\caption{Branching ratios and total widths for ``purple'' points at different $\SPP$ masses and different values of $\xi$ and $\Lambda/M_S$.}
\label{tab:BRpurple}
\end{table}

\begin{table}[!ht]
\tiny
\centering\begin{tabular}{c|cc|c|ccccccc|c}
\toprule
\midrule
\multicolumn{12}{c}{\bf\scriptsize Red Points}\\
\midrule\midrule
\multirow{2}{*}{$M_S$ [GeV]} & \multicolumn{2}{c|}{\multirow{2}{*}{$\xi$}} & \multirow{2}{*}{$\Lambda/M_S$} & \multicolumn{7}{c|}{\bf Branching ratios} & \multirow{2}{*}{width [GeV]} \\
& \multicolumn{2}{c|}{} & & $WW$ & $ee$ & $e\mu$ & $e\tau$ & $\mu\mu$ & $\mu\tau$ & $\tau\tau$ & \\
\midrule
$\begin{array}{c} 164.5 \\ \mbox{(best-fit)} \end{array}$ & \multicolumn{2}{c|}{5.02} & 5.50 & 0.581 & 0 & 0.418 & 0 & 0 & 0 & 0 & 0.000506941 \\
\midrule
\multirow{17}{*}{200} & \multirow{3}{*}{$\begin{array}{c} \mbox{small} \\ (\xi\gtrsim0) \end{array}$} 
& 0.5 & 5 & 0.001 & 0 & 0.999 & 0 & 0 & 0 & 0 & 0.00618056 \\
&& 0.8 & 6 & 0 & 0 & 0.999 & 0 & 0 & 0 & 0 & 0.00719317 \\
&& - & 7,\dots,10 & - & - & - & - & - & - & - & - \\
\cmidrule{2-12}
&\multirow{6}{*}{$\begin{array}{c} \mbox{medium} \\ (\xi \simeq 2\pi) \end{array}$}
& 6 & 5 & 0.904 & 0 & 0.096 & 0 & 0 & 0 & 0 & 0.00115604 \\
&& 6 & 6 & 0.558 & 0 & 0.442 & 0 & 0 & 0 & 0 & 0.000628695 \\
&& 6 & 7 & 0.22 & 0 & 0.78 & 0 & 0 & 0 & 0 & 0.000633603 \\
&& 6 & 8 & 0.035 & 0 & 0.965 & 0 & 0 & 0 & 0 & 0.00179675 \\
&& 6 & 9 & 0.008 & 0 & 0.992 & 0 & 0 & 0 & 0 & 0.00379126 \\
&& 6 & 10 & 0.002 & 0 & 0.998 & 0 & 0 & 0 & 0 & 0.00710341 \\
\cmidrule{2-12}
&\multirow{6}{*}{$\begin{array}{c} \mbox{large} \\ (\xi\lesssim 4\pi) \end{array}$}
& 12 & 5 & 0.995 & 0 & 0.005 & 0 & 0 & 0 & 0 & 0.00422159 \\
&& 12 & 6 & 0.945 & 0 & 0.055 & 0 & 0 & 0 & 0 & 0.00149688 \\
&& 12 & 7 & 0.729 & 0 & 0.271 & 0 & 0 & 0 & 0 & 0.000765041 \\
&& 12 & 8 & 0.35 & 0 & 0.65 & 0 & 0 & 0 & 0 & 0.0007174128 \\
&& 12 & 9 & 0.116 & 0 & 0.884 & 0 & 0 & 0 & 0 & 0.0010678 \\
&& 12 & 10 & 0.038 & 0 & 0.962 & 0 & 0 & 0 & 0 & 0.00171308 \\
\midrule
\multirow{17}{*}{600} & \multirow{3}{*}{$\begin{array}{c} \mbox{small} \\ (\xi\gtrsim0) \end{array}$} 
& 0.5 & 5 & 0 & 0 & 1 & 0 & 0 & 0 & 0 & 0.166463 \\
&& 0.8 & 6 & 0 & 0 & 1 & 0 & 0 & 0 & 0 & 0.19421 \\
&& - & 7,\dots,10 & - & - & - & - & - & - & - & - \\
\cmidrule{2-12}
&\multirow{6}{*}{$\begin{array}{c} \mbox{medium} \\ (\xi \simeq 2\pi) \end{array}$}
& 6 & 5 & 0.029 & 0 & 0.971 & 0 & 0 & 0 & 0 & 0.00307804 \\
&& 6 & 6 & 0.003 & 0 & 0.997 & 0 & 0 & 0 & 0 & 0.00891094 \\
&& 6 & 7 & 0 & 0 & 0.999 & 0 & 0 & 0 & 0 & 0.0225216 \\
&& 6 & 8 & 0 & 0 & 1 & 0 & 0 & 0 & 0 & 0.0500959 \\
&& 6 & 9 & 0 & 0 & 1 & 0 & 0 & 0 & 0 & 0.0855915 \\
&& 6 & 10 & 0 & 0 & 1 & 0 & 0 & 0 & 0 & 0.113601 \\
\cmidrule{2-12}
&\multirow{6}{*}{$\begin{array}{c} \mbox{large} \\ (\xi\lesssim 4\pi) \end{array}$}
& 12 & 5 & 0.318 & 0 & 0.682 & 0 & 0 & 0 & 0 & 0.00110414 \\
&& 12 & 6 & 0.05 & 0 & 0.95 & 0 & 0 & 0 & 0 & 0.00236156 \\
&& 12 & 7 & 0.014 & 0 & 0.986 & 0 & 0 & 0 & 0 & 0.00337974 \\
&& 12 & 8 & 0.002 & 0 & 0.998 & 0 & 0 & 0 & 0 & 0.0125286 \\
&& 12 & 9 & 0 & 0 & 1 & 0 & 0 & 0 & 0 & 0.0254247 \\
&& 12 & 10 & 0 & 0 & 1 & 0 & 0 & 0 & 0 & 0.0243301 \\
\midrule
\multirow{17}{*}{1000} & \multirow{3}{*}{$\begin{array}{c} \mbox{small} \\ (\xi\gtrsim0) \end{array}$} 
& 0.5 & 5 & 0 & 0 & 1 & 0 & 0 & 0 & 0 & 0.773024 \\
&& 0.8 & 6 & 0 & 0 & 1 & 0 & 0 & 0 & 0 & 0.903931 \\
&& - & 7,\dots,10 & - & - & - & - & - & - & - & - \\
\cmidrule{2-12}
&\multirow{6}{*}{$\begin{array}{c} \mbox{medium} \\ (\xi \simeq 2\pi) \end{array}$}
& 6 & 5 & 0.001 & 0 & 0.998 & 0 & 0 & 0 & 0 & 0.0138926 \\
&& 6 & 6 & 0 & 0 & 1 & 0 & 0 & 0 & 0 & 0.0246002 \\
&& 6 & 7 & 0 & 0 & 1 & 0 & 0 & 0 & 0 & 0.0618494 \\
&& 6 & 8 & 0 & 0 & 1 & 0 & 0 & 0 & 0 & 0.233317 \\
&& 6 & 9 & 0 & 0 & 1 & 0 & 0 & 0 & 0 & 0.471533 \\
&& 6 & 10 & 0 & 0 & 1 & 0 & 0 & 0 & 0 & 0.883051 \\
\cmidrule{2-12}
&\multirow{6}{*}{$\begin{array}{c} \mbox{large} \\ (\xi\lesssim 4\pi) \end{array}$}
& 12 & 5 & 0.023 & 0 & 0.977 & 0 & 0 & 0 & 0 & 0.00352694 \\
&& 12 & 6 & 0.003 & 0 & 0.997 & 0 & 0 & 0 & 0 & 0.010437 \\
&& 12 & 7 & 0 & 0 & 1 & 0 & 0 & 0 & 0 & 0.0260958 \\
&& 12 & 8 & 0 & 0 & 1 & 0 & 0 & 0 & 0 & 0.0582806 \\
&& 12 & 9 & 0 & 0 & 1 & 0 & 0 & 0 & 0 & 0.11688 \\
&& 12 & 10 & 0 & 0 & 1 & 0 & 0 & 0 & 0 & 0.221718 \\
\bottomrule
\end{tabular}
\caption{Branching ratios and total widths for ``red'' points at different $\SPP$ masses and different values of $\xi$ and $\Lambda/M_S$.}
\label{tab:BRred}
\end{table}

\begin{table}[!ht]
\tiny
\centering\begin{tabular}{c|cc|c|ccccccc|c}
\toprule
\midrule
\multicolumn{12}{c}{\bf\scriptsize Blue Points}\\
\midrule\midrule
\multirow{2}{*}{$M_S$ [GeV]} & \multicolumn{2}{c|}{\multirow{2}{*}{$\xi$}} & \multirow{2}{*}{$\Lambda/M_S$} & \multicolumn{7}{c|}{\bf Branching ratios} & \multirow{2}{*}{width [GeV]} \\
& \multicolumn{2}{c|}{} & & $WW$ & $ee$ & $e\mu$ & $e\tau$ & $\mu\mu$ & $\mu\tau$ & $\tau\tau$ & \\
\midrule
200 & \multicolumn{2}{c|}{any $\xi$} & 5,\dots,10 & - & - & - & - & - & - & - & - \\
\midrule
\multirow{16}{*}{600} & \multirow{4}{*}{$\begin{array}{c} \mbox{small} \\ (\xi\gtrsim0) \end{array}$} 
& 0.4 & 5 & 0 & 0.828 & 0 & 0.169 & 0.003 & 0 & 0 & 0.00789999 \\
&& 0.5 & 6 & 0 & 0.829 & 0 & 0.168 & 0.003 & 0 & 0 & 0.0152079 \\
&& 0.7 & 7 & 0 & 0.83 & 0 & 0.168 & 0.003 & 0 & 0 & 0.0196076 \\
&& - & 8,9,10 & - & - & - & - & - & - & - & - \\
\cmidrule{2-12}
&\multirow{6}{*}{$\begin{array}{c} \mbox{medium} \\ (\xi \simeq 2\pi) \end{array}$}
& - & 5 & - & - & - & - & - & - & - & - \\
&& 6 & 6 & 0.218 & 0.648 & 0 & 0.132 & 0.002 & 0 & 0 & 0.000133869 \\
&& 6 & 7 & 0.042 & 0.794 & 0 & 0.162 & 0.002 & 0 & 0 & 0.000277175 \\
&& 6 & 8 & 0.009 & 0.821 & 0 & 0.168 & 0.003 & 0 & 0 & 0.000596307 \\
&& 6 & 9 & 0.002 & 0.827 & 0 & 0.168 & 0.003 & 0 & 0 & 0.00121088 \\
&& 6 & 10 & 0 & 0.827 & 0 & 0.17 & 0.003 & 0 & 0 & 0.00225081 \\
\cmidrule{2-12}
&\multirow{5}{*}{$\begin{array}{c} \mbox{large} \\ (\xi\lesssim 4\pi) \end{array}$}
& - & 5,6 & - & - & - & - & - & - & - & - \\
&& 12 & 7 & 0.41 & 0.489 & 0 & 0.099 & 0.002 & 0 & 0 & 0.000113483 \\
&& 12 & 8 & 0.124 & 0.725 & 0 & 0.148 & 0.002 & 0 & 0 & 0.000169039 \\
&& 12 & 9 & 0.033 & 0.801 & 0 & 0.163 & 0.002 & 0 & 0 & 0.000311038 \\
&& 12 & 10 & 0.01 & 0.821 & 0 & 0.166 & 0.003 & 0 & 0 & 0.000570635 \\
\midrule
$\begin{array}{c} 626.01 \\ \mbox{(best-fit)} \end{array}$ & \multicolumn{2}{c|}{3.39} & 8.14 & 0 & 0.986 & 0 & 0.014 & 0 & 0 & 0 & 0.0285773 \\
\midrule
\multirow{18}{*}{1000} & \multirow{4}{*}{$\begin{array}{c} \mbox{small} \\ (\xi\gtrsim0) \end{array}$} 
& 0.3 & 5 & 0 & 0.829 & 0 & 0.169 & 0.003 & 0 & 0 & 0.0652925 \\
&& 0.5 & 6 & 0 & 0.829 & 0 & 0.168 & 0.003 & 0 & 0 & 0.0703724 \\
&& 0.7 & 7 & 0 & 0.83 & 0 & 0.168 & 0.003 & 0 & 0 & 0.0907368 \\
&& - & 8,9,10 & - & - & - & - & - & - & - & - \\
\cmidrule{2-12}
&\multirow{6}{*}{$\begin{array}{c} \mbox{medium} \\ (\xi \simeq 2\pi) \end{array}$}
& 6 & 5 & 0 & 0.999 & 0 & 0 & 0 & 0 & 0 & 0.0343403 \\
&& 6 & 6 & 0 & 0.999 & 0 & 0 & 0 & 0 & 0 & 0.10269 \\
&& 6 & 7 & 0 & 0.991 & 0 & 0.009 & 0 & 0 & 0 & 0.0297808 \\
&& 6 & 8 & 0 & 0.998 & 0 & 0.002 & 0 & 0 & 0 & 0.104177 \\
&& 6 & 9 & 0 & 0.827 & 0 & 0.17 & 0.003 & 0 & 0 & 0.00552482 \\
&& 6 & 10 & 0 & 0.828 & 0 & 0.169 & 0.003 & 0 & 0 & 0.010444 \\
\cmidrule{2-12}
&\multirow{6}{*}{$\begin{array}{c} \mbox{large} \\ (\xi\lesssim 4\pi) \end{array}$}
& 12 & 5 & 0.009 & 0.99 & 0 & 0 & 0 & 0 & 0 & 0.00862917 \\
&& 12 & 6 & 0.001 & 0.998 & 0 & 0 & 0 & 0 & 0 & 0.0256054 \\
&& 12 & 7 & 0 & 0.999 & 0 & 0 & 0 & 0 & 0 & 0.0648168 \\
&& 12 & 8 & 0 & 0.999 & 0 & 0 & 0 & 0 & 0 & 0.143218 \\
&& 12 & 9 & 0 & 0.934 & 0 & 0.065 & 0.001 & 0 & 0 & 0.00858164 \\
&& 12 & 10 & 0 & 0.998 & 0 & 0.002 & 0 & 0 & 0 & 0.0986548 \\
\bottomrule
\end{tabular}
\caption{Branching ratios and total widths for ``blue'' points at different $\SPP$ masses and different values of $\xi$ and $\Lambda/M_S$.}
\label{tab:BRblue}
\end{table}

\clearpage
\bibliographystyle{./JHEP}
\bibliography{./Nu_LHC}

\end{document}